\documentclass[11pt,a4paper]{article}
\pdfoutput=1

\usepackage{bm}
\usepackage{amsmath,amssymb}
\usepackage{cite}

\usepackage{graphicx}
\usepackage{color}

\usepackage{hyperref} 
\hypersetup{colorlinks=true, citecolor=blue, filecolor=black, linkcolor=blue, urlcolor=blue, pdfpagemode=UseNone}

\numberwithin{figure}{section}
\numberwithin{equation}{section}

\newcommand{\be}{\begin{equation}}
\newcommand{\ee}{\end{equation}}
\newcommand{\bea}{\begin{eqnarray}}
\newcommand{\eea}{\end{eqnarray}}

\usepackage[normalem]{ulem}

\newcommand{\vp}{\varphi}

\newcommand{\Vh}{\hat V}
\newcommand{\Kh}{\hat K}
\newcommand{\ph}{\hat \phi}
\newcommand{\kh}{\hat k}
\newcommand{\gmn}{g_{\mu\nu}}

\newcommand{\bgmn}{\bar g_{\mu\nu}}

\newcommand{\dclnf}{\,\partial_\chi\! \ln\! f \,}

\newcommand{\dv}{\mathfrak{v}}
\newcommand{\dw}{\mathfrak{w}}
\newcommand{\dV}{\mathcal{V}}
\newcommand{\dK}{\mathcal{K}}

\newcommand{\cR}{\mathcal{R}}
\newcommand{\qR}{\mathcal{R}_{\rm quant}}
\newcommand{\F}{\mathcal{F}^{(s)}}
\newcommand{\Fp}{\mathcal{F}^{(s)\prime}}

\newcommand\ie{\textit{i.e.}\ }
\newcommand\eg{\textit{e.g.}\ }
\newcommand\cf{\textit{cf.}\ }

\newcommand{\aka}{{a.k.a.}\ }

\newcommand{\half}{\tfrac{1}{2}}

\newcommand{\eps}{\varepsilon}

\begin{document}

\begin{titlepage}

\begin{center}
{\huge \bf Fixed point structure of the conformal factor field in quantum gravity} 
\end{center}
\vskip1cm


\begin{center}
{\bf Juergen A. Dietz, Tim R. Morris and Zo\"e H. Slade }
\end{center}

\begin{center}
{\it STAG Research Centre \& Department of Physics and Astronomy,\\  University of Southampton,
Highfield, Southampton, SO17 1BJ, U.K.}\\
\vspace*{0.3cm}
{\tt  J.A.Dietz@soton.ac.uk,
T.R.Morris@soton.ac.uk, Z.Slade@soton.ac.uk}
\end{center}

\abstract{The $\mathcal{O}(\partial^2)$
background independent flow equations for conformally reduced gravity are shown to be equivalent to flow equations naturally adapted to scalar field theory with a wrong sign kinetic term. This sign change is shown to have a profound effect on the renormalization group properties, broadly resulting  in a continuum of fixed points supporting both a discrete and a continuous eigenoperator spectrum, the latter always including relevant directions. The properties at the Gaussian fixed point are understood in particular depth, but also detailed studies of the local potential approximation, and the full $\mathcal{O}(\partial^2)$ approximation are given.
These results are related to evidence for asymptotic safety found by other authors.}

\end{titlepage}

\tableofcontents

\newpage

\section{Introduction}
\label{sec:Intro}

When applied to quantum gravity, asymptotic safety is the idea that  the renormalization group (RG) flow of gravitational couplings approaches a viable interacting non-perturbative fixed point in the far ultraviolet, such that physical observables are rendered ultraviolet finite despite perturbative non-renormalisability \cite{Weinberg:1980}.
Ever since a functional (\aka ``exact'' \cite{Wilson:1973})
RG equation adapted to this case, was put forward in ref. \cite{Reuter:1996}, a steady increase of interest in the asymptotic safety programme for quantum gravity has produced a wealth of results which so far paint an overall promising picture. For reviews and introductions see \cite{Reuter:2012,Percacci:2011fr,Niedermaier:2006wt,Nagy:2012ef,Litim:2011cp}. 

One apparent advantage of such an approach was already pointed out in ref. \cite{Reuter:1996}.  The Euclidean signature functional integral for the Einstein-Hilbert action suffers from the well known conformal factor problem \cite{Gibbons:1978ac}, which is that the negative sign for the kinetic term of the conformal factor, $\phi(x)$, yields a wrong-sign Gaussian destroying convergence of the integral. On the other hand providing the cutoff is adapted, 
the change in sign ``seems not to pose any special problem'' for the exact RG flow equation \cite{Reuter:1996}. As we will see in this paper, this one sign change however has profound consequences for the RG properties of the solutions, broadly resulting  in a continuum of fixed points supporting both a discrete and a continuous eigenoperator spectrum, the latter always including relevant directions. In the following we will review the exact RG approaches to asymptotic safety only in as much as to highlight how these effects have been overlooked until now and to highlight the technical developments that have been necessary in order to clearly uncover them.



Given that at first sight there is no special problem, the complexity of the extra technology and approximations necessary to make progress with such a functional RG approach to quantum gravity (many already developed in ref. \cite{Reuter:1996}) obscures these effects.
In brief, in order to adapt the infrared cutoff employed in constructing the flow equation \cite{Nicoll1977,Wetterich:1992,Morris:1993}, the background field method is employed and thus the full metric $g_{\mu\nu}$ and a background metric $\bar{g}_{\mu\nu}$ are introduced. Gauge fixing and infrared cutoff terms are introduced in a way that leaves the diffeomorphism invariance for the background metric undisturbed. The gauge fixing requires ghosts, which must themselves be regulated with background covariant cutoff terms. In almost all works further fields are then introduced in order to re-express  the fluctuation in a transverse-traceless decomposition which facilitates the computation of the inverse Hessian involved in constructing the flow equations, and these fields must be similarly treated. Also to facilitate this computation, the cutoff terms are introduced typically in some way which is adapted to the form of the Hessian. In standard fashion, diffeomorphism invariance of the total metric $\gmn$ becomes BRS invariance, which however is broken by the cutoff. In principle it can be recovered once the flow is complete, providing modified Ward identities are satisfied \cite{Reuter:1996}. 

Physics should depend only on the full metric, and not also on the background metric $\bgmn$ that was introduced by hand as part of the background field technique. We will call this the requirement of \emph{background independence}. In the literature the construction of the effective action about a general background metric $\bgmn$, and thus computing in effect on all backgrounds simultaneously, is also referred to as background independence. As explained in ref. \cite{Reuter:2008qx}, this usage follows that in loop quantum gravity \cite{Ashtekar:1991hf,Ashtekar:2004eh,Rovelli:2004tv,Thiemann:2007zz}. However as emphasised in ref. \cite{Dietz:2015owa}, background independence in the sense we mean it, is much more than this, and in fact is a strong \emph{extra} constraint. This requirement can in principle also be recovered providing certain msWI (modified split Ward identities) are satisfied \cite{Pawlowski:2005xe,Litim:2002hj,Bridle:2013sra,Reuter:1997gx,Litim:1998nf,Litim:2002ce,Manrique:2009uh,Manrique:2010mq,Manrique:2010am,Becker:2014qya} (see however the discussion in the conclusions of ref. \cite{Labus:2016lkh}). 

Finally in order to actually calculate anything, some approximations have to be made. Of these, of most interest to our discussion are the so-called single-metric approximation, and what we will refer to as `polynomial truncations'. The former approximation amounts to identifying $\gmn$ and $\bgmn$ at an appropriate point in the calculation. The latter approximation results from retaining only a finite number of operators in the effective action. These two approximations are not always made, but almost without exception one or other approximation is  made, and both contribute to obscuring the consequences of the wrong sign kinetic term for the conformal factor.

Clearly in order to expose these consequences, it helps to concentrate on this component of the metric alone. This is known in the literature as conformally reduced quantum gravity. A small number of works have studied this using the exact RG, starting with ref.  \cite{Reuter:2008wj}. In fact in this reference only the CREH (Conformally Reduced Einstein-Hilbert) truncation was actually computed. This is an example of a polynomial truncation. As we show explicitly in sec. \ref{sec:truncations}, the problem with this type of truncation is that by construction they can only give isolated fixed points with a quantized eigenoperator spectrum, and it is thus very difficult to ascertain the true situation this way. In ref. \cite{Reuter:2008qx} a full LPA (Local Potential Approximation) is derived for the conformal factor field. This functional truncation keeps a general potential for the field and incorporates infinitely many operators. In the Taylor expansion these are all positive integer powers of the field. Such an approximation therefore overcomes the limitations of the polynomial truncations but however, background independence (in the sense we mean it) was not incorporated, which means that the equations have a separate dependence on two fields: $\phi$ and also its background value $\chi$.\footnote{The situation is further obscured by there being no unique way to fix the anomalous dimension, $\eta$, leading to dependence on some arbitrary value of the field $\phi_1$. This should be contrasted with the treatment here and in ref. \cite{Dietz:2015owa} as discussed later.} Nevertheless some indication of there being an infinite number of relevant directions was uncovered \cite{Reuter:2008qx}. Finally in ref. \cite{Manrique:2009uh} not only was an LPA approximation derived but also the msWI that imposes background independence. 
Unfortunately, as discussed in ref. \cite{Dietz:2015owa} (see also \cite{Labus:2016lkh}), the msWI and flow equation derived there were not compatible with each other and furthermore again only polynomial truncations were actually computed. In a separate development \cite{Machado:2009ph,Percacci:2011uf,Codello2013,Pagani:2013fca}, 
the conformal factor is involved although not explicitly. Instead the degrees of freedom are duplicated by introducing a ``dilaton'' (also known as a spurion or compensator field) in order investigate the r\^ole of Weyl invariance, and also all other components of the metric are included. Furthermore,  single metric type approximations are made, and apart from the non-local Riegert action \cite{Machado:2009ph}
(which reproduces the trace anomaly) only polynomial truncations are considered. Finally in ref. \cite{Bonanno:2012dg} the single metric type approximation is again considered, and furthermore the exact RG is replaced with a ``proper time flow''. Again the focus is on the CREH, but the significance of the flow equations for the conformal factor being of backward parabolic type is realised and investigated within a more general LPA setting. We will come back to this observation in sec. \ref{sec:sft}. As we will see however, the wrong-sign kinetic term actually has more immediate and profound effects on the properties of the fixed points themselves and their eigen-spectra, as we have already mentioned.

In fact a continuum of fixed points and a continuous eigenoperator spectrum have already been found in full quantum gravity calculations in the so-called $f(R)$ approximation \cite{DietzMorris:2013-1}. But it was possible to blame this on a break-down of such an LPA-type approximation \cite{DietzMorris:2013-2} and on the use of the single field approximation \cite{Bridle:2013sra,Dietz:2015owa}. Furthermore, as we discuss in the conclusions, the resulting background dependence of this type of approximation obscures the significance of the large $R$ asymptotic behaviour, and from the scalar field study in ref.  \cite{Bridle:2013sra} it is particularly clear that the single field approximation introduces spurious effects and leads to significantly more complicated equations which obscure the basic structure. 

These problems are overcome for conformally truncated gravity in ref. \cite{Dietz:2015owa}. Guided by a remnant of background diffeomorphism invariance, flow equations and msWI are derived which can be compatible with each other after derivative expansion approximation \cite{Labus:2016lkh}. Indeed compatibility is shown to hold for general anomalous dimension, $\eta$,  if and only if power-law cutoff profiles are used \cite{Labus:2016lkh}. Such cutoff profiles have another advantage in that they preserve a reparametrisation invariance \cite{WR,Bell1975,Riedel:1986re} turning the fixed point equations into non-linear eigenvalue equations for $\eta$ \cite{WR,Morris:1994ie,Morris:1994jc,Morris:1998,Bervillier:2013kda,Delamotte:2015aaa}, and thus removing one further arbitrariness in these approximations. Finally, background independence is achieved for conformally truncated gravity for slow background field $\chi$ (equivalent to LPA) and fully at $\mathcal{O}(\partial^2)$ for the fluctuation field, and for general choice of parametrisation, $f(\phi)$, of the conformal factor. Taking the approximation fully to $\mathcal{O}(\partial^2)$ in this way avoids the traps of polynomial truncations, and ensures that break down of the LPA \cite{DietzMorris:2013-2} is avoided, although as we will see, such a break down does not in any case take place here. It also allows us to explore  the equations for general $\eta$ in a context where the corresponding scalar equations result in unique values for critical exponents including $\eta$, and the scaling equation of state \cite{Morris:1996xq}, which are in good agreement with other approaches \cite{Morris:1994ie,Morris:1994jc,Morris:1998}. As shown in ref. \cite{Dietz:2015owa}, the resulting equations can be combined together and recast in terms of background-independent variables, whereupon not only does all dependence on $\chi$ disappear but also all dependence on the form of parametrisation $f$. 

The underlying simplicity of the result is highlighted by the fact that the background independent flow equations are in fact equivalent to flow equations naturally adapted to scalar field theory with a wrong sign kinetic term, as we will see in sec. \ref{sec:sft}. 
The consequences for the eigenspectrum are then particularly transparent for the Gaussian fixed point (and furthermore independent of cutoff profile). Likewise the reason for a continuum of fixed points  is particularly clear from an asymptotic analysis of the LPA fixed point equation. (Furthermore this is demonstrated for general cutoff profile, spacetime and field dimensions.)
For this reason secs. \ref{sec:gaussian} and \ref{sec:LPA} form central parts of the paper.  
The LPA with $\eta=0$ can be solved completely analytically. We derive the continuum of fixed points, each supporting continuous eigenspectra, in sec. \ref{sec:confLPA}. This behaviour is established for the full $\mathcal{O}(\partial^2)$ equations in sec. \ref{sec:fpbeyondLPA} by a combination of numerical analysis and analytical asymptotic analysis, except for a small region $\eta\in\cR$ where likely there are no solutions. Asymptotic analysis is  used in sec. \ref{sec:eigen-asymptotics} to establish that also at $\mathcal{O}(\partial^2)$ the eigenoperator spectrum has a continuous part. We see this already in sec. \ref{sec:gaussian}, but here it is established for all the continuum of fixed point solutions at this level. In sec. \ref{sec:truncations}, polynomial truncations are considered, and finally in sec. \ref{sec:conclusions} we present our conclusions. This last section is likewise one of the central parts of the paper. It starts with a potted summary of the main findings,  discussing also their significance and highlighting possible extensions, and ends with a detailed discussion of how these findings fit with the existing literature. 

We start however with a miniature review of the results from refs. \cite{Dietz:2015owa,Labus:2016lkh}, sufficient for the rest of the paper.

\section{Review}
\label{sec:review}

In this section we very briefly review some of the results from refs. \cite{Dietz:2015owa,Labus:2016lkh} so that we can set out the notation and equations we will need. We will set out the equations in $d$ dimensions, but we will then mostly specialise to $d=4$ dimensions (in particular whenever we derive concrete solutions). We work in Euclidean signature with a conformally truncated metric $\gmn = f(\phi)\, \delta_{\mu\nu}$.  Here $\phi$ is the total conformal factor field and $f$ is some choice of parametrisation which is left arbitrary. 
The background field method is employed, with the background metric set equal to $\bgmn = f(\chi)\, \delta_{\mu\nu}$. The fluctuation conformal factor field is $\vp = \phi-\chi$. 
In order to truncate the effective action of the conformal factor $\Gamma_k[\vp,\chi]$ we make use of a derivative expansion. This is an expansion scheme that has proved successful in applications of the functional RG to other quantum field theories such as scalar field theory, \eg \cite{Morris:1994ie}. 

We simplify matters by specialising to 
a slow background field such that $\partial_\mu\chi$ is neglected.
We are then able to preserve a remnant diffeomorphism invariance (scaling of the coordinates) which is sufficient at this level of approximation to fix how $f(\chi)$ must appear, in much the same way that appearances of the background metric $\bgmn$ are fixed by full diffeomorphism invariance in the flow equation for full quantum gravity \cite{Dietz:2015owa}. The effective action then takes the form:
\be
\label{equ:ansatzGamma}
\Gamma_k[\vp,\chi] = \int d^dx f(\chi)^\frac{d}{2}\left(-\frac{1}{2}\frac{K(\vp,\chi)}{f(\chi)}\left(\partial_\mu \vp\right)^2
		      +V(\vp,\chi)\right)
\ee
in which we keep a general scalar potential $V(\vp,\chi)$ at zeroth order of the derivative expansion and a general scalar function $K(\vp,\chi)$ at $\mathcal{O}\big(\partial^2\big)$ for the fluctuation field $\vp$. It is understood that both of them depend on the RG time $t=\ln(k/\mu)$ ($\mu$ is a fixed physical mass scale). 

The infrared cutoff $R(p^2/f)$ depends on the background field $\chi$ as required by (remnant) background diffeomorphism invariance.\footnote{Here and in the ensuing, $f$ without qualification means $f(\chi)$.} However this means that the flow equation and thus effective action depend separately on both $\chi$ and $\vp$. The fact that the underlying theory depends only on the total field $\phi$ is expressed through a modified split Ward identity (msWI) which is derived through considering the breaking of the `split' invariance $\varphi(x)\mapsto\varphi(x)+\epsilon(x)$, $\chi(x)\mapsto\chi(x)-\epsilon(x)$. Providing the flow equations and the msWI are compatible,  imposing the msWI at any scale $k$ then ensures that split invariance is recovered in the limit $k\to0$. At the exact level the equations are automatically compatible but this needs to be verified once approximations are made \cite{Labus:2016lkh}.

After performing the derivative expansion of the flow equation and the msWI, 
we tidy up the equations a little by redefining
\begin{equation}
\label{redefine}
 V \mapsto \frac{\tilde \Omega_{d-1}}{2}V, \qquad K \mapsto \frac{\tilde \Omega_{d-1}}{2}K, \qquad R\mapsto \frac{\tilde \Omega_{d-1}}{2}R\,,
\end{equation}
where the constant $\tilde \Omega_{d-1}= \Omega_{d-1}/(2\pi)^d$, and $\Omega_{d-1}$ is the surface area of the $(d-1)$-dimensional sphere.
Then the flow equation and msWI for the potential can be written
\begin{subequations}
\label{equ:sysV-resc}
\begin{align}
 \partial_t V(\vp,\chi) &= f(\chi)^{-\frac{d}{2}}\int dp \,p^{d-1}\,Q_0\,\dot R, \label{equ:flowV-resc}\\
 \partial_\chi V- \partial_\vp V+\frac{d}{2}\dclnf V
&= f(\chi)^{-\frac{d}{2}}\int dp\, p^{d-1} \,Q_0\left[\partial_\chi R+\frac{d}{2}\dclnf R\right], \label{equ:sWIV-resc}
\end{align}
\end{subequations}
where we have made use of the Hessian at zeroth order of the derivative expansion:
\be 
\label{equ:Q0}
Q_0 = \left[\partial_\vp^2 V - K p^2/f+R(p^2/f) \right]^{-1} \,.
\ee
In the same way, the flow equation and msWI for the kinetic term are
\begin{subequations}
\label{equ:sysK-resc}
\begin{align}
 f^{-1}\partial_t K(\vp,\chi) &= 2 f^{-\frac{d}{2}}\int dp\, p^{d-1}\, P\left(p^2,\vp,\chi\right) \dot R, \label{equ:flowK-resc}\\
 f^{-1}\left\{\partial_\chi K - \partial_\vp K +\frac{d-2}{2}\dclnf K\right\}
&= 2f^{-\frac{d}{2}}\int dp\, p^{d-1} \,P(p^2,\vp,\chi)\left[\partial_\chi R+\frac{d}{2}\dclnf R\right]\,, \label{equ:sWIK-resc}
\end{align}
\end{subequations}
where $P$ is given by
\begin{align}
\label{equ:P}
 P =& -\frac{1}{2}\frac{\partial_\vp^2 K}{f}Q_0^2 + \frac{\partial_\vp K}{f}\left(2\partial_\vp^3 V-\frac{2d+1}{d}\frac{\partial_\vp K}{f}p^2\right) Q_0^3 \\ \notag
			  &  -\left[\left\{\frac{4+d}{d}\frac{\partial_\vp K}{f}p^2-\partial_\vp^3V\right\}\left(\partial_{p^2}R-\frac{K}{f}\right) + \frac{2}{d}p^2\partial_{p^2}^2 R\left(\frac{\partial_\vp K}{f}p^2-\partial_\vp^3 V\right)\right]\left(\partial_\vp^3 V -\frac{\partial_\vp K}{f}p^2\right) Q_0^4 \\ \notag
 &-\frac{4}{d}p^2\left(\partial_{p^2} R-\frac{K}{f}\right)^2\left(\partial_\vp^3V-\frac{\partial_\vp K}{f}p^2\right)^2 Q_0^5\,.
\end{align}
In fact the msWIs in the derivative expansion approximation above are compatible with the flow equations if and only if one of two conditions are met: (1) either the anomalous scaling dimension of the fields happens to vanish or (2) as functions of $p$ \cite{Labus:2016lkh},
\be 
{\dot R} \propto \left[\partial_\chi R+\frac{d}{2}\dclnf R\right]\,.
\ee
Furthermore there are no solutions to the combined system unless the msWIs are compatible with the flow  \cite{Labus:2016lkh}. Since we will mostly be interested in the case where the anomalous dimension is non-vanishing we will restrict the cutoff profile to satisfy the above condition.\footnote{The alternative case (1)  is studied in ref. \cite{Labus:2016lkh} and shown to lead to closely similar results for general cutoff profile.}
Using dimensional analysis \cite{Labus:2016lkh,Dietz:2015owa}, we find that this is ensured if 
\begin{equation}
\label{equ:cutoff}
 R\left(p^2/f\right) = - k^{d-\eta-\frac{d}{2}d_f}\,r\!\left(\frac{p^2}{k^{2-d_f}f}\right)\,,
\end{equation}
with 
\begin{equation}
 \label{equ:cutoff-pwrlw}
 r(z) = \frac{1}{z^n}\,,
\end{equation}
where 
$n$ is chosen to be an integer. We have taken the scaling dimension of $f$ to be $d_f$. Although classically the conformal factor is  naturally dimensionless, we allow for an anomalous scaling dimension $[\vp]=[\chi]=\eta/2$.
To ensure finiteness of the integrals on the right hand sides of \eqref{equ:sysV-resc} and \eqref{equ:sysK-resc}, the exponent $n$ has to be chosen such that $n>d/2-1$, \cf \cite{Morris:1994ie}. From \eqref{equ:cutoff} we also need to ensure that
\be
\label{bad-n}
n\ne\frac{\eta}{2-d_f}-\frac{d}{2}\,,
\ee
otherwise $R$ becomes independent of $k$. The respective flow equations and msWIs can then be combined into linear partial differential equations which can be solved to yield background independent variables \cite{Dietz:2015owa}:
\begin{equation}
 \label{equ:chvars}
 V(k,\vp,\chi) = f(\chi)^{-\frac{d}{2}}\tilde V(\tilde k,\phi), \qquad K(k,\vp,\chi) = f(\chi)^{-\frac{d}{2}+1}\tilde K(\tilde k,\phi), \qquad \tilde k = k f(\chi)^\frac{1}{\alpha}\,,
\end{equation}
where $\phi=\vp+\chi$ is the total field. The constant $\alpha$ is given by
\begin{equation}
\label{eq:alpha}
 \alpha = 
 2\left(1-\frac{\eta}{d+2n}\right)-d_f\,,\qquad\alpha\ne0\,,
\end{equation}
where the inequality follows from \eqref{bad-n}. Defining a mass dimension one background independent scale
\be
\label{khat}
\kh = \tilde k^\frac{\alpha}{\alpha+d_f}=k^\frac{\alpha}{\alpha+d_f}f(\chi)^\frac{1}{\alpha+d_f}\,,
\ee
we can define dimensionless background independent quantities 
via
\begin{equation}
\label{equ:dimless}
 \tilde V = \kh^{d}\,\Vh, \qquad \tilde K = \kh^{d-2-\eta}\Kh, \qquad \phi = \kh^{\frac{\eta}{2}}\ph, \qquad p = \kh\,\hat p\,.
\end{equation}
Note that if $\big[\tilde k    \big]=0$ we cannot make variables dimensionless by using $\tilde{k}$, and the definition of $\hat k$ then makes no sense. Since making the variables dimensionless is equivalent to the blocking step in the Wilsonian RG framework \cite{Morris:1994ie,Morris:1998}, in this case the Wilsonian RG framework breaks down. We therefore need to utilise the freedom to choose the cutoff exponent $n$ so that
\be
\label{bad-n2}
\alpha\ne-d_f\qquad\hbox{or equivalently}\qquad\eta \ne d+2n\,,
\ee
using \eqref{eq:alpha}.
%
In terms of these dimensionless variables the flow equations and msWIs then collapse to two background-independent flow equations which will be the subject of our study from now on:
\begin{subequations}
\label{equ:sys-red-final}
\begin{align}
 \partial_{\hat t} \Vh +d\Vh-\frac{\eta}{2}\ph \Vh' &= -\left(d-\eta+2n\right)\int d\hat p\, \hat p^{d-1} \,\hat Q_0 \,r\big(\hat p^2\big), \label{equ:flowV-final} \\
 \partial_{\hat t} \Kh +(d-2-\eta)\Kh-\frac{\eta}{2}\ph \Kh' &= -2(d-\eta+2n)\int d\hat p\, \hat p^{d-1} \,\hat P\big(\hat p^2,\ph\big) \,r\big(\hat p^2\big). \label{equ:flowK-final}
\end{align}
\end{subequations}
where $\hat t = \ln (\hat k/\mu)$ and where now
\begin{align}
 \hat Q_0 = &\left[\Vh''-\Kh \hat p^2-r\big(\hat p^2\big)\right]^{-1}\,, \label{Q0final} \\ 
 \hat P = &-\frac{1}{2}\Kh'' \hat Q_0^2 + \Kh'\left(2\Vh'''-\frac{2d+1}{d}\Kh' 
 			\hat p^2\right)\hat Q_0^3    \label{Pfinal} \\
    & +\left[\left\{\frac{4+d}{d}\Kh' \hat p^2 - \Vh'''\right\}\left(r'(\hat p^2)+\Kh\right)+\frac{2}{d}\hat p^2 r''\big(\hat p^2\big)\left(\Kh' \hat p^2-\Vh'''\right)\right]\!\left(\Vh'''-\Kh' \hat p^2\right)\!\hat Q_0^4
     \notag \\
    & -\frac{4}{d}\hat p^2 \left(r'\big(\hat p^2\big)+\Kh\right)^2 \left(\Vh'''-\Kh' \hat p^2\right)^2 \hat Q_0^5 \, , \notag 
\end{align}
and primes denote derivatives with respect to $\ph$ (except on $r$ where they denote derivatives with respect to its argument ${\hat p}^2$). 
We see that \eqref{equ:sys-red-final} takes the form of two partial differential equations for $V_{\kh}(\ph)$ and $\Kh_{\kh}(\ph)$ with respect to the RG scale $\kh$ and the total conformal factor field $\ph$.

When we specialise these equations to the case of most interest, namely $d=4$ dimensions,
we will set $n=2$, since this is the smallest possible (integer) choice for the exponent in the power law cutoff to ensure convergence, \cf below \eqref{equ:cutoff-pwrlw}. A third virtue of the power-law cutoff \eqref{equ:cutoff-pwrlw} besides ensuring compatibility of the msWI and flow equations, and facilitating the combination of the flow equation with the msWI, is that the flow equations \eqref{equ:sys-red-final} 
enjoy a (non-physical) scaling symmetry (which thus preserves the quantisation of the anomalous dimension in non-gravitational systems, \eg scalar field theory \cite{Morris:1994ie,Morris:1994jc,Morris:1998}). In $d=4$ this is  characterised by the following scaling dimensions:
\begin{equation}\label{scalings}
[\Vh]=4, \qquad [\Kh]=-6,\qquad [\ph]=4, \qquad [{\hat p}]=1.
\end{equation}

\section{Comparison to scalar field theory}
\label{sec:sft}

As noted in ref. \cite{Dietz:2015owa}, the background independent flow equations \eqref{equ:sys-red-final} bear a very close resemblance to those of scalar field theory. If the potential is renamed as $\Vh \mapsto - \Vh$, and the anomalous dimension is reparametrised as
\be  
\label{eta-scalar}
\eta = d-2+\eta^{(s)}\,, 
\ee 
then the resulting flow equations are exactly the ones that would be derived at $O(\partial^2)$ for scalar field theory in $d$ dimensions with power-law cutoff profile \eqref{equ:cutoff-pwrlw} \cite{Morris:1994ie}, except for an overall sign on the right hand side of the flow equations. The background independent definitions result in the cutoff term turning into:
\begin{equation}
\label{cutoff-kbar}
f^{\frac{d}{2}}\, R\big(p^2/f\big) =  - \, \kh^{d-\eta}\, r\big(p^2/\kh^2\big) =: R_{\kh}(p)\,,
\end{equation}
 which after the translation \eqref{eta-scalar} is the form expected for scalar field theory, except for the sign that is needed to match the wrong sign kinetic term of the conformal factor field.

As we will shortly verify, it follows that the flow equations can be derived from the $O(\partial^2)$ expansion of 
\be
\label{equ:FRGE}
\partial_{\hat t} \Gamma_{\kh} = -\frac{1}{2}\,\mathrm{Tr}\left[\frac{\delta^2\Gamma_{\kh}}
				  {\delta \phi \delta \phi}- R_{\kh}\right]^{-1}\!\!\! \partial_{\hat t} R_{\kh},
\ee
where the change of variables \eqref{equ:chvars} together with \eqref{khat} turns the effective action \eqref{equ:ansatzGamma} into
\begin{equation}
\label{equ:Gamma-hat}
 \Gamma_{\kh}[\phi] = \int d^dx \left(-\frac{1}{2} \tilde K_{\kh}(\phi) \left(\partial_\mu \phi\right)^2+ \tilde V_{\kh}(\phi)\right)\,.
\end{equation}
Note that the explicit dependence on background has also disappeared from the action, with the action depending on the background field $\chi$ only through $\hat k$. 

To verify this we start by noting that the two signs on the right hand side of \eqref{equ:FRGE} are absent in scalar field theory. However making the indicated change ${\tilde V}\mapsto-{\tilde V}$ and recognising that \eqref{equ:Gamma-hat} is then minus the effective action $\Gamma^{(s)}_{\kh}[\phi]$ of scalar field theory at $O(\partial^2)$, we see that \eqref{equ:FRGE} indeed turns into 
\be
\label{equ:FRGEs}
\partial_{\hat t} \Gamma^{(s)}= -\frac{1}{2}\,\mathrm{Tr}\left[\frac{\delta^2\Gamma^{(s)}}{\delta \phi \delta \phi}+ R_{\kh}\right]^{-1}\!\!\! \partial_{\hat t} R_{\kh}\,,
\ee
the flow equation for scalar field theory except for an overall sign on the right hand side, as claimed.

As it turns out, a more fruitful way to make the comparison is to `Wick rotate' the field $\phi = i\phi^{(s)}$ where the latter is a real field and thus the functional integral over the conformal factor field is to be done along a path along the imaginary axis. This is exactly the cure for the conformal factor problem in the functional integral, as proposed in ref. \cite{Gibbons:1978ac}, and indeed is nothing but the required choice of contour (steepest descents) that follows from the assumption of analyticity. Making the identifications that $K^{(s)}(\phi^{(s)}) = {\tilde K}(i\phi)$ and $V^{(s)}(\phi^{(s)}) = {\tilde V}(i\phi)$, both the action \eqref{equ:Gamma-hat} and the flow equation \eqref{equ:FRGE}, and hence also the $O(\partial^2)$ flow equations \eqref{equ:sys-red-final}, turn into the standard ones for scalar field theory (\ie now with all signs correct).

At first sight the extra signs  in \eqref{equ:FRGE} and \eqref{equ:Gamma-hat}, or equivalently the overall sign  on the right hand side of \eqref{equ:FRGEs}, 
are harmless since unlike the original functional integral, the flow equations seem well defined even with the wrong sign kinetic term \cite{Reuter:1996}. In fact such RG flows are then backward-parabolic meaning that the Cauchy problem for flow towards the infrared is not well posed. Instead, for a general `initial' effective action, well-defined RG flows only exist towards the ultraviolet \cite{Bonanno:2012dg}. Strictly speaking this already undermines the Wilsonian interpretation \cite{Dietz:2015owa}, but as we will see the sign difference actually has  more immediate and profound effects on the properties of both fixed points themselves and their eigen-spectra.

\section{The Gaussian fixed point and its eigen-operator spectrum}
\label{sec:gaussian}

\emph{At the risk of some confusion, for typographical clarity from here on we drop all the hats, however we emphasise that all quantities will still refer to scaled background independent variables except where explicitly stated.}

These equations are already very informative if we analyse the properties of the Gaussian fixed point. For a Gaussian fixed point we want to find a solution where $V=V^{\rm GFP}_*$ and $K=K^{\rm GFP}_*>0$ are constants, independent of both $\phi$ and $t$, so that \eqref{equ:Gamma-hat} amounts to a free massless field theory. Substituting such constant values into \eqref{equ:flowK-final}, we see from \eqref{Pfinal} that $P$ vanishes and thus  at $O(\partial^2)$ consistency demands that $\eta=d-2$. Thus at the Gaussian fixed point the background independent version of the conformal factor naturally acquires the scaling dimension $[\phi]= (d-2)/2$ of a Gaussian scalar field.

We now set $d=4$, so now $\eta=2$ and $[\phi]=1$. As we noted at the end of sec. \ref{sec:review}, we set $n=2$.
By the scaling symmetry \eqref{scalings}, we can choose the canonical value $K^{\rm GFP}_*=1$. From \eqref{equ:flowV-final} we thus find 
\be 
\label{GaussianVstar}
V^{\rm GFP}_* = \frac32 \int_0^\infty\!\!\! dp \,\frac{p^3}{1+p^6}= \frac{\pi}{2\sqrt{3}}\approx 0.9069\,.
\ee
Linearising the flow equations \eqref{equ:sys-red-final} about these values by writing 
\be 
\label{lineariseGaussian}
V(\phi,t)=V^{\rm GFP}_*+\delta V(\phi,t)\,,\qquad {\rm and}\qquad K(\phi,t) = 1+\delta K(\phi,t)\,,
\ee 
we have by separation of variables that 
\be 
\label{separateVars}
\delta V(\phi,t) =\epsilon\, \dV(\phi)\,{\rm e}^{-\lambda{t}}\,,\qquad{\rm and} \qquad\delta K(\phi,t) =\epsilon\, \dK(\phi)\,{\rm e}^{-\lambda{t}}\,, 
\ee 
where $\epsilon$ is a small proportionality factor. Thus we find
\begin{align}
(4-\lambda) \dV -\phi \dV' &= 
\left(\dV'' -2\dK\right)/{2a^2}\,,\notag\\
-\lambda \dK -\phi \dK' &= 
{\dK''}/{2a^2}\,,\label{gaussian-perturbations}
\end{align}
where we have written $a^2 = 3\sqrt{3}/4\pi$.
Solving these equations yields the eigen-operators 
\be 
-\half\dK(\phi) (\partial_\mu\phi)^2 + \dV(\phi)
\ee 
with their associated RG eigenvalues $\lambda$. Providing the linearised analysis is valid \cite{Morris:1996nx,Morris:1996xq,Morris:1998,Bridle:2016nsu}, we deduce that the eigenvalue is the scaling dimension of the associated coupling $\epsilon \mu^\lambda$, while the scaling dimension of $\dV(\phi)$ is $4-\lambda$ and the scaling dimension of $\dK(\phi)$ is $-\lambda$.

Before analysing the perturbations further, we recall that in scalar field theory about the Gaussian fixed point \cite{Morris:1996nx,Bridle:2016nsu} or in fact any fixed point \cite{Morris:1996xq,Morris:1998}, the eigen-perturbations divide into two classes: the quantized perturbations that grow as a power for large $\phi$, and the non-quantized perturbations that grow like the exponential of a fixed power of $\phi$ for large $\phi$. Any function of the field that grows slower than the latter, can be expanded uniquely as a series in the power-law perturbations. This may be proven by using Sturm-Liouville theory \cite{Ince1956} to show that power-law perturbations  are orthogonal and complete with respect to the appropriate measure. Furthermore for the power-law perturbations the scaling dimension deduced from linearised analysis is valid, and thus for RG eigenvalue $\lambda\ge0$ the perturbations can be associated to renormalised couplings $\epsilon \mu^\lambda$; on the other hand the non-power-law perturbations cannot be associated to renormalised couplings but instead follow mean-field evolution for large $\phi$ which, under any evolution to the infrared, falls back into the space of functions that can be expanded in terms of the power-law perturbations \cite{Morris:1996nx,Bridle:2016nsu,Morris:1996xq,Morris:1998}.

\begin{figure}[ht]
\centering
\includegraphics[scale=0.25]{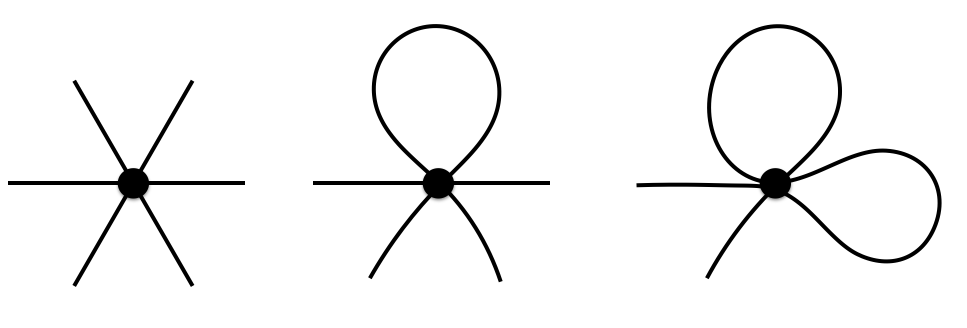}
\caption{The eigen-operators at the Gaussian fixed point are linear in an $n$-point interaction (here $n=6$), with lower powers of $\phi$ being generated by successive tadpole corrections.}
\label{fig:tadpoles}
\end{figure}

In the current case, just as for the Gaussian fixed point in scalar field theory, quantized solutions are related to Hermite polynomials $H_n$ (equivalent for even $n$ to the generalised Laguerre polynomials $L^{-\half}_{n/2}$ analysed in ref. \cite{Bridle:2016nsu}). More specifically, let us write 
\be 
\label{Hermite}
{\cal O}_n(\phi)  := H_n(ia\phi)/\left(2ia\right)^{n}=\phi^n +n(n-1)\phi^{n-2}/4a^2+\cdots\,,
\ee
with $n$ a non-negative integer. The potential perturbations  are then:
\be
\label{GaussianPotentialOp}
\dV_n(\phi) =  {\cal O}_n(\phi)\,,\quad \lambda=4-n\,,
\ee
(with $\dK(\phi)=0$), and perturbations with a non-vanishing kinetic term contribution take the form:
\be 
\label{GaussianKineticOp}
\dK_n(\phi) = {\cal O}_n(\phi)\quad{\rm with}\quad \dV^{(kin)}_n(\phi) = -4 {\cal O}_n(\phi)/{a^2}\,,\quad \lambda=-n\,.
\ee
We see that the RG eigenvalues and corresponding dimensions are just the engineering ones expected at the Gaussian fixed point. The polynomials are of the expected form: generated by an integer power of the field plus successive tadpole corrections
\cite{Altschul:2004yq}, see fig. \ref{fig:tadpoles}. The only new feature is the presence of $i$ in the definition \eqref{Hermite}. Its only effect is to remove the alternating signs we would otherwise have had in the sum over lower powers, for example in scalar field theory the $\phi^{n-2}$ term in \eqref{Hermite} appears with a minus sign. This is a direct consequence of the wrong-sign kinetic term for the conformal factor, which provides an extra minus sign for every propagator in fig. \ref{fig:tadpoles}.

These adaptations so far seem innocuous, nevertheless the sign change has far reaching consequences. 
Firstly the polynomials \emph{only formally} satisfy orthonormality relations:
\be
\label{completeness}
\int^\infty_{-\infty}\!\!\!\! d\phi\,\, {\rm e}^{a^2\phi^2} {\cal O}_n(\phi) {\cal O}_m(\phi) = -\frac{i}{a}\left(-\frac{1}{2a^2}\right)^n\! n!\sqrt{\pi}\,\delta_{nm}\,,
\ee
which we can justify only by Wick rotation of the conformal factor to an integral along the imaginary axis, as already introduced at the end of sec. \ref{sec:sft},
and then defining the result by analytic continuation back to an integral along the real line.
Since these integrals do not converge as integrals along the real line, they cannot be used to derive a completeness relation for  functions over the real line. For example for a perturbation $\dV(\phi)$ other than a polynomial, it is not possible to write $\dV(\phi) = \sum_m \dV_m {\cal O}_m(\phi)$ for some coefficients $\dV_m$ because there is no notion of convergence for real $\phi$ of the sum 
\be 
\label{partial-sum}
\sum_{m=0}^N \dV_m {\cal O}_m(\phi)
\ee  
to $\dV(\phi)$ as $N\to\infty$. Indeed a translation of the usual proof of completeness would demonstrate that 
\be 
\label{completeness-proof}
\int^\infty_{-\infty}\!\!\!\! d\phi\,\, {\rm e}^{a^2\phi^2} \left( \dV(\phi) - \sum_{n=0}^N \dV_n {\cal O}_n(\phi)\right)^{\!2} 
\ee
tends to zero as $N\to\infty$, but clearly this can make sense in general only if the integral is taken along the imaginary axis.
In fact, since \eqref{partial-sum}
is a polynomial, the integral can only converge along the real line if $\dV(\phi)$ is already the same polynomial plus a term that decays exponentially fast. 

\begin{figure}[ht]
\begin{center}
$
\begin{array}{ccc}
\includegraphics[width=0.45\textwidth]{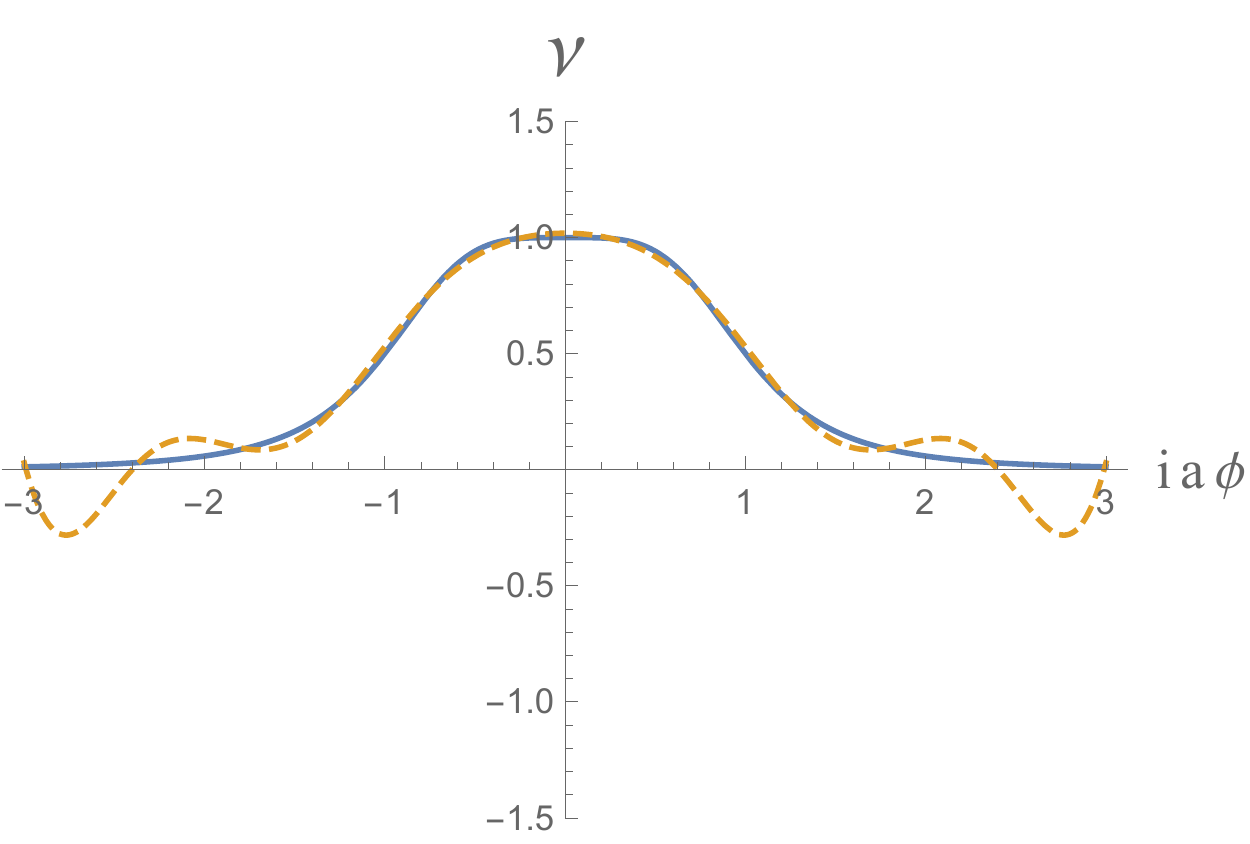} &
\includegraphics[width=0.45\textwidth]{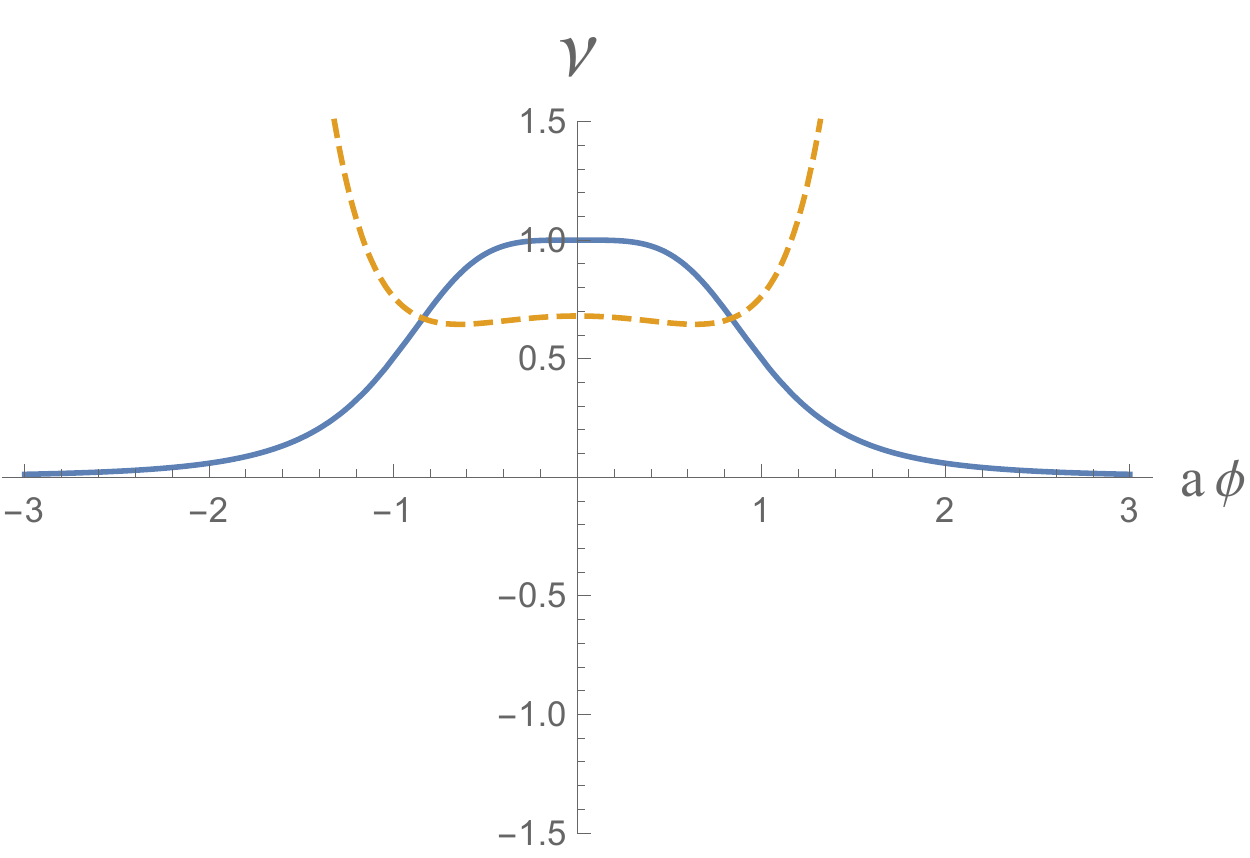} \\
\end{array}
$
\end{center}
\caption{The example $\dV$ in \eqref{example-dV} is plotted in blue, and the partial sum \eqref{partial-sum} with $N=15$, plotted as a dashed line in orange. They are plotted  along the imaginary axis in the left panel and along the real axis in right panel.}
\label{fig:completeness}
\end{figure}

Let us illustrate these issues with a simple example. Suppose that the linearised perturbation (at $t=0$) is given by
\be 
\label{example-dV}
\dV(\phi) = \frac{1}{1+a^4\phi^4}\,.
\ee
Integrating along the $\phi$ imaginary axis, the $\dV_m$ can be computed using \eqref{completeness}. We see from fig. \ref{fig:completeness}, that the resulting sum, \eqref{partial-sum}, approximates the original function well along the imaginary axis, for $N$ sufficiently large, as required by the vanishing of the norm \eqref{completeness-proof} as $N\to\infty$. Note that differences visible at large $\pm ia\phi$ are exponentially damped in the integrand of the norm-squared of the difference \eqref{completeness-proof}. In fact with $N=15$, this integrand is never more than $5\times10^{-4}$. However as seen in fig. \ref{fig:completeness}, the approximation breaks down completely for real $\phi$.

%
%

There are two related issues. Firstly we can no longer cast the equation in terms of a Sturm-Liouville operator \cite{Ince1956} that is self-adjoint in an appropriate space and thus we are also unable to show that the eigenvalues $\lambda$ must be real. Since the differential equations are real, we would then have a complex pair of solutions associated to a complex pair of eigenvalues. (This type of situation was analysed in ref. \cite{DietzMorris:2013-1}.) Although the analysis we present can be extended to the case of complex $\lambda$ we will in the ensuing only analyse the subset of perturbations with real $\lambda$.

Secondly, the non-quantized solutions to \eqref{gaussian-perturbations} are no longer excluded.
In the following we consider only the potential perturbations with $\dK\equiv0$ which thus from \eqref{gaussian-perturbations} satisfy
\be 
\label{gaussian-perturbations-V}
(4-\lambda) \dV -\phi \dV' =\dV''/{2a^2}\,.
\ee
(It is straightforward to adapt the arguments to the case of kinetic term perturbations \eqref{GaussianKineticOp}.) It is worth  noting that up to scaling, eqn. \eqref{gaussian-perturbations-V} is universal, independent of the choice of cutoff profile just as is true for scalar field theory \cite{Bridle:2016nsu}.

In scalar field theory, the non-quantized perturbations grow exponentially for large $\phi$, in particular about the Gaussian fixed point  as $\phi^{\lambda-5}\exp (a^2\phi^2)$, invalidating the linearised approximation \eqref{lineariseGaussian} for sufficiently large $\phi$, no matter how small we set $\epsilon$ in \eqref{separateVars}. In turn this tells us that these non-quantized perturbations cannot be associated with renormalised couplings and instead follow at large $\phi$ a mean-field evolution that instantly collapses the interaction back into the space spanned by the quantized operators as the perturbation is evolved under the RG towards the IR \cite{Morris:1996nx,Morris:1996xq,Morris:1998,Bridle:2016nsu}.

On the contrary here, for any real $\lambda$, neither solution can be ruled out by such arguments applied to its large $\phi$ behaviour. The general solution is given by the linear combination
\be 
\label{gaussian-perturbation-general}
\dV = C_1\, \phi \,M\left(\frac{\lambda}{2}-\frac{3}{2}, \frac{3}{2}, -a^2\phi^2\right) + C_2\, M\left(\frac{\lambda}{2}-2,\frac{1}{2},-a^2\phi^2\right)\,,
\ee
in terms of the Kummer $M$-function \cite{AbramowitzStegun} and constants $C_i$. The linearly independent solutions are smooth (in fact entire) functions of $\phi$, with the first being an odd function of $\phi$ and the second an even function of $\phi$. For general $\lambda$ and for large $\phi$ both of these behave as a power law 
\be 
\label{GaussianEigenasymptotic}
\dV\propto\phi^{4-\lambda}+\frac{(4-\lambda)(3-\lambda)}{4a^2}\phi^{2-\lambda}+
\mathcal{O}(\phi^{-\lambda})\,,
\ee
which is an asymptotic series with exponentially decaying corrections $\sim\phi^{\lambda-5} \exp (-a^2\phi^2)$. Only for $\lambda=4-n$ ($n$ a non-negative integer) is there the additional possibility to exclude the decaying exponential corrections and arrive at the polynomial solutions \eqref{Hermite}, and only for $\lambda= 5+n$ is it possible to exclude the power-law part and have a solution that for  $\phi\to\pm\infty$ decays as $\sim\phi^{\lambda-5} \exp (-a^2\phi^2)$. These latter `super-relevant' perturbations take the form of polynomials times the exponential factor. The first two are
\be 
\label{super-relevant}
\dV = \exp (-a^2\phi^2)\,,\quad \lambda=5\qquad{\rm and}\qquad \dV = \phi \exp (-a^2\phi^2)\,,\quad \lambda=6\,.
\ee
They clearly fulfil the linearisation approximation \eqref{lineariseGaussian} ever more accurately for large $\phi$, confirming that they evolve as an operator of scaling dimension $4-\lambda$ associated to a renormalised coupling  $\epsilon \mu^\lambda$. If a solution is taken with asymptotics \eqref{GaussianEigenasymptotic}, then since the non-linear terms depend only on $\dV''$, it likewise continues to fulfil the linearisation approximation for $2-\lambda\le0$  (for $2-\lambda<0$ ever more accurately) as $\phi\to\pm\infty$, while for $2-\lambda>0$,  mean-field evolution takes over for large $\phi$ but does allow the RG-time dependence of the leading term to be associated to evolution of $g=\epsilon \mu^\lambda$.


To justify this last conclusion, we note that for $2-\lambda>0$, the linearisation approximation is invalid for sufficiently large $\phi$. Following the analysis of refs. \cite{Morris:1996nx,Morris:1996xq,Morris:1998,Bridle:2016nsu}, we add the perturbation \eqref{lineariseGaussian} but recognise that separation of variables as in \eqref{separateVars} is no longer justified. Instead we set $\delta V(\phi,0) = \epsilon\, \dV(\phi)$, and using this boundary condition determine the correct $t$-evolution at large $\phi$. Since the perturbation is then no longer small, we need to work with the full flow equation which from \eqref{equ:flowV-final}  for $V(\phi,t)$ reads
\be 
\label{non-p-Vflow}
\partial_t V -\phi V' + 4V = 3\,\mathcal{F}(V'')\,,
\ee
where
\be
\label{F}
\mathcal{F}\!\left(z\right) := 2 \int_0^\infty\!\!\!\! dp \,\frac{p^3}{1-p^4z+p^6}\,.
\ee
We notice that $\mathcal{F}$
is a strictly positive monotonically increasing function
where the allowed range for its argument is 
\begin{equation} \label{convbound}
-\infty<z<3\cdot 2^{-2/3}
\end{equation}
to guarantee finiteness of the integral. Since $\dV(\phi)$ is a finite smooth solution for all finite $\phi$, providing we add it with small enough $\epsilon$ we can ensure the integral is well defined, providing we also choose the $C_i$ so that $\dV''<0$ for $\phi\to\pm\infty$. Then we see from \eqref{GaussianEigenasymptotic} that actually $\dV''\to-\infty$ asymptotically, which forces $\mathcal{F}\to0$. Therefore for large $\phi$ only the left hand side of \eqref{non-p-Vflow} matters, which is solved by mean field evolution $\delta V(\phi,t) = {\rm e}^{-4t}\, \delta V(\phi\,{\rm e}^{\,t},0)$. 
Now using \eqref{GaussianEigenasymptotic}, we see that its power-law behaviour ensures that the $t$ dependence factorises so that we get back the linearised result $\delta V(\phi,t) = \epsilon\,\dV(\phi)\,  {\rm e}^{-\lambda t}$, but here holding even when $\epsilon\,\dV(\phi)$ is no longer small.


Thus we conclude that for conformally reduced gravity, around the Gaussian fixed point there are \emph{two independent relevant couplings for every real positive} $\lambda$. 
In the next sections we will uncover another RG consequence of the conformal factor instability: that fixed points are likewise no longer isolated but instead form continuous sets.

\section{Fixed points in the Local Potential Approximation}
\label{sec:LPA}

In order to show very clearly that it is the change in sign of the kinetic term which forces fixed points to form continuous sets, in this section we will treat the Local Potential Approximation (LPA) for the flow equation of standard single-component scalar field theory. We will show that the large field dependence forces the fixed point equation to have at most a discrete set of fixed points. On changing the sign of the kinetic term we will see that the same large field behaviour is mapped to one that allows a continuum of fixed points. 

We will show this without specifying the precise form of the cutoff profile $r$, in order to emphasise that these effects are independent of this choice. Therefore the existence of a continuum of fixed points is universal, as we showed also to be the case for the continuous spectrum of perturbations around the Gaussian fixed point (\cf below \eqref{gaussian-perturbations-V}).

Furthermore we will prove that the effects are independent of any specific value for the space-time dimension $d$ or the scaling dimension $d_\phi$ of the field $\phi$. We will only need that $d_\phi>0$ and that $d/d_\phi>2$. Although we will not specify them beyond these inequalities, let us note that at LPA level it would be typical to neglect the anomalous dimension $\eta^{(s)}$ and thus set $d_\phi = (d-2+\eta^{(s)})/2=(d-2)/2$. In this case $d_\phi>0$ for $d>2$, after which $d/d_\phi>2$ holds automatically.

The restriction to $d_\phi>0$ is necessary, since a continuum of fixed points is anyway found for standard scalar field theory when $d_\phi=0$ (the critical Sine-Gordon models) \cite{Morris:1994jc}. We will see in the next section that $d_\phi=0$ still leads to a continuum of fixed points if the sign of the kinetic term is reversed (this is also shown for optimised cutoff profile in ref. \cite{Labus:2016lkh}),
but moreover they support continuous eigenoperator spectra. It is possible straightforwardly to extend the analysis in this section to show that in LPA, there exists a continuous eigenoperator spectrum also for $d_\phi>0$, again without specifying $r, d$ and $d_\phi$. 
For a general cutoff profile we have already seen this at the Gaussian fixed point in the previous section, while in sec. \ref{sec:eigen-asymptotics} we will establish the continuous spectrum for power-law cutoff at $\mathcal{O}(\partial^2)$.

It is of course no surprise that the fixed point equations for standard scalar field theory yield only discrete fixed points when $d_\phi>0$ and indeed also only discrete spectra of eigen-operators. 
Specifically with the power-law cutoff profile as used in this paper, it has long been established that the fixed point equations have no fixed singularities, but yield only discrete fixed points and spectra, as can be understood by counting parameters in the large field behaviour \cite{Morris:1994ie,Morris:1994jc}. Nevertheless the analysis in this section closes one small gap in that these studies were not performed in $d=4$ dimensions, and indeed close a gap by all of this, including the effects of negative kinetic term, is actually insensitive to $d$, $d_\phi$ and choice of cutoff profile $r$ (conditions for which are supplied below).

After tidying up the equations by redefining in a similar fashion to \eqref{redefine}, the LPA equation for the fixed point potential $V_*(\phi)$ in standard single component scalar field theory can be written as
\be 
\label{+s}
d V_* - d_\phi \phi V'_* = \F(V''_*)\,,
\ee
where ($r'\equiv \partial_{p^2} r$):
\be 
\label{Fs}
\F\!\left(z\right) :=  -\int_0^\infty\!\!\!\! d(p^2) \,\frac{p^{d-2}r'(p^2)}{z+p^2+r(p^2)}\,.
\ee
We will only require that the cutoff profile $r(p^2)$ is  positive, monotonically decreasing, and ensures a finite integral for $z>0$. Notice that these properties imply that $\F(z)$ is positive and monotonically decreasing, and has limit $\F(z)\to0$ as $z\to\infty$. 

For example the properties hold true for the optimised cutoff  $r = (1-p^2)\theta(1-p^2)$ \cite{opt1,opt3}, and for the choice in this paper ($d=4$ and the power-law cutoff $r=1/p^4$).  With these choices one finds $\F(z)=\frac2d\frac1{1+z}$ and $\F(z) \sim \frac32 \ln(z)/z$ respectively. (The notation $g(z)\sim f(z)$ means asymptotically equal, \ie $\lim_{z\to\infty} g(z)/f(z) =1$.)

The fixed point equation for scalar field theory with wrong-sign kinetic term is related by the transformation at the beginning of sec. \ref{sec:sft} and thus at the LPA level is simply\footnote{For $d=4$, $d_\phi=(d-2)/2=1$ and $r(p^2) =1/p^4$, this equation plus \eqref{Fs} appear already as \eqref{non-p-Vflow} and \eqref{F} (after rescaling the right hand side by $3/2$, by replacing $K=1$ with $K=2/3$ and sending  $V\mapsto 2V/3, p^2\mapsto 2p^2/3$).}
\be 
\label{-s}
d V_* - d_\phi \phi V'_* = \F(-V''_*)\,.
\ee

Starting with the standard scalar field theory equation \eqref{+s} we now recover the conclusions in refs. \cite{Morris:1994ki,Morris:1994ie,Morris:1994jc}. We note that  at large $\phi$, the equation is solved by solving the left hand side only, with thus $V_*\sim V_A$, where
\be 
\label{classical}
V_A(\phi) := A\, |\phi|^{d/d_\phi}\,,
\ee
and $A>0$ is a parameter. To see this, 
note that since $d/d_\phi>2$, $V''_{A}$ diverges for large $|\phi|$, and thus we must have $A>0$ in order for \eqref{Fs} to be well defined; furthermore $\F$ then supplies corrections that vanish in the limit of large $|\phi|$. 

Since \eqref{+s} is a second order differential equation, we would expect to find two parameters in the solution.  To discover what happened to the other parameter, we linearise \eqref{+s} around the solution \eqref{classical} by writing 
\be 
\label{linear}
V_* = V_A(\phi) +B\, v(\phi)\,, 
\ee
where $B$ is some small parameter, and thus 
\be 
d v -d_\phi \phi v' = v''\,\Fp(V''_A) \,.
\ee
One solution to this equation is of course $v= |\phi|^{d/d_\phi}$ corresponding to $A\mapsto A+B$. The other solution is given by 
\be 
\label{diverging}
v = \exp f(\phi)\,, 
\ee
where $f'$ is diverging for large $\phi$. In this regime $v''\sim (f')^2 v$ and thus we find
\be 
f' \sim -d_\phi \phi/\Fp(V''_A)\,.
\ee
From $d_\phi>0$ and the properties of $\F$ set out below \eqref{Fs}, we see that $f'>0$ and diverges faster than linearly as $\phi\to\infty$. Thus 
\be 
f \sim -d_\phi \int\!\!d\phi\, \phi/\Fp(V''_A)
\ee
 is also positive and diverging faster than $ \phi^2$. What we have found therefore is that in the neighbourhood of the solution \eqref{classical}, the other parameter is associated to an exponentially growing perturbation. However for any $B$, and for sufficiently large $\phi$, $B v$ is no longer small compared to $V_A$, ruling out the linearisation used to find it. Therefore asymptotically, the fixed point solution takes the form of an isolated one-parameter set $V_*\sim V_A$ in both regimes $\phi\to\pm\infty$ (with \textit{a priori} different parameters $A=A_\pm$). This thus provides two constraints, \aka boundary conditions, fixing the solution space to a discrete set.\footnote{Alternatively one can require $\phi\mapsto-\phi$ invariance, then $V'_*(0)=0$ provides one of the boundary conditions.}
 As we know, what happens to the $Bv$ perturbations is that once the non-linear terms become important the solution ends in a moveable singularity \cite{Morris:1994ki,Morris:1994ie,Morris:1994jc}. 

Now let us see what changes if we flip the sign of the kinetic term. For the fixed point equation \eqref{-s}, the only change is that the argument of $\F$ picks up a sign. This implies that asymptotically the equation can be solved now by $V_*\sim -V_A$ (where we keep $A>0$). Linearising as in \eqref{linear}, the only change is then an overall sign on the right hand side:
\be 
d v -d_\phi \phi v' = -v''\,\Fp(V''_A)\,.
\ee
Thus the other solution is in this case again of form \eqref{diverging} where
\be 
f \sim +d_\phi \int\!\!d\phi\, \phi/\Fp(V''_A)
\ee
diverges faster than $\phi^2$ but is now \emph{negative}. Therefore $v$ is now an exponentially decaying solution which fulfils the linearised approximation ever more accurately as $\phi\to\pm\infty$. We see that asymptotically the solution therefore has two parameters and thus no longer constrains the solution space. Indeed asymptotically it is of the form \eqref{linear} where $B$ is a free parameter, since $Bv$ is exponentially smaller than $V_A$  for sufficiently large $\phi$.

From here on we return to power-law cutoff profile as required for the validity of the background-independent flow equations \eqref{equ:sys-red-final}.

\section{Local Potential Approximation with vanishing anomalous dimension}
\label{sec:confLPA}

Before delving into aspects of an analysis of the full system of flow equations, it is also instructive to derive the space of fixed points in one other version of the LPA.

We specialise to the lowest order of the derivative expansion by setting $\hat K=1$ and discarding the second equation in \eqref{equ:sys-red-final}. Although a non-vanishing anomalous dimension can be justified in the Local Potential Approximation (LPA)  \cite{Bervillier:2013},  and we have already seen that at $O(\partial^2)$ the Gaussian fixed point requires $\eta = d-2$, we will adopt the traditional stance  in this section and set  $\eta=0$ \cite{Nicoll:1974zz,Wilson:1973,Morris:1994ki}. This simplifies the equations sufficiently to allow for an exact analysis, from which we will gain yet more insight.

Untying the changes of variables \eqref{equ:dimless} and \eqref{equ:chvars} with \eqref{khat} shows that this corresponds to setting the original $K=k^{d-2} k^{(1-d/2)d_f}$ in the original effective action \eqref{equ:ansatzGamma}, \ie we indeed obtain the LPA as characterised by a field independent coefficient of the kinetic term. In contrast to scalar field theory this coefficient is here the appropriate power of the RG scale due to the vanishing classical scaling dimension of the conformal factor field $\ph$ and the non-vanishing scaling dimension of $d_f=[f(\chi)]$

With these provisions \eqref{equ:flowV-final} in $d=4$ dimensions becomes
\be 
\label{LPA}
\partial_t V(\phi,t) + 4 V = 4\, \mathcal{F}\!\left(V''\right)\,,
\ee
where we have already introduced $\mathcal{F}$ in \eqref{F}.
The fixed point potential is therefore a solution of $V_*(\phi) = \mathcal{F}\!\left(V_*''\right)$.
Forming first order perturbations $V(\phi,t)=V_*(\phi)+\delta V(\phi,t)$, we have by separation of variables that $\delta V(\phi,t) =\epsilon \dV(\phi)\,{\rm e}^{-\lambda{t}}$, where $\epsilon$ is a small proportionality factor, and thus from \eqref{LPA},
\be
\label{eigenLPA}
\left(4-\lambda\right) \dV(\phi) = \dV''(\phi) / \rho(\phi)\,,
\ee
where we have introduced for later purposes,
\be
\label{rho}
\frac1{\rho(\phi)} := 4 \mathcal{F}'\!\left(V_*''\right)
=8\int_0^\infty\!\!\! dp \, \frac{p^7}{\left(1-p^4V_*''+p^6\right)^2}\,.
\ee

\subsection{The LPA Gaussian fixed point and its operator spectrum}
\label{sec:gaussLPA}

Let us again analyse the Gaussian fixed point, \ie the simple exact solution where $V_*$ is a constant independent of $\phi$, but this time with $\eta=0$. 
Evaluating the integral, we have in this case that
\be 
\label{GFPLPA}
V_*={\cal F}(0)=\frac{2\pi}{3\sqrt{3}}=1.2092\,,
\ee
($(d+2n)/(2+2n)$ times \eqref{GaussianVstar}, the $O(\partial^2)$ answer).
Although this is a Gaussian fixed point, the eigen-operators again have unusual properties compared to standard quantum field theory. From \eqref{eigenLPA}
\be
\label{eigenGaussian}
\left(4-\lambda\right) \dV(\phi) = \frac{8\pi}{9\sqrt{3}}\, \dV''(\phi)\,.
\ee
As before, apparently none of the solutions are forbidden, and thus we find that $\lambda$ is continuous, being any real number, with two independent eigen-operators for each $\lambda$.
Define  
\be 
\label{omegaGFP}
\omega(\lambda) := \sqrt{\frac{9\sqrt{3}}{8\pi}|\lambda-4|}\,. 
\ee
For $\lambda>4$,
\be
\label{lambdag4}
\dV = \cos(\omega\phi)\quad{\rm and}\quad \dV=\sin(\omega\phi)\,;
\ee
for $\lambda<4$,
\be 
\label{lambdal4}
\dV = \cosh(\omega\phi)\quad{\rm and}\quad \dV=\sinh(\omega\phi)\,;
\ee
and finally when $\lambda=4$, 
\be
\label{lambda4}
\dV = 1 \quad{\rm and}\quad \dV=\phi\,.
\ee
In sec. \ref{sec:LPA-RGprops}, we  show that there are no restrictions on this eigen-operator spectrum coming from the leading large $\phi$ behaviour.

\subsection{The plane of fixed points at the LPA level}
\label{sec:planeLPA}

First we will show that there is a continuous set of fixed points solutions at the LPA level.
We follow ref. \cite{Morris:1994jc}
to solve the fixed point equation $V_*(\phi) = \mathcal{F}\!\left(V_*''\right)$. We proceed
as for Newton's equation for a particle in one dimension 
and find a first integral by inverting the function $\mathcal{F}$ and solving $dU/dV_* = -\mathcal{F}^{-1}\!\left(V_*\right)$ so that the solutions are labelled by one parameter $E$ and satisfy $E=1/2 (V_*')^2 + U(V_*)$. Note that in this Newtonian analogy, $V_*$ plays the r\^ole of the position of the particle and $\phi$ plays the r\^ole of the time.
From the properties of $\mathcal{F}$, we see that the `Newtonian potential' $U$ exists only for positive `position' $V_*$, and has a maximum  $U=U_\mathrm{max}$ at $V_* = \mathcal{F}(0)$.  Here the particle can just sit stationary at the top of the potential, 
corresponding the constant potential Gaussian fixed point solution \eqref{GFPLPA} we have already discussed.

\begin{figure}[ht]
\centering
\includegraphics[scale=0.65]{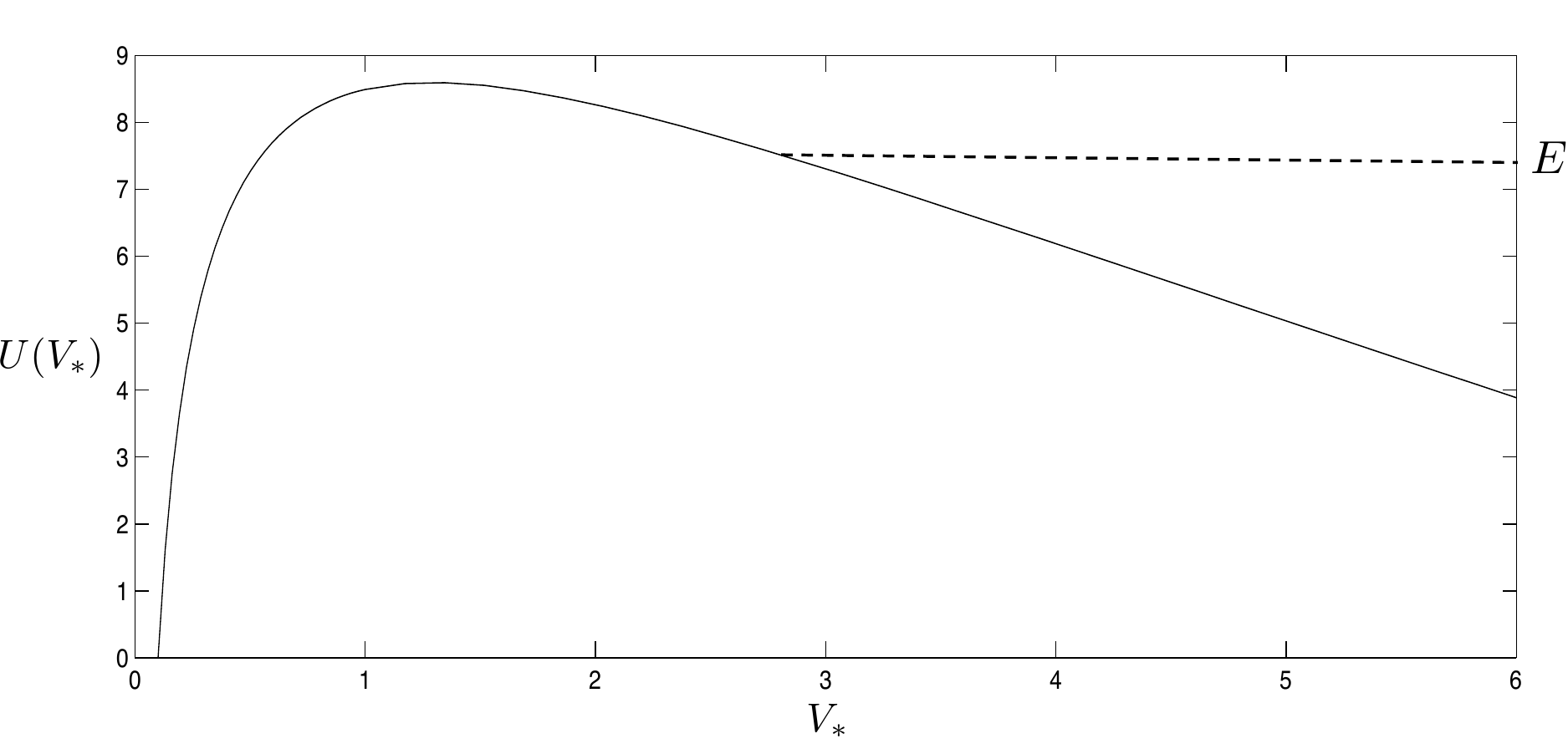}
\caption[The `Newtonian potential' used in solving the LPA for the conformal factor.]{Characterising the solutions of  $V_*(\phi) = \mathcal{F}\!\left(V_*''\right)$, with the `Newtonian potential' $U(V_*)$. At value $E$ for the first integral, $V_*(\phi)$ ranges over values indicated by the dashed line.}
\label{fig:NewPot}
\end{figure}

The potential $U$ can be determined numerically and takes the form shown in fig. \ref{fig:NewPot}.  
Globally valid solutions are obtained only if $E$ is not greater than  $U_\mathrm{max}$, and if $V_*$ accordingly takes values corresponding to the region to the right of ${\cal F}(0)$ and in fact at or above the lower bound provided by the intersection of the horizontal $E$ line with $U$, as illustrated in fig. \ref{fig:NewPot}. All other solutions end at the singularity $V_* \to 0 \Leftrightarrow V_*'' \to -\infty$ at some finite `time' $\phi=\phi_c$. If a solution is globally defined we see that $V_*(\phi)\to \infty$ asymptotically which entails $V_*''(\phi)\to 3\cdot 2^{-2/3}$, \ie $V_*''$ is asymptotically approaching the upper limit of the convergence range \eqref{convbound}. From the second derivative tending to a constant, one already expects a two parameter set of solutions. The additional parameter besides $E$ is slightly hidden in the approach using the first integral $U$ here but it can be recovered by exploiting 
the `time' translation symmetry $V_*(\phi) \mapsto V_*(\phi +c)$ for any solution at fixed $E$ of the fixed point equation $V_*(\phi) = \mathcal{F}\!\left(V_*''\right)$. This symmetry can be exploited to implement $\phi \mapsto -\phi$ symmetry of the solutions $V_*(\phi)$, corresponding to time reflection symmetry in the Newtonian analogy.\footnote{Note here the distinction between a symmetry of the fixed point equation and a symmetry of its solutions.}

\begin{figure}[ht]
\centering
\includegraphics[scale=0.4]{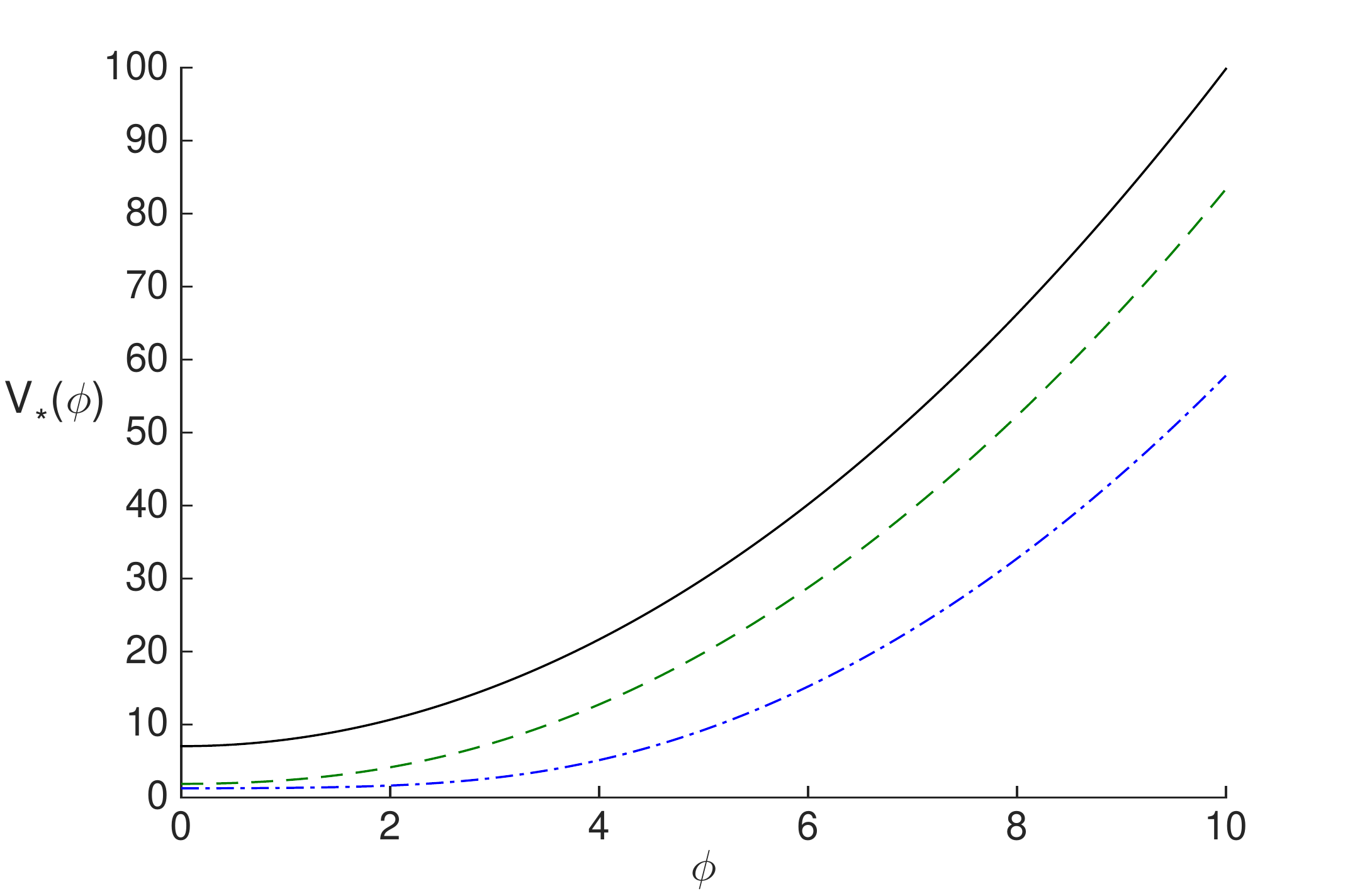}
\caption{Potentials $V_*(\phi)$ translated so that $V'_*(0)=0$ with the other initial condition being $V_*(0) = 7.03, 1.86$ and $1.25$; to be compared to the Gaussian solution $V_*(\phi)=1.2092$ in \eqref{GFPLPA}.}
\label{fig:Potentials}
\end{figure}

Hence, the fixed point equation  has a two parameter set of solutions that can be thought of as parametrised by $V_*'(0)$, or equally $\phi_0$ defined by $V_*'(\phi_0)=0$, and $E\le U_\mathrm{max}$. 
If we choose to implement the condition $V_*'(0)=0$ to obtain even solutions, this set reduces to a single ray as given by $E\leq U_\mathrm{max}$. Illustrative solutions in this case are displayed in fig. \ref{fig:Potentials}.
Note that as $E\to U_\mathrm{max}$, $V_*(0)\to {\cal F}(0)$ from above and the potential takes longer and longer to reach its asymptotic regime. In the limit $E=U_\mathrm{max}$ we reach the Gaussian fixed point $V_*(\phi) \equiv {\cal F}(0)$. 

\subsection{Eigen-operator spectrum about a general fixed point in the LPA}
\label{sec:eigenLPA}

We first note that \eqref{eigenLPA} continues to have the exact solutions \eqref{lambda4} if $\lambda=4$. For $\lambda\ne4$, it is straightforward to derive the general properties that we will need from the following observations. We note that $\rho(\phi)$, defined in \eqref{rho}, is a positive function of $\phi$. For all fixed point solutions apart from the Gaussian fixed point, we also know that $\rho\to0$ as  $\phi\to\pm\infty$,  since $V''_*(\phi)$ tends towards the upper limit of the convergence range \eqref{convbound}. Rewriting \eqref{eigenLPA} as
\be
\label{Schrodinger}
-\dV''(\phi) + (4-\lambda) \rho(\phi)\, \dV(\phi) =0\,,
\ee
we recognise that $\dV(\phi)$ may interpreted as the zero energy wavefunction solution for the Schr\"odinger equation of a particle at `position' $\phi$ in a `potential' $(4-\lambda) \rho(\phi)$.   
If $\lambda>4$ the potential is negative for all finite $\phi$, implying that a zero energy solution must  have positive kinetic energy and therefore generically asymptotically the two solutions can be chosen to obey $\dV(\phi) \to \cos(\omega\phi)$ and $\dV(\phi) \to \sin(\omega\phi)$, for some positive function ${\omega}(E,\lambda)$, where the parameter $E<U_{\rm max}$ labels the choice of fixed point as discussed in the previous section, and by linearity we can normalise so that the leading term has unit amplitude as shown. For some discrete value of $\lambda$, we may also expect to find a zero energy bound state with an exponentially fast fall off for $\dV(\phi)$ at infinity.
We note that the other fixed point parameter $\phi_0$ has no effect on $\omega$ since it just labels the position of the peak of $\rho(\phi)$. If $\lambda<4$ the potential is positive at all finite $\phi$, therefore the zero energy solution must have negative kinetic energy, implying that asymptotically as $\phi\to\pm\infty$, the two solutions can be chosen as $\dV(\phi) \to \cosh \omega\phi$  and $\dV(\phi) \to \sinh \omega\phi$, for some positive function ${\omega}(E,\lambda)$.\footnote{If $\phi_0=0$ it is clear by $\phi$ reflection symmetry that the solutions $\dV(\phi)$ can be chosen to be even or odd and thus behave asymptotically as stated for both $\phi\to\infty$ and $\phi\to-\infty$. Since the $\phi_0\ne0$ case is just a shift of these solutions by $\phi\mapsto\phi+\phi_0$, we see that by linear combinations, it is again true that the solutions can be chosen to behave asymptotically as stated in both regimes.} Comparing to \eqref{lambdag4} -- \eqref{lambda4},
we see that the asymptotic behaviour is the same, indeed we identify the explicit solution in \eqref{omegaGFP} as nothing but $\omega(\lambda) \equiv \omega(U_{\rm max},\lambda)$.

\subsection{RG properties of the eigen-operators}\label{sec:LPA-RGprops}

As we will see the RG properties of the eigen-operators about the general $\eta=0$ LPA fixed point turn out to be the same as the RG properties of the eigen-operators about the $\eta=0$ LPA Gaussian fixed point. To determine these properties we consider the perturbation  
\be 
\label{LPAperturbation}
V(\phi,t)=V_*(\phi)+\epsilon\, \dV(\phi)\,{\rm e}^{-\lambda{t}}
\ee
more carefully than is usually done.

For the exact solutions at $\lambda=4$, namely \eqref{lambda4}, the right hand side of \eqref{LPA} is still $4\mathcal{F}\!\left(V_*''\right)$. Since the left hand side of \eqref{LPA} is already linear, these solutions therefore are exact even when $\epsilon \dV(\phi)\,{\rm e}^{-4{t}}$ is not small. Therefore we can safely conclude that these solutions are legitimately associated to renormalised dimension 4 couplings $g = \epsilon\, {\rm e}^{-4{t}}$.

The leading large field behaviour of the other eigen-operators determines the RG properties of their associated couplings \cite{Morris:1996nx,Morris:1996xq,Morris:1998,Bridle:2016nsu}, an observation used already in sec. \ref{sec:gaussian}. 

For $\lambda>4$ we have seen that the large field behaviour is (at worst) oscillatory with fixed amplitude (normalised to unity). Therefore if $\epsilon$ is small enough in \eqref{LPAperturbation} to justify linearisation in \eqref{LPA} at finite $\phi$, it remains small enough to justify this step for all $\phi$. It follows that the RG time dependence is really given by \eqref{LPAperturbation} as the perturbation exits the fixed point, and that it is therefore legitimate to regard the combination $\epsilon \,{\rm e}^{-\lambda{t}}$ as the associated renormalised coupling, with scaling dimension $\lambda$. Note that this already means we have uncovered a two-fold continuous infinity of relevant directions. 

Finally for $\lambda<4$ we have seen that the behaviour as $\phi\to\infty$, is exponential; for the moment we will concentrate on the even solution $\dV\sim\cosh\omega\phi$. Now, no matter how small we choose $\epsilon$, the large $\phi$ behaviour ensures that the perturbation in \eqref{LPAperturbation} is no longer small and therefore the linearised equation \eqref{eigenLPA} is no longer justified. Fortunately for sufficiently large $\phi$ we can solve the original flow equation \eqref{LPA} instead. Choosing the boundary condition at $t=0$ as $V(\phi,0)= V_*(\phi)-\epsilon\, \dV(\phi)$, where for the moment the minus sign is required to stay within the range \eqref{convbound} as $\phi\to\pm\infty$, we have for large $\phi$ that the right hand side of \eqref{LPA} can be neglected since $V''(\phi)\to-\infty$. Since only the left hand side remains, we can solve to find that $\dV(\phi,t) \sim  {\rm e}^{-4t}\,\cosh\omega\phi$. We see that since $\phi$ carries no scaling dimension, we can absorb the $t$ dependence into a renormalised coupling $g = \epsilon\, {\rm e}^{-4{t}}$ which however has dimension $4$ and not the dimension $\lambda$ we find from the linearised analysis. Since at large $\phi$, $\dV(\phi,t)$ grows exponentially with $t$ as we flow towards the IR, the perturbation is relevant, even if $\lambda<0$. But $\dV(\phi,t)$ does not evolve with a single well-defined scaling dimension. For $\mathcal{O}(1)$ values of $\phi$, $\dV(\phi,t)$ is $\mathcal{O}(\epsilon)$ and its $t$ dependence is given ${\rm e}^{-\lambda{t}}$ (which may be growing or decaying depending on the sign of $\lambda$). At a cross-over region $\omega \phi\sim \ln(1/\epsilon)+4t$, the $t$ dependence changes, and then for much larger $\phi$, the perturbation always grows as ${\rm e}^{-4{t}}$. Since $\lambda<4$, it is this behaviour that gives the dominant amplitude eventually as $t$ reduces, and also we see that the cross-over region moves in towards the origin. 

Note that in general, the existence of the RG flow \eqref{LPA} near $t=0$ requires through \eqref{convbound} that if operators are added with $\lambda<4$,  then the operators added with the lowest $\lambda$ (most positive $\omega$) are such that the one that behaves as $\cosh\omega\phi$ for large $\phi$, has a negative coupling (larger in magnitude than the operator that behaves as $\sinh\omega\phi$ if this is also present). With this restriction in place a general sum over $\lambda<4$ of both even and odd operators can be considered, and we would still establish that this combination is relevant, scaling as ${\rm e}^{-4{t}}$ for large $\phi$.
This does not establish that the flow exists for all $t$, but we would expect that some solutions do exist for all $t$, for example if we choose to restrict to the $\cosh\omega\phi$-type operators and give all these negative couplings. 

Even though the oscillatory perturbations form a continuous spectrum they do form a complete set for the Schr\"odinger equation \eqref{Schrodinger} (for the LPA $\eta=0$ Gaussian these are just the Fourier modes \eqref{lambdag4}) however they span the space of functions that are bounded as $\phi\to\pm\infty$. The exponential modes we have just discussed, lie outside this space and stay outside this space under RG evolution to the infrared, again in contrast to the situation for scalar field theory in $d>2$ dimensions. (We will contrast with the situation for scalar field theory in $d=2$ dimensions in sec. \ref{sec:sft2d} below.)

In summary then, there is a continuous spectrum of perturbations about any fixed point in the line of fixed points, with two eigenoperators per RG eigenvalue $\lambda$. For $\lambda=4$ these are the ones given in \eqref{lambda4}, and might equally be classified as discrete, but they are fully embedded in the continuous spectrum. For  $\lambda\ge4$ there are two operators for each $\lambda$ and they have renormalised relevant couplings with the expected scaling dimension $\lambda$. For all $\lambda<4$,  the two perturbations do not have well defined scaling dimensions but nevertheless are relevant, the latter being in contradiction with the na\"\i ve answer for $\lambda<0$. This is reflected in their large $\phi$ dependence, which eventually takes over the whole function, where it grows as ${\rm e}^{-4{t}}$ characteristic of an associated coupling of scaling dimension $4$ and independent of the value of $\lambda<4$.

\subsection{Comparison to scalar field theory in two dimensions}\label{sec:sft2d}

As we have reviewed in sec. \ref{sec:sft}, 
the background independent flow equations \eqref{equ:sys-red-final}  are closely related to those of scalar field theory. After the map $V\mapsto-V$ and a change in parametrisation of the scaling dimension, the result is the flow equations for scalar field theory except for an overall sign on the right hand side of the flow equations, and with however important differences in physical interpretation. At the level of the form of LPA discussed in this section, the differences in physical interpretation also lead to a mathematical difference since we here assume the field to have zero overall scaling dimension. Therefore despite the fact that the momentum integral on the right hand side is fundamentally four-dimensional, the flow equation most closely resembles the LPA description of scalar field theory in two dimensions. There also, since the $\phi V'$ term is missing from the left hand side, the fixed point solutions can be studied by means of an effective Newtonian potential $U$ \cite{Morris:1994jc}. Since $U$ was bounded below, this resulted in a semi-infinite line of periodic solutions for $V_*(\phi)$, corresponding to critical sine-Gordon models \cite{Morris:1994jc}. With the periodicity of the field thus fixed, the eigen-operator is discrete since it must have the same periodicity.
The overall sign difference on the right hand side however maps $U\mapsto-U$ and as we have seen, this then results instead in a semi-infinite line of fixed point solutions that support an eigen-operator spectrum which remains continuous, lacking the quantization that comes from either periodicity or leading large field RG constraints \cite{Morris:1996nx,Morris:1996xq,Morris:1998,Bridle:2016nsu}.

%
%


\section{Fixed points and eigen-operators at order derivative squared}
\label{sec:fpbeyondLPA}
The fixed point equations pertaining to the full system \eqref{equ:sys-red-final} in $d=4$ dimensions and with $n=2$ in the cutoff \eqref{equ:cutoff} are
\begin{subequations}
\label{equ:sysfp}
\begin{align}
 4V_*-\frac{\eta}{2}\phi V_*' &= -\left(8-\eta\right)\int_0^\infty \frac{dp}{p}\,Q_0 , \label{equ:fpV} \\
(2-\eta)K_*-\frac{\eta}{2}\phi K_*' &= -2(8-\eta)\int_0^\infty \frac{dp}{p}\,P\big(p^2,\phi\big) , \label{equ:fpK}
\end{align}
\end{subequations}
where we continue to omit the hats and $Q_0$ and $P$ are now given by the corresponding versions of \eqref{Q0final} and \eqref{Pfinal},
\begin{subequations}
\label{Q0Pfp}
\begin{align}
 Q_0 = &\left[V_*''-K_*p^2-\frac{1}{p^4}\right]^{-1} \label{Q0fp} \\
 P = &-\frac{1}{2}K_*'' Q_0^2 + K_*'\left(2V_*'''-\frac{9}{4} K_*' p^2\right)Q_0^3  \notag \\
    & +\left[\left\{2K_*' p^2 - V_*'''\right\}\left(K_*-\frac{2}{p^6}\right)+ \frac{3}{p^6}\left(K_*' p^2-V_*'''\right)\right]\left(V_*'''-K_*' p^2\right) Q_0^4 \label{Pfp} \\
    & -p^2 \left(K_*-\frac{2}{p^6}\right)^2 \left(V_*'''-K_*' p^2\right)^2 Q_0^5, \notag
\end{align}
\end{subequations}
and we now allow for a non-vanishing anomalous dimension.

As to the general structure of the system of fixed point equations, differentiating \eqref{equ:fpV} once, solving for the third derivative $V_*'''$ and substituting the result into \eqref{equ:fpK} reveals that \eqref{equ:sysfp} is of second order in both $V_*$ and $K_*$ and therefore admits a four dimensional space of local solutions around any generic initial value $\phi=\phi_0$. For the present purpose of finding global solutions valid on the whole real line $-\infty<\phi<\infty$, we can take $\phi_0=0$ and start with the local parameter space spanned by $V_*(0), V_*'(0), K_*(0), K_*'(0)$. Since there are no explicit appearances of the field $\phi$ in \eqref{Q0Pfp} the fixed point equations do not feature any fixed singularities. However from \eqref{Q0fp}, the generalisation of \eqref{convbound} now takes the form,
\begin{equation}
\label{convboundfull}
V_*''(\phi) < 3\left(\frac{K_*(\phi)}{2}\right)^{2/3} \qquad \text{and} \qquad K_*(\phi)>0\, ,
\end{equation}
and any violation of these inequalities on a finite range for $\phi$ will lead to a moveable singularity, thus placing a restriction on parameter space.

Indeed the integrals converge for $p\to\infty$ if and only if $K_*\ne0$, while $Q_0$ diverges at finite positive $p$ if $K_*$ is negative. If $K_*>0$, the polynomial $p^4/Q_0 = p^4 V_*'' -K_*p^6-1$ reaches a maximum at $p^2 = 2V_*''/3K_*$, where it takes the value
\be
\label{peak}
\frac{p^4}{ Q_0} =  \frac{4}{27} \frac{{V_*''}^3}{ K_*^2} -1\,. 
\ee
Clearly if this maximum is negative, $Q_0$ is negative and finite over the whole integration range, and if it
is otherwise then $p^4/Q_0$ crosses or touches the axis, and again $Q_0$ will diverge for finite $p$. This gives the first inequality in \eqref{convboundfull}.


Rescaling all quantities in \eqref{equ:sysfp} with the power of a real number as given in \eqref{scalings} 
leaves the fixed point equations unchanged. This can be exploited to eliminate one parameter of solution space. Note however that the scaling prescriptions do not allow to change the sign of either $V_*$ or $K_*$. From the inequalities in \eqref{convboundfull} it is therefore convenient to eliminate the parameter $K_*(0)$ by fixing it to $K_*(0)=2$.

Finally, since the fixed point equations \eqref{equ:sysfp} are symmetric under $\phi \mapsto -\phi$ one may choose to impose $V_*'(0)=K_*'(0)=0$ to restrict to even fixed point solutions. It has to be emphasised however that at this point requiring either $V_*$ or $K_*$ or both to be even is an additional assumption. We will address this further in the discussion and conclusions, sec. \ref{sec:conclusions}.

Following this route and regarding the anomalous dimension as just an additional parameter, we so far find from parameter counting that we are left with only the two parameters $V_*(0)$ and  $\eta$. In general however, an asymptotic analysis of the fixed point equations \eqref{equ:sysfp} is needed to capture possible constraints on parameter space as $\phi \to \infty$ and to arrive at conclusive results for parameter counting \cite{Morris:1994ie,Morris:1994jc,DietzMorris:2013-1}. 

From the structural similarity of \eqref{equ:sysfp} to standard scalar field theory, as discussed in ref. \cite{Dietz:2015owa} and sec. \ref{sec:sft},
one may be led to investigating the corresponding asymptotic behaviour given to leading order by solving the left hand sides of the fixed point equations, 
\begin{equation}
\label{sfasy}
V_*(\phi)=A\phi^{8/\eta} + \dots \qquad \text{and} \qquad K_*(\phi) = B\phi^{4/\eta-2} +\dots \, ,
\end{equation}
for constants $A,B$ (assuming $\eta\ne0$). From this one finds that for $0<\eta<8$ the dominant term at large field in the first inequality in \eqref{convboundfull} is $V_*''$ and the only way to avoid a moveable singularity is therefore to have $A<0$, leading to a potential unbounded below. We will discuss the implications of this in sec. \ref{sec:conclusions}. In fact our numerical investigations uncovered only fixed point potentials that are bounded below, and this is also what we found in
our LPA study in sec. \ref{sec:confLPA}. Expanding \eqref{Q0fp} in $V''_*$ we have:
\begin{equation}
\label{expQ0}
Q_0 = -\frac{p^4}{K_* p^6+1}-\frac{p^8}{(K_*p^6+1)^2}\,V_*''-\frac{p^{12}}{(K_*p^6+1)^3}(V_*'')^2-\dots \,,
\end{equation}
The fixed point equation \eqref{equ:fpV} for the potential then evaluates to
\begin{equation}\label{odeVexp}
4V_*-\frac{\eta}{2}\phi V_*' = \frac{(8-\eta)\pi}{3\sqrt{3}}\left(\frac{1}{K_*^{2/3}}
+\frac{1}{3}\frac{V_*''}{K_*^{4/3}}+ \frac{\sqrt{3}}{4\pi}\frac{(V_*'')^2}{K_*^2}+ \dots \right).
\end{equation}
We see \textit{a posteriori} that this expansion is useful as long as, asymptotically for large $\phi$, $V_*''/K_*^{2/3}$ is small. With the assumed asymptotic form \eqref{sfasy}, this is the case precisely for $\eta<0$ or $\eta>8$. A brief calculation shows however, that for these ranges of $\eta$ the right hand side in \eqref{odeVexp} cannot be neglected compared to the left hand side and thus that the asymptotic behaviour \eqref{sfasy} is not consistent. 
Since $\eta=8$ is excluded by \eqref{bad-n2},
standard scalar field theory asymptotic behaviour 
seems therefore to be completely excluded. 
%
%
We therefore assume that the leading asymptotic behaviour of solutions to \eqref{equ:sysfp} is not determined by scaling dimensions, meaning that
the quantum corrections on the right hand side of \eqref{equ:sysfp} cannot be neglected in the large field regime. While this is surprising from the point of view of scalar field theory, the same situation was encountered in ref. \cite{DietzMorris:2013-1} for the asymptotic behaviour in the $f(R)$ truncation.
A much more comprehensive asymptotic analysis is therefore required in the present case and we come back to this in sec. \ref{sec:asymptotics}.

\subsection{Numerical solution}\label{sec:numerical}

\begin{figure}[ht]
\begin{center}
$
\begin{array}{cc}
\includegraphics[width=0.45\textwidth]{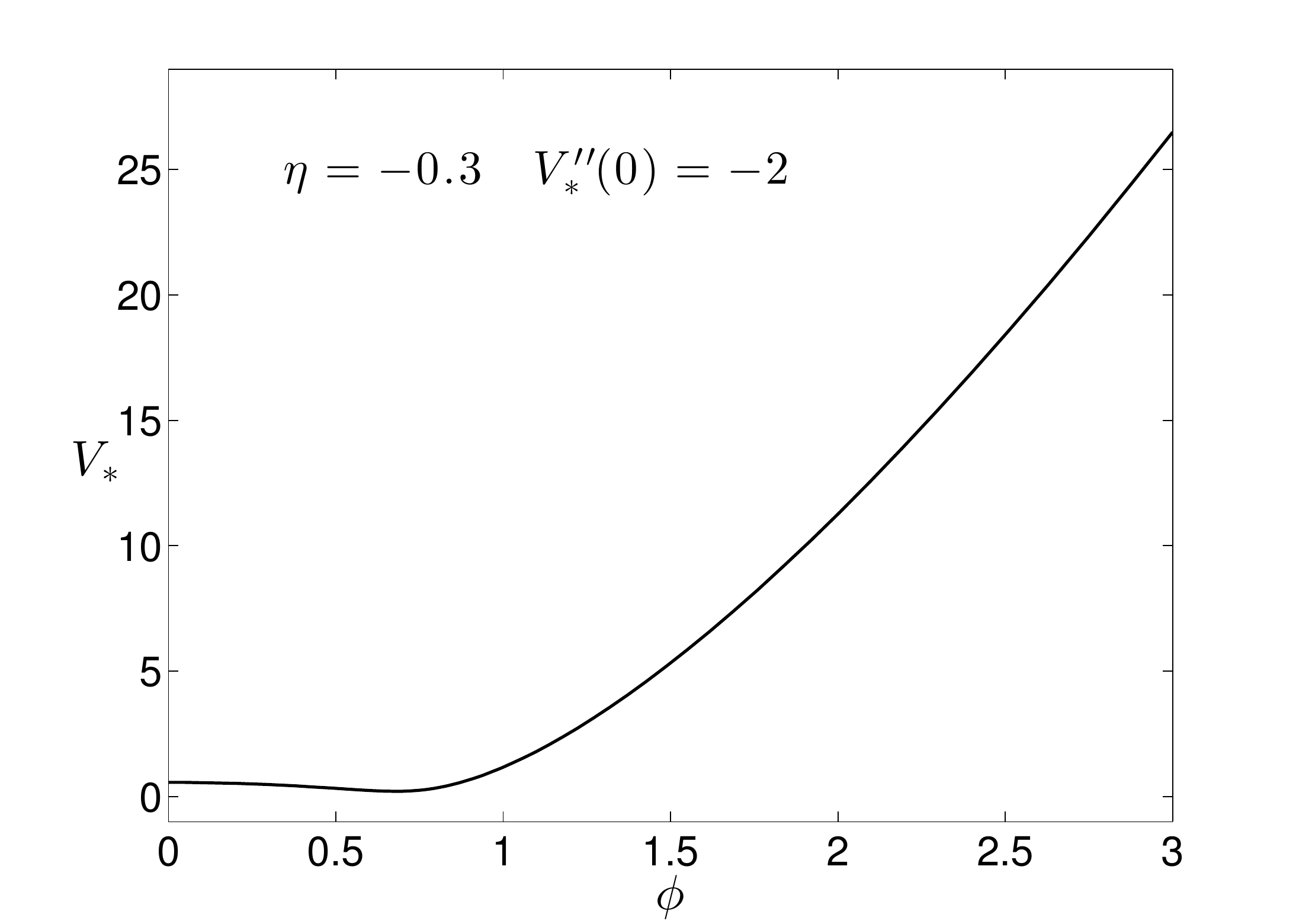} &
\includegraphics[width=0.45\textwidth]{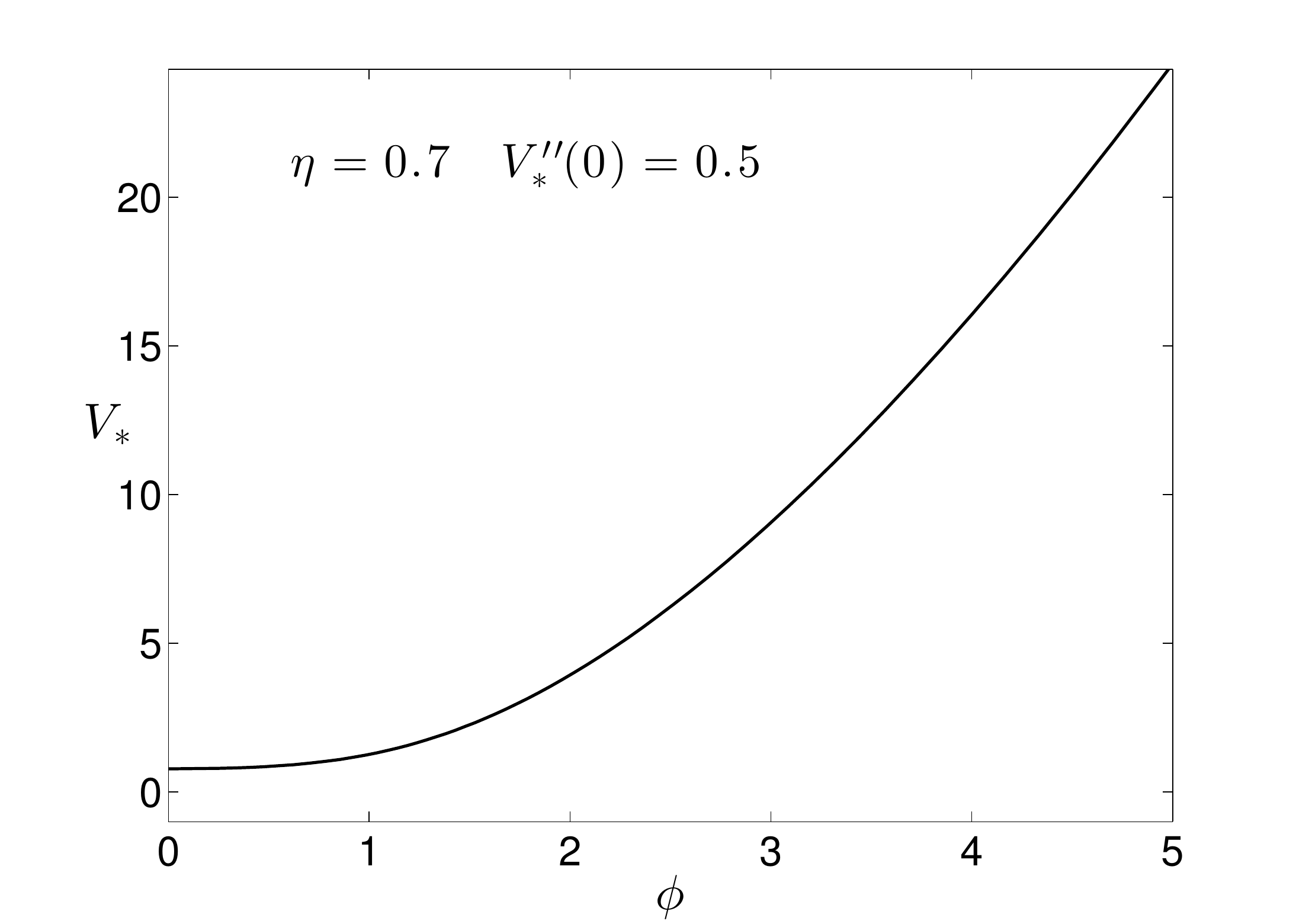} \\
\includegraphics[width=0.45\textwidth]{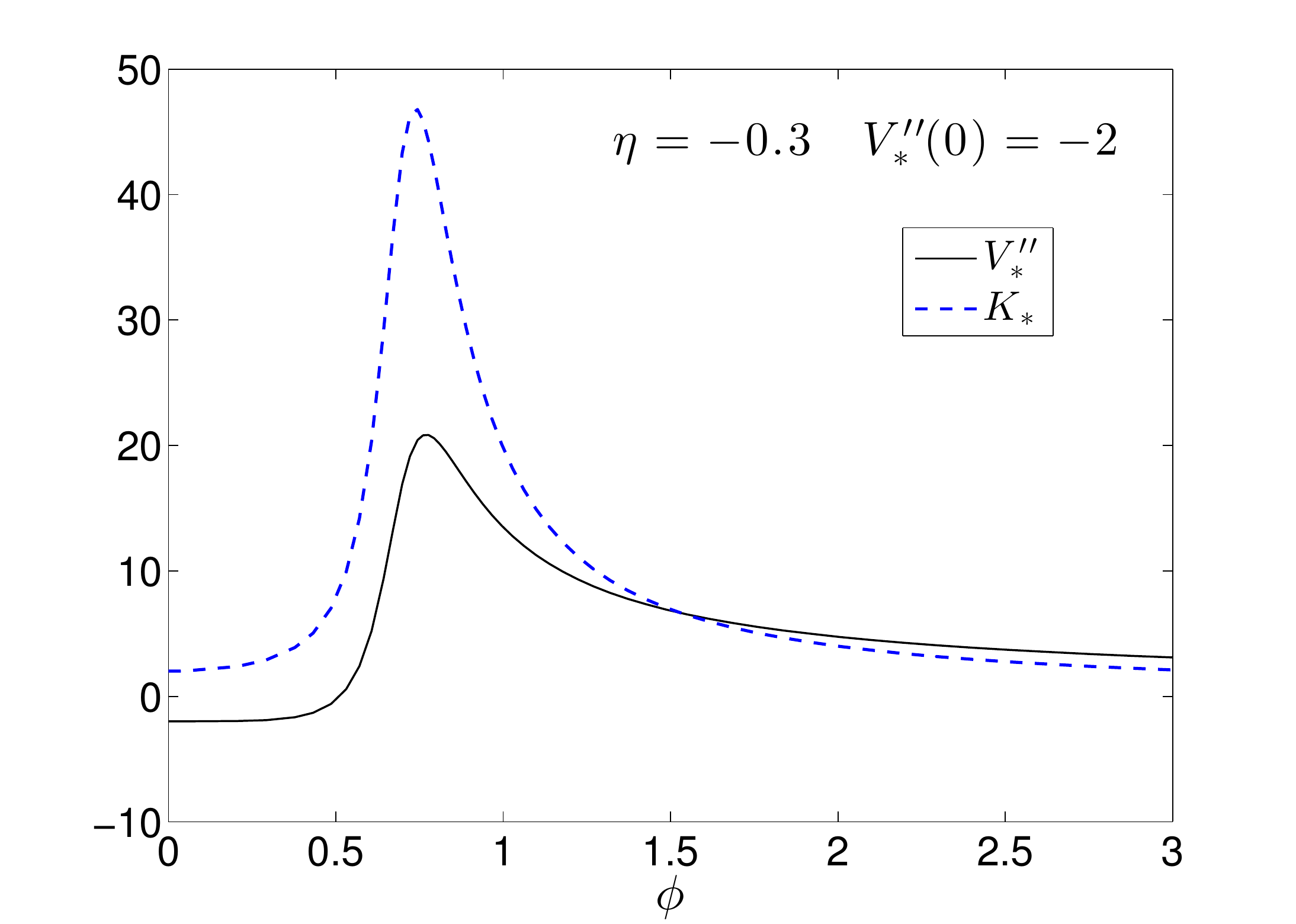} &
\includegraphics[width=0.45\textwidth]{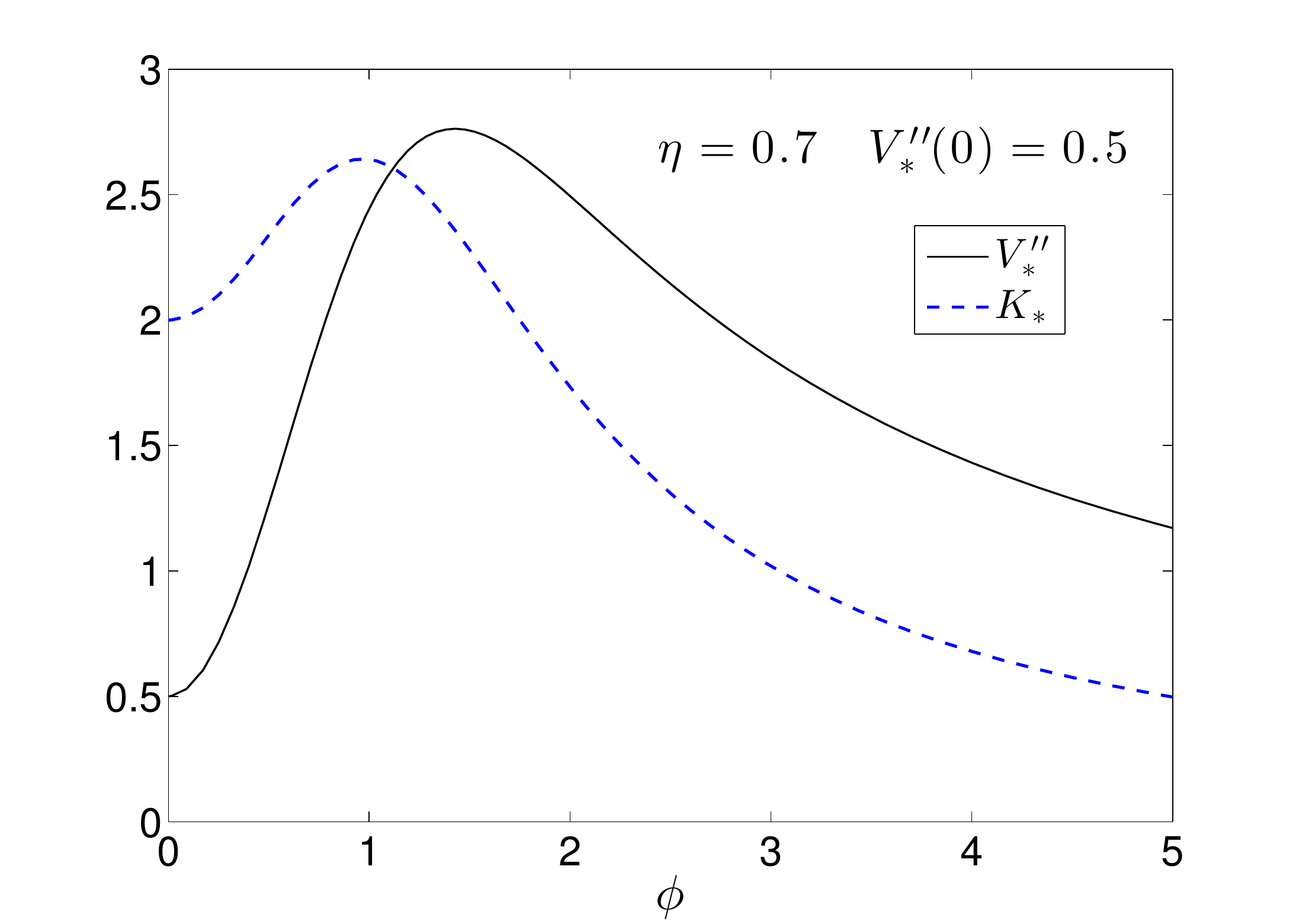}
\end{array}
$
\end{center}
\caption[Numerical integration of conformal fixed point equations]{Numerical integration of \eqref{equ:sysfp} for $\eta=-0.3, \,V_*''(0)=-2$ in the left panel and $\eta=0.7,\, V_*''(0)=0.5$ on the right. The top panels display $V_*(\phi)$ while the bottom panels display both $V''_*(\phi)$ and $K_*(\phi)$. The other initial conditions have been fixed as discussed in the text.}
\label{fig:confexsols}
\end{figure}

In principle, the integrals on the right in \eqref{equ:sysfp} can be evaluated using contour integration in the complex plane. The length of the resulting expressions is however such that they become unmanageable. For integrating \eqref{equ:sysfp} numerically, it is therefore advisable to also perform a numerical evaluation of the integrals at each step of the solver. To bring the system \eqref{equ:sysfp} into normal form for actual computations, we solve the differentiated version of \eqref{equ:fpV} for $V_*'''$ and trade the initial condition $V_*(0)$ for $V_*''(0)$, while eqn. \eqref{equ:fpK} is easily solved for the highest derivative $K_*''$. 

Fig. \ref{fig:confexsols} shows one example integration for negative anomalous dimension on the left and a second for positive anomalous dimension on the right.
In both cases the numerical integration can be carried out to arbitrarily large field, limited only by the efficiency of the solver, and it is interesting to note that the constraints \eqref{convboundfull} seems to saturate asymptotically (ruling out an expansion of the form \eqref{odeVexp} in fact). This numerical evidence will provide the clue to solving the asymptotic behaviour, as we will see in the next section.

By varying the two parameters $\eta$ and $V_*''(0)$ one finds that solving the fixed point equations \eqref{equ:sysfp} numerically is in general not hampered by the appearance of moveable singularities as caused by violation of \eqref{convboundfull} at finite field. However, as for parameter counting, an effective and comprehensive numerical analysis of the system \eqref{equ:sysfp} has to build on a thorough understanding of the fixed point solutions at large field. 

\subsection{Asymptotic regime of the fixed point equations}
\label{sec:asymptotics}
Numerically one finds that for many initial conditions at $\phi=0$ the system of fixed point equations \eqref{equ:sysfp} leads to solutions that for large $\phi$ tend to saturate the first inequality in \eqref{convboundfull} by forcing the maximum of $p^4/Q_0$ ever closer to the axis.  This can be exploited to expand in an asymptotically small function so that computing the integral on the right hand side of the fixed point equations results in a comparatively simple expression. To this end we proceed by setting
\begin{equation}
\label{uv}
K_*(\phi) = 2\,u(\phi)^3, \qquad \text{and} \qquad V_*''(\phi)=3\,u(\phi)^2\left[1-v(\phi)^2\right],
\end{equation}
where we omit the corresponding asterisk on the the new functions $u(\phi)$ and $v(\phi)$, and we note from \eqref{convboundfull} that $u(\phi)>0$. From \eqref{peak}, $p^4/Q_0$ touches the axis asymptotically if
\be
\label{limitv}
v(\phi)  \to 0 \qquad \text{for} \qquad \phi \to \infty\,.
\ee
As mentioned before, the integrals on the right hand side of \eqref{equ:sysfp} can be evaluated using contour integration in the complex plane. Suppose in general that we want to compute the integral $\mathcal{I}=\int_0^\infty dx\, H(x)$, where $H$ is assumed to have the necessary properties for the following discussion. We then consider the auxiliary function $F(z)=H(z)\ln(z)$ in the complex plane with the branch cut of the logarithm located on the ray of non-negative real numbers. Using the appropriate contour for this branch cut and the property $\lim_{\eps\to0} \ln(x+i\eps) = 2\pi i + \lim_{\eps\to 0} \ln(x-i\eps)$ for real $x>0$, the residue theorem leads to
\begin{equation}\label{intstrategy}
\mathcal{I} = \int_0^\infty \!\!\!dx\, H(x) = -\sum_n \mathrm{Res}\left\{H(z)\ln(z), z_n\right\},
\end{equation}
with the sum encompassing all residues of the function $F(z)$ in the complex plane. The location of each residue in this formula is given by $z_n$.

We now use this technique to derive the asymptotic form of the fixed point equations \eqref{equ:sysfp} valid in the regime characterised by \eqref{limitv}. By analysing these equations we will gain crucial insights into the large field behaviour of the fixed point solutions that will allow us to reliably apply the parameter counting method \cite{Morris:1994ie,Morris:1994jc,DietzMorris:2013-1}. The locations $z_n$ of the residues needed in \eqref{intstrategy} for the evaluation of the integrals in \eqref{equ:sysfp}, follow from  \eqref{Q0fp}, and are given by the following cubic complex polynomial equation (which is nothing but $p^4/Q_0$ after the change of variables \eqref{uv} and with $z=p^2$):
\begin{equation}\label{Q0poly}
1+2\,u^3 z^3+3u^2(v^2-1)z^2=0.
\end{equation} 

The zeros of \eqref{Q0poly} and the corresponding residues in \eqref{intstrategy} needed for the integrals in  \eqref{equ:sysfp} can then be expanded as a series in $v(\phi)$. Differentiating \eqref{equ:fpV} once, and using \eqref{uv}, thus allows $V'_*(\phi)$ to be expressed as a series in $v(\phi)$. Substituting this back into \eqref{equ:fpV} allows $V_*(\phi)$ itself to be expressed as a series in $v(\phi)$. On the other hand differentiating \eqref{equ:fpV} twice allows all occurrences of $V_*$ (and $K_*$) to be eliminated in favour of $u$ and $v$ via \eqref{uv}. Doing the same with \eqref{equ:fpK}, and keeping track of different orders by introducing the book-keeping parameter $\eps$ via the replacement $v(\phi)\mapsto\eps v(\phi)$, the system  then takes the following form:
\begin{subequations}
\label{sys-asy}
\begin{align}
3u\left[(\eta-4)u+\eta\phi u'\right] +\mathcal{O}\!\left(\eps^2\right) =&\, \frac{(\eta-8)\pi}{6u^4 v^3\,\eps}\left(2v'^2u^2 -2u''v^2u+4v'u'uv-vv''u^2+6u'^2v^2\right) \notag \\
 & +\frac{(\eta-8)(3+\ln(2))}{9 u^4}(u''u-3u'^2) + \mathcal{O}\!\left(\eps\right) \,, \label{veq}\\
u^2\left[2(\eta-2)u+3\eta\phi u'\right] =&\, \frac{(\eta-8)\pi}{72v^5u^3\,\eps^3}\left(12u''v^2u+15v'^2u^2-30v'u'vu-41u'^2v^2\right) \notag \\ &+\mathcal{O}\!\left(1/\eps\right). \label{ueq}
\end{align}
\end{subequations}
As indicated in these equations, the left hand side of the first equation does not have an $\mathcal{O}\!\left(\eps\right)$ piece and the left hand side of the second equation is exact, while its right hand side has a vanishing $\mathcal{O}\!\left(\eps^{-2}\right)$ term. These equations can easily be derived to higher orders in $\eps$ but we have displayed only the terms needed in the following.

Using power law ans\"atze in the system \eqref{sys-asy}, one finds that
\begin{equation}
\label{u0v0}
u(\phi)=u_0(\phi) := A\,\phi^{-\frac{1}{4}(2+q)}\,,\qquad v(\phi)=v_0(\phi) := -\frac{\pi q(q-2)(\eta-8)}{18 A^4((q-2)\eta+16)}\,\phi^q 
\end{equation}
with the exponent
\begin{equation}
\label{q}
q=-\frac{86}{331}-\frac{8}{331}\sqrt{219} \approx -0.6175
\end{equation}
and $A$ a real parameter, solves \eqref{veq} by balancing the left hand and right hand sides but without the $\mathcal{O}\!\left(\eps^0\right)$ term on the right hand side, and solves just the right hand side of \eqref{ueq}. We first note that $v_0(\phi) \to 0$ 
for $\phi \to \infty$ as required. Furthermore, one easily confirms that $u_0$ and $v_0$ are indeed valid leading terms for $u$ and $v$ by verifying that the left hand side of \eqref{ueq} and the second term on the right of \eqref{veq} are subleading. One can also explicitly confirm that higher orders in $\eps$ are also subleading, as expected, but this will become evident in a moment.

Converting the solutions \eqref{u0v0} back to work out the leading asymptotic behaviour for $K_*$ and $V_*$ we find
\be 
\label{VKleading}
K_* \sim 2 A^3 \phi^{-\frac{3}{4}(2+q)}\qquad{\rm and}\qquad V_* \sim \frac{12A^2}{q(q-2)}\phi^{1-q/2}\,.
\ee
Numerically the exponents are given by $-\frac{3}{4}(2+q)\approx-1.037$ and $1-q/2\approx1.309$.

We now proceed to work out explicitly the sub-leading terms. We immediately see from \eqref{u0v0} that the expansion in $v(\phi)$ will become an expansion in $\phi^q/A^4$ and thus that we should regard
$u(\phi)^4$ as accompanied by a factor of $1/\eps$. This observation motivates the further change of variables
\begin{equation}
\label{w}
w(\phi)=1/u(\phi)^4, \qquad w(\phi) \mapsto \eps w(\phi).
\end{equation}
Eliminating $u(\phi)$ in favour of $w(\phi)$ and $\eps$ in this way does not change the relative orders as expressed by powers of $\eps$ on the right hand sides of \eqref{equ:sysfp}. This can be seen from the two terms on the right hand side of \eqref{veq} and can be confirmed for higher orders not displayed for both \eqref{veq} and \eqref{ueq}. At the same time, the change \eqref{w} leads to the left hand side of \eqref{veq} being of the same order in $\eps$ as the first term on the right hand side, as implied by the solution \eqref{u0v0}.

Since the relevant terms of the system \eqref{sys-asy} are second order differential equations in $u$ and $v$, or equivalently in $w$ and $v$, the equations have three additional 
solutions beyond \eqref{u0v0}. Due to the non-linearity of the relevant terms in \eqref{sys-asy} it is difficult to find explicit expressions for these additional solutions. However, for the purposes of the asymptotic analysis here, a full investigation of the non-linear leading terms of \eqref{sys-asy} is not necessary, since we will already find from analysing the leading corrections to \eqref{u0v0}, that  it is part of a solution for which the dimension of parameter space is unrestricted and which for large $\phi$ match on well to  the numerical solutions we have found, such as those in fig. \ref{fig:confexsols}. 

\begin{figure}[ht]
\centering
\includegraphics[width=0.7\textwidth]{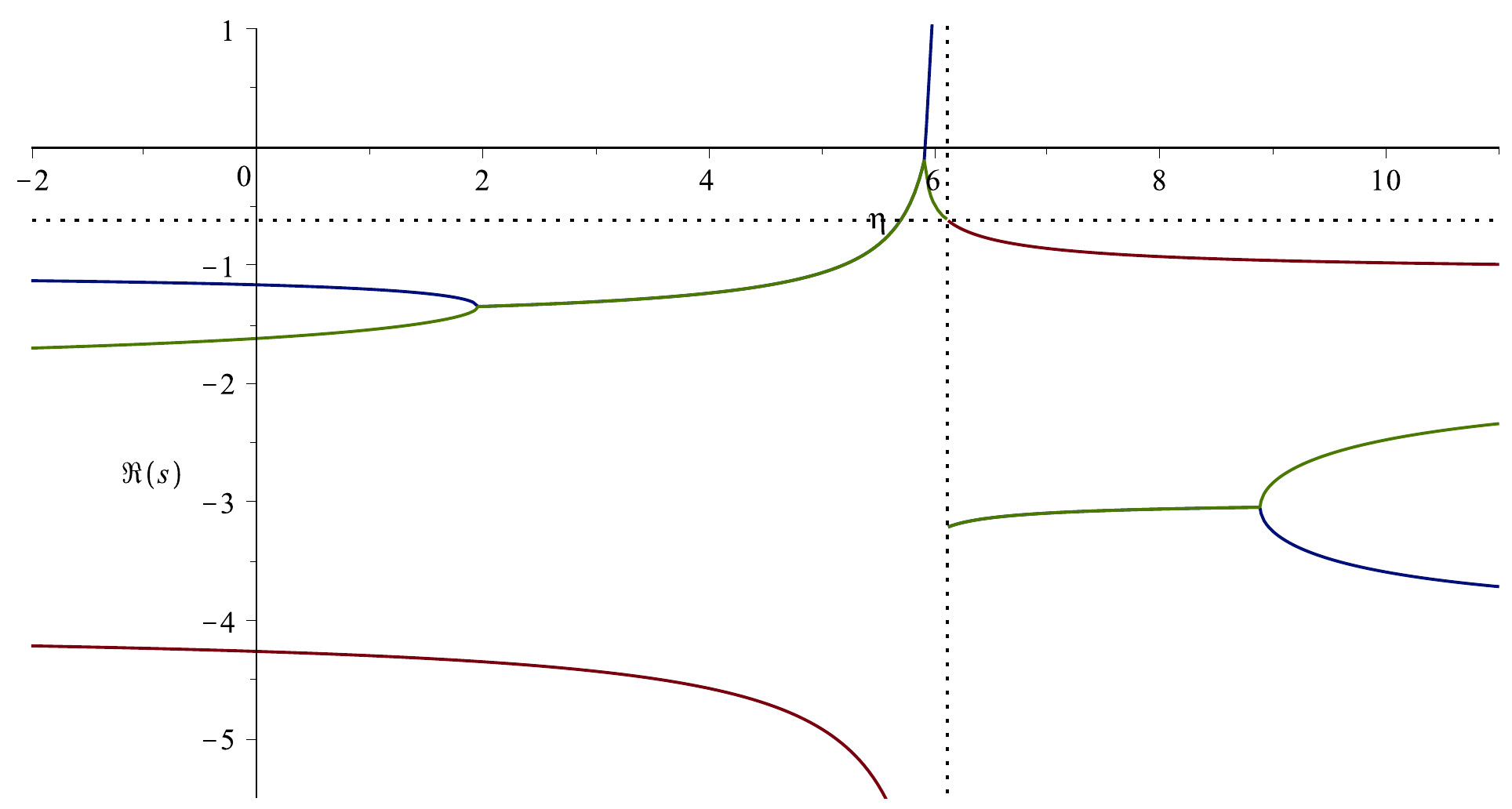} 
\caption{The three exponents $s_1,s_2,s_3$ of the solution \eqref{solshom} as a function of the anomalous dimension $\eta$ in the range $-2\leq \eta \leq 11$. The  dotted lines mark  $s=q$ and $\eta=16/(2-q)$. For $\eta$ outside the plotted range, all exponents are real and satisfy $s_i<q$. Whenever the exponents are complex, the plot shows only the real part. This happens for the upper curves for 
$2.0\lesssim \eta\lesssim5.9$, and for the lower curves from $16/(2-q)\le \eta\lesssim8.9$.}
\label{fig:exponents}
\end{figure}

To find the leading corrections, we implement the change of variable \eqref{w} and substitute
\begin{equation}
\label{linwv}
w(\phi) = w_0(\phi) + \eps w_1(\phi) \qquad \text{and} \qquad v(\phi) = v_0(\phi) +\eps v_1(\phi),
\end{equation}
in \eqref{sys-asy}, where $w_0=1/u^4_0$ and $v_0$ are the leading solutions from \eqref{u0v0}. 
As we will see, this next order in $\eps$ gives us both the next terms in the expansion in powers of $\phi^q$ and new powers $\phi^{s_i(\eta)}$ which for a range of $\eta$ are all sub-leading compared to $\phi^q$.
All terms in \eqref{veq} now contribute to give the result:
\begin{multline}
\label{v1eq}
2\phi^2[\eta(q-2)+16]\, w_1'' 
- \phi[\eta (7q+4)(q-2)+32(3q+2)]\, w_1' \\
+ 2[\eta(3q+2)(q-2)(q+1)+8(5q^2+8q+2)]\, w_1 \\
+\frac{18(\eta(q-2)+16)^2 \phi^2}{q(q-2)(\eta-8)\pi}\left\{4\phi^2 \,v_1''-4\phi(3q-2)\, v_1' +3q(3-2) \,v_1\right\} \\
+\frac{3+\ln 2}{54 \sqrt{A}}(\eta-8)q^2(q-2)(q+2)\phi^{2q+2} =0.
\end{multline}
After the change \eqref{w}, the left hand side of \eqref{ueq} becomes $\mathcal{O}\!\left(1/\eps\right)$ and thus still does not contribute. The linearisation of the right hand side is:
\begin{multline}
\label{w1eq}
24\phi^2 \,w_1'' - \phi(79q+38)\,w_1' + (q+2)(55q+14)\,w_1 \\
+\frac{27(\eta(q-2)+16)\phi^2}{q(q-2)(\eta-8)\pi}\left\{40\phi(5q+2)\,v_1'-(531q^2+252q-20)\,v_1\right\}=0.
\end{multline}
The first equation \eqref{v1eq} is non-homogeneous due to the last term  on its left hand side. The associated particular solution (of both equations), which does not contain a new parameter, is found to take the form:
\begin{subequations}
\label{partsol}
\begin{align}
w_1(\phi)&= \frac{c}{36 \sqrt{A}}(131q^2+92q-20) \,\phi^{2q+2}, \\
v_1(\phi)&= -\frac{c}{972\sqrt{A}}\frac{\pi q^2(q-2)(7q-34)(\eta-8)}{(q-2)\eta+16}\, \phi^{2q},
\end{align}
\end{subequations}
where the constant $c$ is a function of the exponent $q$ and the anomalous dimension:
\begin{equation}
c= \frac{(q-2)(q+2)(3+\ln 2)(\eta-8)}{(q-2)(7q-34)(q+2)\eta - 8(379q^2+316q+76)}.
\end{equation}
Using this solution to eliminate the non-homogeneous term in \eqref{v1eq}, the four dimensional solution space of the remaining homogeneous system is then made up of the already known leading solutions $w_0$ and $v_0$ as well as three additional power law solutions: 
\begin{subequations}
\label{solshom}
\begin{align}
w_1(\phi) &= \left(B_1 \phi^{\,s_1} + B_2 \phi^{\,s_2} +B_3 \phi^{\,s_3}\right)\phi^2	\\
v_1(\phi) &= \kappa_1 B_1 \phi^{\,s_1} + \kappa_2 B_2 \phi^{\,s_2} +\kappa_3 B_3 \phi^{\,s_3}
\end{align}
\end{subequations}
Here, the $B_i$ are free parameters, the $\kappa_i$ are relative normalisation constants  and are lengthy functions of the anomalous dimension $\eta$, the leading exponent $q$, and the power $s_i$. The powers $s_i$ are the three roots of the polynomial
\begin{align}
\label{s-poly}
0= & \, 192[\eta(q-2)+16]\,s^3 - 8[\eta(q-2)+16](277q+2)\, s^2 \notag \\ 
&+ [2(q-2)(3969q^2-16q-164)\eta+32(3669q^2-136q-164)]\, s \\
&-3(q-2)(5q+2)(637q^2-364q+20)\eta-96(1327q^3-399q^2-304q+20). \notag
\end{align}
While for the particular solution \eqref{partsol} the anomalous dimension appears only in one of the coefficients and the exponents are independent of $\eta$, the exponents of the solutions \eqref{solshom} all depend on the anomalous dimension. Their values are plotted in fig. \ref{fig:exponents}. In the indicated ranges, two of them become a complex conjugate pair $a \pm ib$, in which case the corresponding real solutions are 
\be \label{complex-exponent}
\phi^a\cos(b\ln\phi)\qquad{\rm and} \qquad\phi^a\sin(b\ln \phi)\,.
\ee

Collecting the solutions \eqref{u0v0} with \eqref{w}, \eqref{partsol} and \eqref{solshom}, as well as re-naming $B_0=1/A^4$, the generic asymptotic fixed point behaviour in the regime \eqref{limitv} is therefore:
\begin{subequations}
\label{solsasy}
\begin{align}
w(\phi) &= \phi^2 \left\{ B_0\,\phi^{q} +\frac{cB_0^2}{36}(131q^2+92q-20) \,\phi^{2q}+B_1 \phi^{\,s_1} + B_2 \phi^{\,s_2} +B_3 \phi^{\,s_3}\right\}\,, \\
v(\phi)&= -\frac{\pi q(q-2)(\eta-8)B_0}{18((q-2)\eta+16)}\,\phi^q -\frac{cB_0^2}{972}\frac{\pi q^2(q-2)(7q-34)(\eta-8)}{(q-2)\eta+16}\, \phi^{2q}\\ 
&  \hspace{5cm} +\kappa_1 B_1 \phi^{\,s_1} + \kappa_2 B_2 \phi^{\,s_2} +\kappa_3 B_3 \phi^{\,s_3}\,,\notag
\end{align}
\end{subequations}
where the $s_i$ are the three roots of \eqref{s-poly}. These sub-leading contributions are included only if they are genuinely subleading, \ie the exponents satisfy  $\Re(s_i)<q$. As shown in the plot fig. \ref{fig:exponents}, we find that for $\eta \in \mathcal{R}=[\eta_c,\frac{16}{2-q})\approx[5.7003, 6.113)$ two roots violate this condition. At the upper limit  the $s^3$ and $s^2$ terms in \eqref{s-poly} simultaneously vanish, and the only root is $s=q$. A more detailed analysis is therefore needed for this one point, which we will not further pursue here. Otherwise we see that for $\eta\in\mathcal{R}$ we have only one sub-leading correction $\phi^{s_1}$ ($s_1<q$ and real), while for $\eta\notin\mathcal{\bar R}=[\eta_c,\frac{16}{2-q}]$, \ie outside the closed interval, all three roots satisfy $\Re(s_i)<q$.

 For $\eta\in\mathcal{R}$ there are thus in total two free parameters: $B_0,B_1$. Since four parameters are expected in general, we see that asymptotically two restrictions are placed on the parameter space.
 As discussed before, the fixed point equations \eqref{equ:sysfp} enjoy the scaling symmetry \eqref{scalings} which can be used as a condition on parameter space. Furthermore we chose to restrict to even solutions by imposing the conditions
$V_*'(0)=K_*'(0)=0$. Since we can be confident that these five conditions act independently, this over-constrains the parameter space at fixed $\eta$, or alternatively provides exactly the right number of conditions for $\eta$ a free parameter. In other words, for $\eta\in\mathcal{R}$ we can expect at most a discrete set of fixed point solutions with quantized value of $\eta$. We emphasise however that the counting argument does not guarantee that this discrete set is non-empty. Especially since $\eta$ is already restricted to the small range $\mathcal{R}$, it seems likely that there are in fact no solutions in this range.

\begin{figure}[ht]
\begin{center}
$
\begin{array}{cc}
\includegraphics[width=0.45\textwidth, height=0.25\textheight]{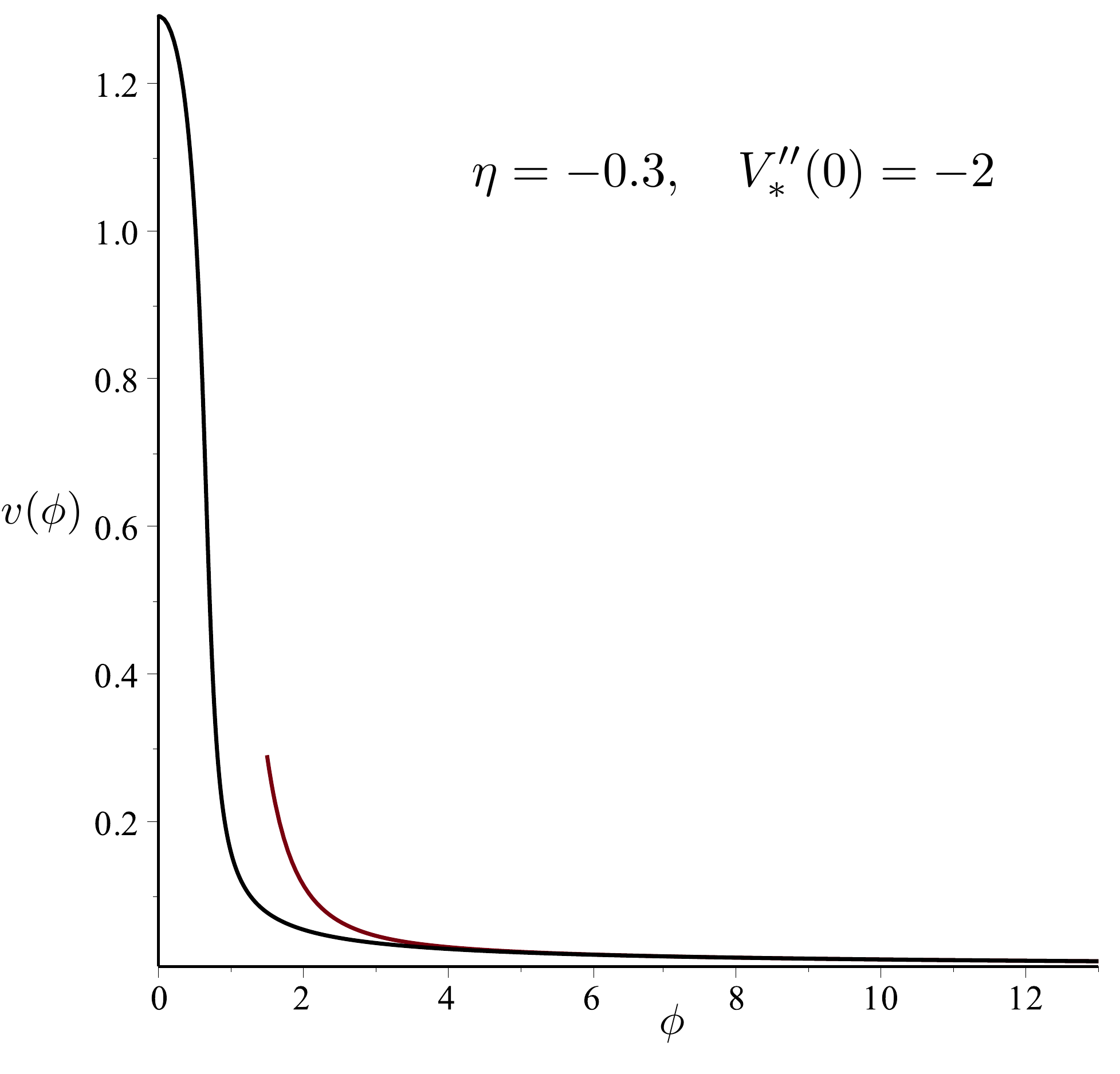} &
\includegraphics[width=0.45\textwidth, height=0.25\textheight]{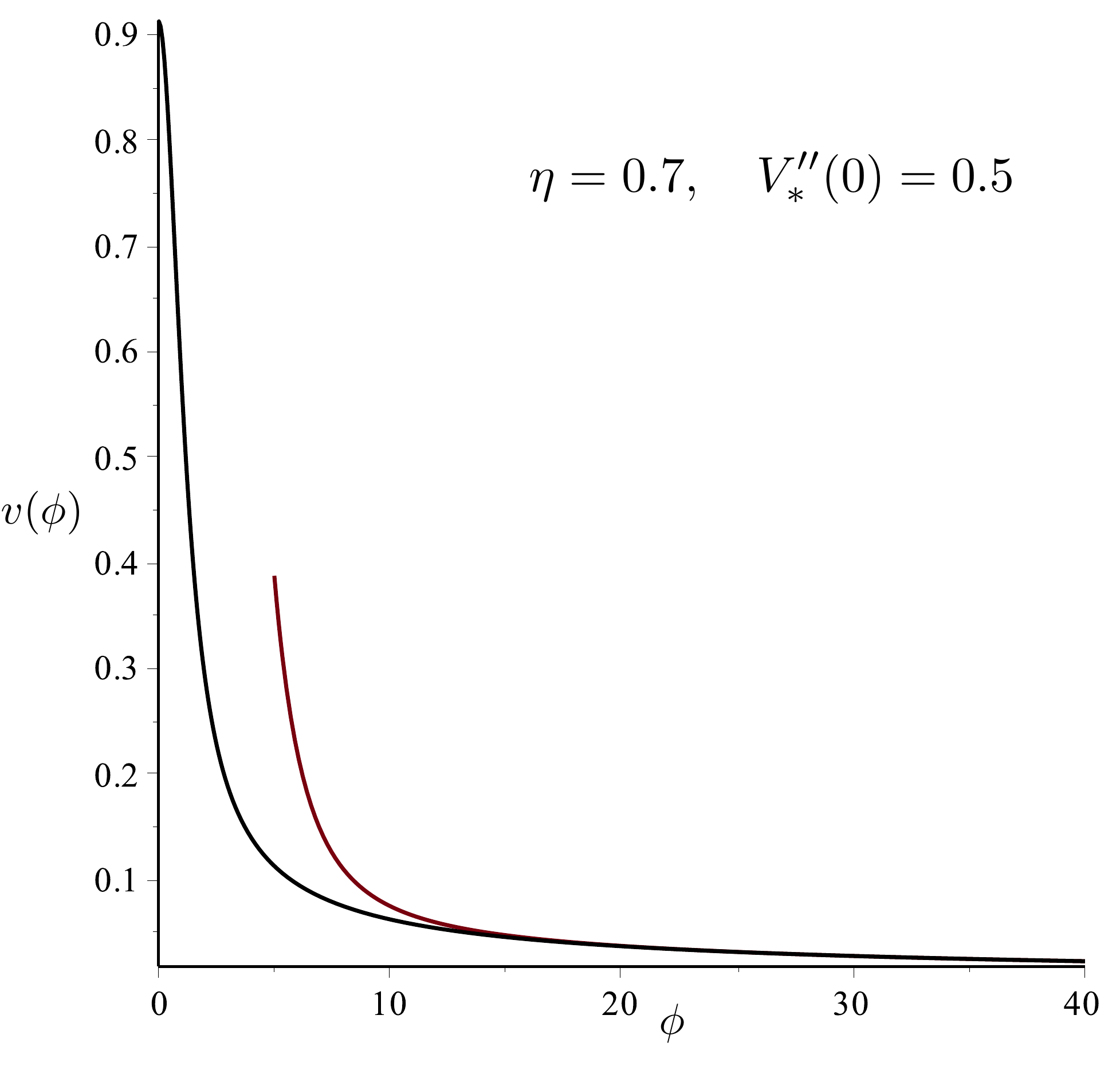} \\
\includegraphics[width=0.45\textwidth, height=0.25\textheight]{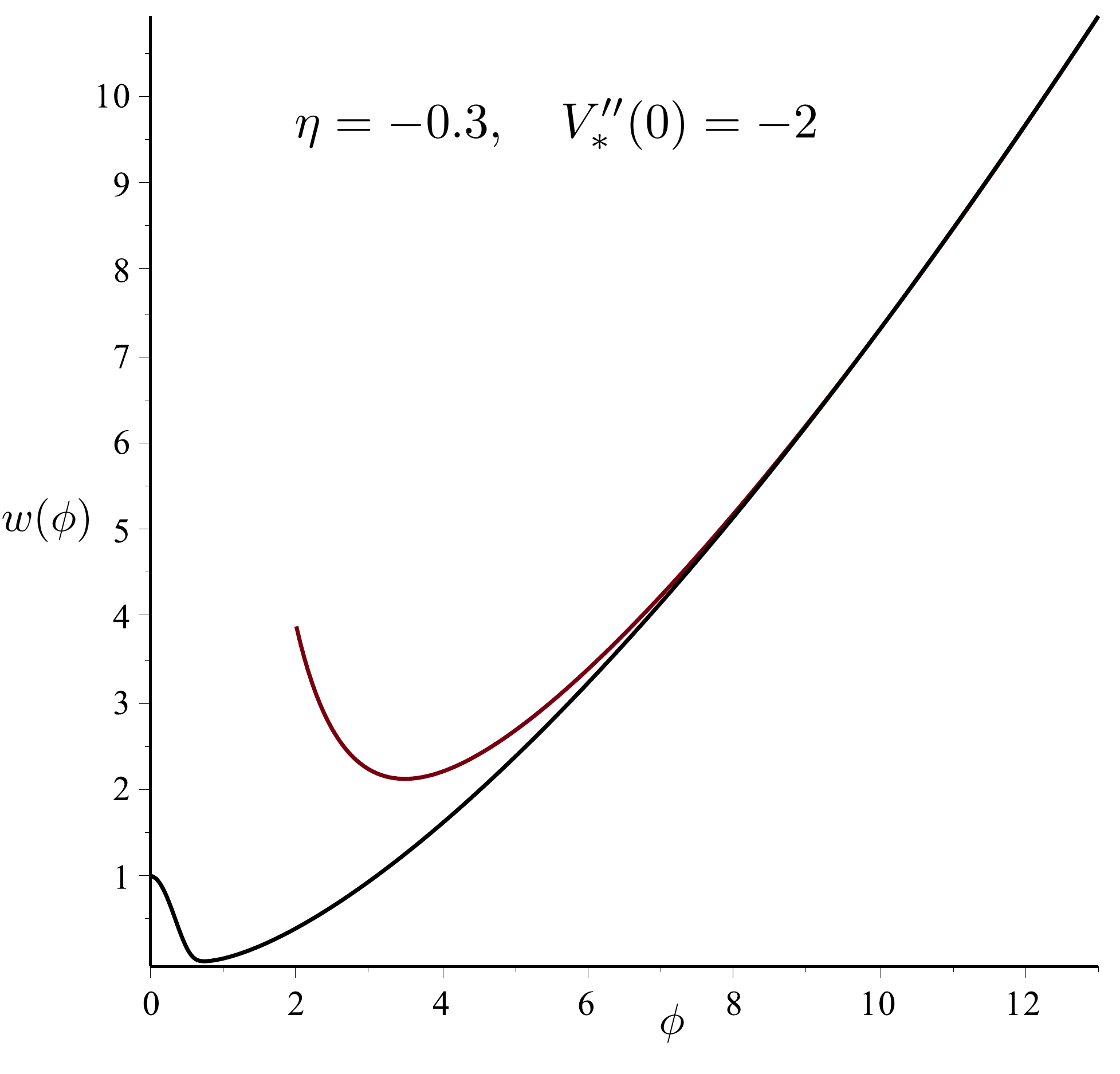} &
\includegraphics[width=0.45\textwidth, height=0.25\textheight]{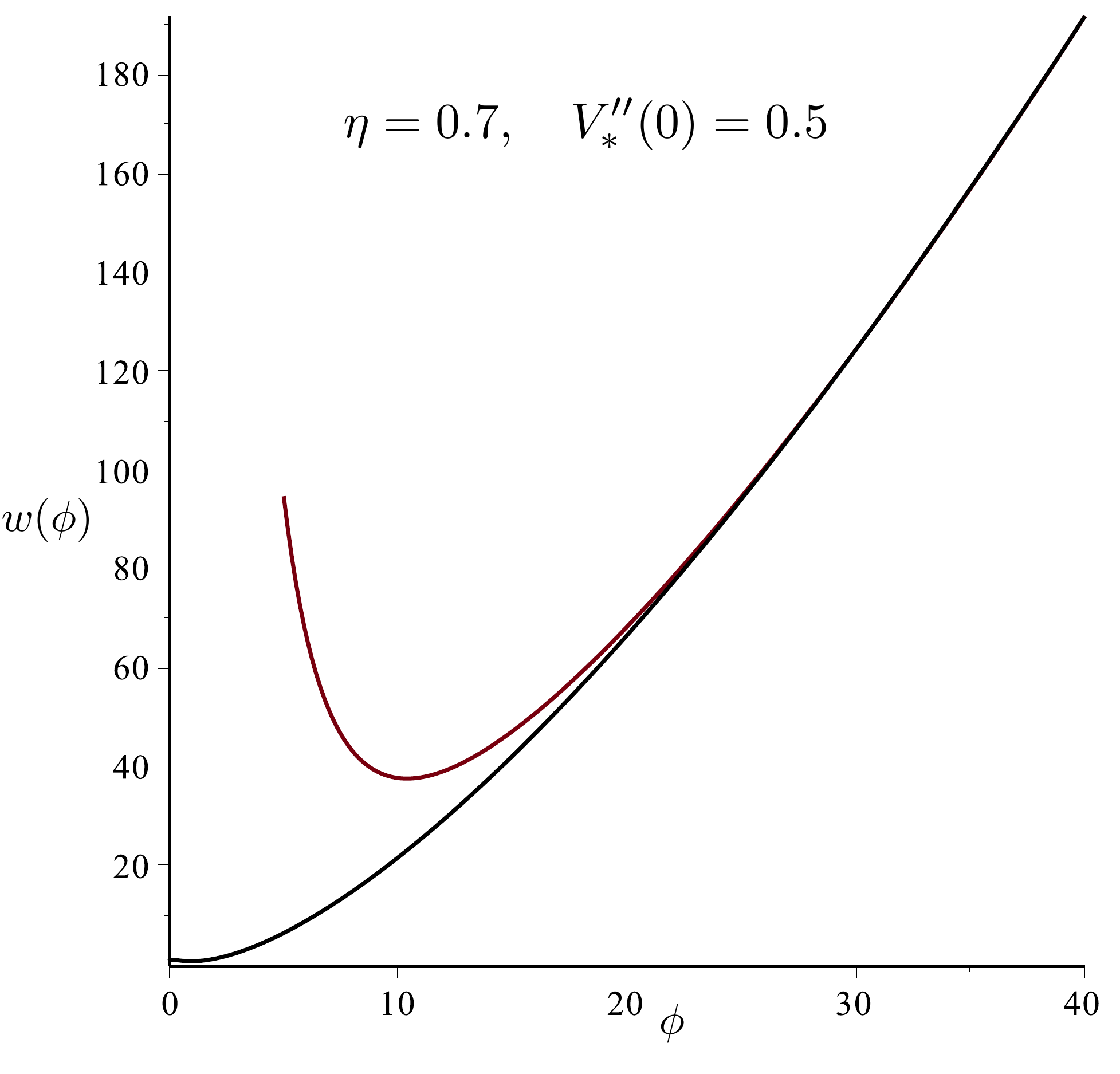}
\end{array}
$
\end{center}
\caption{The solutions displayed in fig. \ref{fig:confexsols} expressed in terms of the variables $v(\phi)$ and $w(\phi)$ in black, and the corresponding asymptotic solutions \eqref{solsasy} in brown.}
\label{fig:solsasy}
\end{figure}

On the other hand, for $\eta\notin\mathcal{\bar R}$  there are four free parameters $B_0,\dots,B_3$. In this case, requiring that the fixed point solutions exist for arbitrarily large field does not place any restrictions on parameter space for the asymptotic regime \eqref{limitv}. We thus have only the three conditions from evenness and scaling, leaving us with a line of fixed points \emph{for each} $\eta$, or in other words a two-dimensional space of global solutions including ranges of $\eta$.

Of course, the expressions  \eqref{solsasy} only contain the first terms of an infinite asymptotic series for $w(\phi)$ and $v(\phi)$. One could proceed to the next order in $\eps$ by continuing \eqref{linwv} with an appropriate $\mathcal{O}\!\left(\eps^2\right)$ term and by taking into account the corresponding higher order terms of the asymptotic differential equations \eqref{sys-asy}. However the order we have taken it to is already sufficient numerically.
The two example solutions plotted in fig. \ref{fig:confexsols} both show the asymptotic behaviour characterised by \eqref{limitv}. 
Hence the asymptotic solutions \eqref{solsasy} apply, and since $\eta\notin\mathcal{\bar R}$, with all four parameters $B_i$. The result of matching the numerical solution to the asymptotic solution is displayed in fig. \ref{fig:solsasy}. It can be seen that the asymptotic expansions agree very well with numerical solutions for sufficiently large $\phi$.

Varying the parameters $V_*''(0)$ and $\eta$ around their values for the example solutions of fig. \ref{fig:confexsols}, the solution can be integrated unhampered by moveable singularities to the large field regime, where it can again be matched onto the asymptotic solution \eqref{solsasy}. Although we do not report the details, we have also confirmed solutions for $\eta=0$. In this way one obtains numerical confirmation, including ranges of negative as well as positive values for the anomalous dimension, of the prediction  from parameter counting that the fixed point equations \eqref{equ:sysfp} admit continuous two-dimensional sets of global solutions for $\eta\notin\mathcal{\bar R}$.


\subsection{Asymptotic behaviour of eigenoperators}
\label{sec:eigen-asymptotics}

In this section we will demonstrate that the eigenoperator spectrum, about any of these fixed points, has again quantized and continuous components. 
In standard fashion, we find the eigenoperators by writing $V=V_*(\phi)+\delta V(\phi,t)$ and $K=K_*(\phi)+\delta K(\phi,t)$, and then linearising the flow equations \eqref{equ:sys-red-final}.  By separation of variables we factorise the exponential $t$ dependence as in \eqref{separateVars}, thus introducing the RG eigenvalue $\lambda$, eigenoperator components $\dV$ and $\dK$, and converting the equations to a coupled pair of ordinary differential equations.
Since these equations are linear with well-behaved coefficients, solutions exist for any $\lambda$. Whether they are acceptable or not, crucially depends on their large field behaviour \cite{Morris:1996nx,Morris:1996xq,Morris:1998,Bridle:2016nsu}, an observation used already in secs. \ref{sec:gaussian} and \ref{sec:LPA-RGprops}. 
We therefore concentrate on the asymptotic solution of these eigenoperators. 

The asymptotic form of the perturbation equations can be derived directly from the variation of \eqref{sys-asy} ($u\mapsto u+\delta u$, $v\mapsto v+\delta v$), by extending the change of variables \eqref{uv} to apply now for $t$-dependent functions $K(\phi,t), V''(\phi,t)$ and $u(\phi,t)$ and $v(\phi,t)$, and
remembering to include now $-\delta \dot V''(\phi,t) $ on the left hand side of the first equation and $-\delta\dot K(\phi,t)$ on the left hand side of the second equation.\footnote{This works because the only terms that depend explicitly on the time are the very first terms in \eqref{equ:sys-red-final}. There is an overall sign difference between \eqref{equ:sys-red-final} and \eqref{sys-asy}, and of course we have used $d=4$ and $n=2$.}
Note that in a similar way to before, \cf below \eqref{Q0poly}, the (linearised) flow equation \eqref{equ:flowV-final} and its first $\phi$ differential will allow us to reconstruct $\delta V(\phi,t)$. Recalling the further change of variables \eqref{w} (which is similarly now extended to $t$-dependent quantities) it is thus convenient to re-express the perturbations through
\bea
\label{delta-vars}
\delta K(\phi,t) &=& -\frac{3}{2} w(\phi)^{-7/4}\,\delta w(\phi,t)\,, \\
 \delta V''(\phi,t) &=& -\frac{3}{2}w(\phi)^{-3/2} \left[1-v(\phi)^2\right] \delta w(\phi,t) - 6\,w(\phi)^{-1/2} v(\phi)\,\delta v(\phi,t)\,.\nonumber
\eea
The resulting perturbation equations are too long to display but straightforwardly derived.  By separation of variables we can write
\be 
\label{eigenop-t-dependence}
\delta v(\phi,t) = \dv(\phi)\,{\rm e}^{-\lambda{t}}\qquad{\rm and}\qquad \delta w(\phi,t) = \dw(\phi)\,{\rm e}^{-\lambda{t}}\,.
\ee
Retaining just the leading ($B_0$) term in \eqref{solsasy}, the perturbation equations can be solved with the ansatz
\be 
\label{leading-eigen}
\dv = b_0\, \phi^r\qquad{\rm and}\qquad \dw = \phi^{r+2}\,,
\ee
where we have used linearity to normalise $\dw$. We thus find
\be 
b_0 = {\frac {1}{483840}} {\frac { \left( 8-\eta \right)  \left( 55\,q-24\,r-34 \right) 
 \left(331\,q -50 \right) \pi }{\eta-16/(2-q)}}
\ee
and 
\begin{multline}
\label{lambdar}
\lambda = 
-\frac{1}{11200}(33578+32769q)\left(\eta-\frac{16}{2-q}\right)
\left( {r}^{3} -\frac{277q+2}{24}\,r^2\right) \\
\hskip3cm+
 \left( -{\frac {9357703}{432600}}+{\frac {24523\,\eta}{16800}}-{
\frac {170502403\,q}{865200}}+{\frac {997303\,\eta\,q}{33600}}
 \right) r \\
 +{\frac {8848}{515}}+{\frac {121\,\eta\,q}{10}}-{\frac {12\,
\eta}{5}}-{\frac {34424\,q}{515}}\,.
\end{multline}
Apart from the one value $\eta=\frac{16}{2-q}=\sup\cR$, which we must exclude since the asymptotic behaviour of the fixed point equations themselves needs a more detailed analysis (\cf sec. \ref{sec:fpbeyondLPA}), 
solving this last equation gives three powers $r$ for every real $\lambda$. The three powers are either all real, or one power is real and the other two form a complex pair. In the case of complex $r=a\pm ib$, the real solutions corresponding to $\phi^r$ are given by \eqref{complex-exponent}. In fact as already mentioned in sec. \ref{sec:gaussian}, since the Sturm-Liouville properties \cite{Morris:1996xq}
are broken by the wrong sign in the kinetic term, there is no reason to expect that the eigenvalues are real. Again in this case since the differential equations are real, we would have a complex pair of solutions associated to a complex pair of eigenvalues. In any case we see that there are three solutions with large $\phi$ asymptotics \eqref{leading-eigen} for every RG eigenvalue $\lambda$, and the general solution is a linear combination of these. 

To compute the sub-leading terms for these solutions, we need to reintroduce the sub-leading terms from \eqref{solsasy} and the book-keeping $\eps$ as in \eqref{sys-asy}, \eqref{w} and \eqref{linwv}. Similarly we write
$\dv = \dv_0 +\eps\dv_1$, $\dw=\dw_0+\eps\dw_1$. The leading solutions $\dv_0$ and $\dw_0$ are those in \eqref{leading-eigen}, and solve the $\mathcal{O}\!\left(1/\eps\right)$ of the $\delta V''$ equation and the $\mathcal{O}\!\left(1/\eps^3\right)$ part of the $\delta K$ equation. The subleading pieces $\dv_1$ and $\dw_1$ thus solve the $\mathcal{O}\!\left(1\right)$ and $\mathcal{O}\!\left(1/\eps^2\right)$ parts respectively. However the general solution merely reproduces the solutions \eqref{leading-eigen} as expected. By inspection it can be seen that the particular solutions are linear combinations, with calculable coefficients, of terms 
\be 
\dv_1\sim \phi^{r+q}\qquad{\rm and}\qquad \dw_1\sim\phi^{2+r+q}\,,
\ee
which since $q<0$, are indeed subleading, and 
\be 
\dv_1\sim \phi^{r+s_i-q}\qquad{\rm and}\qquad \dw_1\sim\phi^{2+r+s_i-q}\,,
\ee
which are also subleading since these $s_i$ solutions are included only when  $\Re(s_i)<q$. We see that the asymptotic series can be developed in this way, and no further restrictions arise.

We have thus found by linearisation about the fixed point, that asymptotically there are three independent  solutions of form \eqref{leading-eigen}, for every choice of eigen-value $\lambda$. The next question we must ask is whether the linearisation step remains valid for large $\phi$ \cite{Morris:1996xq,Morris:1998,DietzMorris:2013-1}. 
This is true if and only if $\delta K(\phi,t)/K_*(\phi)$ and $\delta V''(\phi,t)/V''_*(\phi)$ remain small as $\phi\to\pm\infty$. Using \eqref{delta-vars} and \eqref{uv} together with $u=w^{-1/4}$, then substituting the leading terms from \eqref{solsasy} and the linearisation result \eqref{eigenop-t-dependence} and \eqref{leading-eigen}, it is straightforward to show that the large $\phi$ dependence of both ratios is controlled by the ratio $\delta w/w$ and hence that linearisation remains valid if and only if $\Re(r)\le q$. Whether this is satisfied for the solutions $r$ to the cubic \eqref{lambdar}, clearly depends on the value of $\eta$ at the underlying fixed point. Scanning over the possibilities for $\lambda$, we will then find that we are left with $n_r(\lambda)$ solutions \eqref{leading-eigen}, where \textit{a priori} $n_r(\lambda)$ can take any integer value from 0 to 3 inclusive, depending on $\lambda$.

Since the eigen-value equations are equivalent to linear coupled second order ordinary differential equations, there are in fact four independent solutions.  Since the analysis above found a maximum of three, we have in fact determined that there is another linearised solution which not power-law for large $\phi$. Given the behaviours uncovered for the Gaussian fixed point and in $\eta=0$ LPA in secs. \ref{sec:gaussian} and \ref{sec:LPA-RGprops} respectively, it is a reasonable conjecture that the missing solution decays faster than any power and thus is actually always a legitimate  linearised perturbation. However, since the main point we want to demonstrate is that the fixed points we have uncovered do support a continuous spectrum of eigenoperators, we will for the moment assume the most constraining scenario, which is that the missing solution is illegitimate, and then show that even under this assumption we can still uncover a continuous spectrum.  With this assumption we are left with $n_r(\lambda)$ legitimate solutions, and thus we conclude that requiring the linearised solutions to remain valid for $\phi\to\pm\infty$ leads to $4-n_r(\lambda)$ constraints.

Since we have chosen to focus on fixed points that are even under $\phi\mapsto-\phi$, the eigen-perturbations are (or if degenerate can be taken to be) even or odd, and therefore satisfy two constraints namely $\dV'(0)=\dK'(0)=0$ or $\dV(0)=\dK(0)=0$ respectively. Linearity allows us to impose one further constraint for example $\dK(0)=2$ or $\dK'(0)=2$ respectively. In total therefore we have $7-n_r$ constraints. Recalling that for each $\lambda$, we have \textit{a priori} a four-dimensional vector space of linearised solutions, and recalling that  the non-power-law linearised solution could after all be legitimate -- leading then to only $6-n_r$ constraints, we conclude that $n_r(\lambda)=3$ provides us with a continuous spectrum of at least one even and one odd operator for every such eigenvalue $\lambda$; $n_r(\lambda)=2$ will lead to an extra constraint that may be sufficient to quantize the spectrum, \ie such that it can only be satisfied for discrete values of $\lambda$ in this range; while $n_r(\lambda)<2$ leads to a quantized spectrum or no solutions.


\begin{figure}[ht]
\centering
\includegraphics[width=0.73\textwidth]{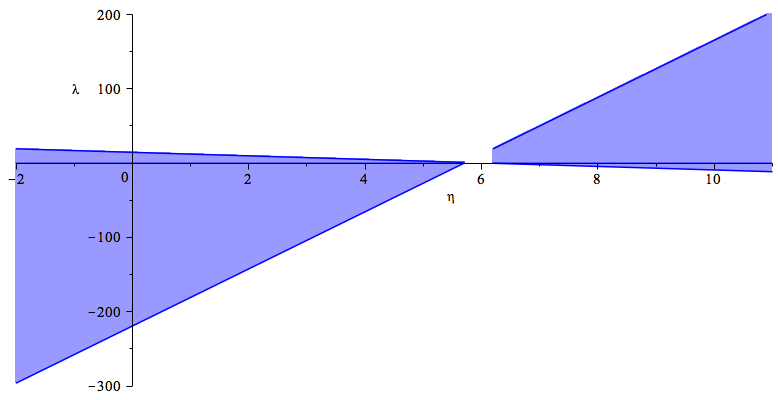} 
\caption{The eigenoperators that follow legitimately from linearisation, form a continuous spectrum over a range of $\lambda$, with at least one odd and one even operator for every real $\lambda$ in the blue shaded regions. The gap is the interval $\eta\in \qR\approx(5.916,6.113)\subset\cR$.}
\label{fig:continuous-spectrum}
\end{figure}

Since linearisation is valid for all the remaining solutions, they are confirmed as eigenoperators which are (ir)relevant if $\Re(\lambda)$ is (negative) positive, and  if relevant their RG evolution can be attributed to  renormalised couplings $g=\epsilon \mu^\lambda$. 
Solving the cubic \eqref{lambdar} for the roots $r$, and scanning over $\eta$, we have computed numerically the range of real $\lambda$ where all $n_r(\lambda)=3$ roots satisfy $\Re(r_i)\le q$, where thus these eigenoperators definitely form a continuous spectrum. The results  are displayed in fig. \ref{fig:continuous-spectrum} (for comparison over the same range as in fig. \ref{fig:confexsols}). We see that the continuous spectra always include relevant directions $(\lambda>0$), and that the range grows ever larger for both negative and positive $\eta$. In the interval $\qR=(\eta_q,\frac{16}{2-q})$, where $\eta_q\approx5.916$, we find that we have only $n_r(\lambda)=2$, and therefore the eigenvalues in principle may only form a discrete spectrum here (depending on the status of the `missing' non-power-law perturbation). It is interesting that $\qR$ is a subset of $\cR$ (proper subset since $\eta_q>\eta_c$) where, by the counting argument, the fixed points themselves must form a discrete set. However as discussed in sec. \ref{sec:fpbeyondLPA} it is unlikely that fixed point solutions actually exist in this range.





(In fact the existence of $\qR$ may be proven as follows. Rearrange  \eqref{lambdar} so that it reads 
\be 
r^3+a_2\,r^2+a_1(\eta) r+a_0(\eta,\lambda) = (r-r_1)(r-r_2)(r-r_3)=0\,.
\ee
Note that $a_2=-(277q+2)/24$ is a constant, and $a_1$ depends only on $\eta$. Since we take $\lambda$ to be real, the roots $r_i$ are either all real, or two form a complex pair. Either way we see that the requirement that all three roots satisfy $\Re(r_i)\le q$ implies the three conditions:
\be 
a_2 \ge-3q\,,\qquad a_1\ge 3q^2\,,\qquad a_0\ge-q^3\,.
\ee
Recalling \eqref{q}, we see that the $a_2$ condition is satisfied. The $a_0$ condition can always be satisfied for suitable $\lambda$. However it is straightforward to show that the $a_1$ condition is violated if and only if $\eta\in\qR$.)

Following our discussion above, depending on the status of the `missing' non-power-law perturbation, the continuous spectrum may extend beyond the limits displayed in fig. \ref{fig:continuous-spectrum} and then, depending also on the value of $n_r(\lambda)<3$, at some larger values of $\lambda$ a discrete spectrum takes over or there are no further eigen-operators. Following the findings at the Gaussian fixed point, \cf sec. \ref{sec:gaussian}, we expect that the eigenoperators that follow legitimately from linearisation do not form a complete set (in contrast to scalar field theory). As in secs. \ref{sec:gaussian} and \ref{sec:gaussLPA}, we therefore expect that there are small finite perturbations that are relevant (\ie grow as $t$ decreases) but which cannot be described by a sum over the legitimate linearised eigenoperators. We saw in secs. \ref{sec:gaussian} and \ref{sec:gaussLPA} that they are also continuous in number.

\section{Polynomial truncations}
\label{sec:truncations}

Of course we are dealing only with conformally reduced gravity, a severe truncation of full quantum gravity. However given that the conformal sector mainly governs the structure of the fixed points \cite{Demmel2015b} and  
 given the level of sophistication incorporated in the equations of ref. \cite{Dietz:2015owa}, one might have hoped that the description of fixed points would be a clear advance on the standard lore \cite{Reuter:2012,Percacci:2011fr,Niedermaier:2006wt,Nagy:2012ef,Litim:2011cp}. Unfortunately the fixed point structure we have uncovered (continuous sets of fixed points supporting continuous eigenspectra) bears no relation to the picture built up in that research, although it does show similarity to the results \cite{Bridle:2013sra} from solving the $f(R)$ approximation developed in ref. \cite{Benedetti:2012}. 
We will discuss this latter similarity further in the conclusions. Here we note that by far the weight of evidence for the standard picture of an isolated fixed point with three relevant eigenoperators, comes from polynomial truncations. By construction however, polynomial truncations can only give isolated fixed points with a quantized eigenoperator spectrum. It is therefore interesting to see to what extent polynomial truncations of \eqref{equ:sys-red-final} reflect the standard lore and/or if there are any imprints of the true situation in the current case. We will find that such polynomial truncations neither support the standard lore nor properly reflect the true situation.
%
%

\subsection{Gaussian fixed point}\label{sec:truncGauss}

The Gaussian fixed point  \eqref{GaussianVstar} ($K_*=1$) itself is still an exact solution in polynomial truncations. For the eigenoperators, polynomial truncations only find the polynomial solutions \eqref{GaussianPotentialOp} and \eqref{GaussianKineticOp} of the equations \eqref{gaussian-perturbations}. They therefore miss entirely the continuum of solutions \eqref{gaussian-perturbation-general}, despite the fact that these are not excluded by their behaviour at large $\phi$.

\subsection{Local Potential Approximation with vanishing anomalous dimension} \label{sec:truncatedLPA}

First we treat the LPA Gaussian fixed point with $\eta=0$, analysed exactly in sec. \ref{sec:gaussLPA}. The corresponding Gaussian fixed point \eqref{GFPLPA} ($K\equiv1$) itself is also an exact solution of polynomial truncations. For the eigenoperators
in this case we find that all the $\lambda\ne4$ operators are invisible in polynomial truncations. Substituting 
\be 
\dV= \sum_{m=0}^\infty \frac{\dV_m}{m!}\phi^m\,,
\ee
into \eqref{eigenGaussian},  gives $\dV_{m+2}=\pm\omega^2\dV_{m}$. As usual a polynomial truncation is imposed by requiring $\dV_m=0$ $\ \forall m>n$ and keeping only the equations for $m\le n$. If $\omega\ne0$, then for $m=n$ we deduce $\dV_n=0$. Likewise for $m=n-1$ we find $\dV_{n-1}=0$. By iteration we thus find all the coefficients vanish. Therefore the only solutions are those we obtain when $\omega=0$. These are in fact the exact $\lambda=4$ solutions  \eqref{lambda4}.


More generally, all fixed point solutions to $V_*(\phi) = \mathcal{F}\!\left(V_*''\right)$ can be derived through translations of fixed point solutions that are even functions of $\phi$. We already know this from the exact solution as derived in sec. \ref{sec:planeLPA}, but here we furnish an alternative proof that does not presuppose knowledge of the exact solution. We first note that since $\phi$ does not appear explicitly, fixed points actually appear as lines of fixed points $V_*(\phi-c)$, parametrised by the translation $c$. Since the integrand cannot change sign without causing the integral to diverge,  the right hand side ${\cal F}$ is a strictly positive function whenever it is defined. Therefore $V_*(\phi)$ is bounded below and has a minimum at some point $c$, where it thus satisfies $V'_*(\phi)|_{\phi=c}=0$. Changing variables $\phi\mapsto\phi+c$, we obtain a solution with $V'_*(0)=0$. Since $V_*(\phi) = \mathcal{F}\!\left(V_*''\right)$ is also symmetric under $\phi\mapsto-\phi$, such a solution is an even function of $\phi$. From here on we will concentrate on the fixed points $V_*(\phi)$ that are even functions.

Similarly to above, to obtain the  truncations to polynomials of rank $2n$, we Taylor expand 
\be 
\label{truncated-FP-LPA}
V_*(\phi) = \sum^n_{m=0} \frac{V_{2m}}{(2m)!} \phi^{2m}\,,
\ee
making the ansatz that the $\phi^{2n+2}$ coefficient vanishes (and likewise all higher coefficients). Taylor expanding $V_*(\phi) = \mathcal{F}\!\left(V_*''\right)$ to power $\phi^{2n}$ gives $n+1$ equations for the $n+1$ free coefficients. Real solutions can thus be found numerically. (The right hand side contains integrals over $p$ of $p^{2r+1} (1-p^4V_2+p^6)^{-s}$ for some positive integers $r$ and $s$, which are straightforward to handle numerically.) 

Since the fixed point potential is an even function, we can assume the eigenoperators to be even or odd. We will concentrate only on the even ones. 
Having obtained the rank $2n$ approximation to the fixed points, we likewise expand the eigenperturbation $\dV(\phi)$ to rank $2n$ and substitute both of these into the Taylor expanded eigenoperator equation \eqref{eigenLPA}. The ansatz $\dV_{2n+2}=0$ then results in a matrix eigenvalue equation determining $n+1$ eigenvalues $\lambda$  and corresponding eigenvectors $(\dV_0,\dV_2,\cdots,\dV_{2n})$. We will label them as $\lambda=\lambda_j$, ordered in decreasing relevance. These coincide with the definition of the  ``critical exponents'' $\theta_j$ that can be found in the literature.

\begin{table}[ht]
\begin{center}
\begin{tabular}{|c|c|c|c|c|c|c|}
\hline
$V_0$ & $V_2$ & $V_4$ & $V_6$ & $V_8$ & $V_{10}$ &  $V_{12}$ \\
\hline
18.2 & 1.88 & 0.00292 & & & &  \\
\hline
1.24 & 0.0695 &0.163 &0.313 &  & &  \\
\hline
1.2156 & 0.0158 & 0.0387 & 0.0910 & 0.178& & \\
\hline
1.2107 & 0.00376 &0.00930 &0.0228 &0.0537 &0.106 &\\
\hline
1.2096 & 0.000915 & 0.00227 & 0.00561& 0.0138 & 0.0325 & 0.0641  \\
\hline
\end{tabular}
\end{center}
\caption{Coefficients for non-trivial fixed point polynomial truncations of rank $n=4,6,8,10$ and 12. These are given to 3 sf (significant figures) except for $V_0$ where the later values are given to 5 sf to compare to the Gaussian fixed point $V_0= 1.2092$ (at 5 sf).}
\label{table:LPA}
\end{table}

The results are displayed in tables \ref{table:LPA} -- \ref{table:LPA2}.   However we exclude the Gaussian fixed point since that was already treated at the beginning of this subsection. Up to rank 12, there is always a non-trivial fixed point solution bounded below, as in fact is true of the exact solutions \cf the argument above \eqref{truncated-FP-LPA} and sec. \ref{sec:planeLPA}. The fixed point couplings are displayed in table \ref{table:LPA}. For ranks 6, 10 and 12, a second non-trivial solution appears which corresponds to a fixed point potential that is instead bounded above. These fixed point couplings are displayed in table \ref{table:LPA2}. There seems to be no sign however of the fact that the fixed points form a continuum. Recall from  sec. \ref{sec:planeLPA} that analysed exactly at the LPA level,  these symmetric $V_*(\phi)$ form a line of fixed points ending at the Gaussian fixed point \eqref{GFPLPA}, this latter characterised by $V_0= 1.2092$ (to 5sf) with all other $V_{2n}=0$. We see from table \ref{table:LPA} that with increasing rank, the non-trivial polynomial truncations are in fact rapidly converging towards the Gaussian fixed point. Likewise the upside-down fixed point potentials in table \ref{table:LPA2} appear also to be converging towards the Gaussian fixed point, though more slowly. 

\begin{table}[ht]
\begin{center}
\begin{tabular}{|c|c|c|c|c|c|c|}
\hline
$\lambda_1$ & $\lambda_2$ & $\lambda_3$ & $\lambda_4$ & $\lambda_5$& $\lambda_6$& $\lambda_7$\\
\hline
 && & &4&-108&-5260 \\
\hline
&&& 5.64 &4& 3.90 & -2.79  \\
\hline
& &\multicolumn{2}{|c|}{$5.66\pm0.331i$} &4& 3.95 &-1.90  \\
\hline
&\multicolumn{2}{|c|}{$5.76\pm1.75i$} &4.76 &4& 3.97 & -1.43\\
\hline
5.63 &\multicolumn{2}{|c|}{$5.26\pm2.58i$} & 4.53 &4& 3.98 &-1.14\\
\hline
\end{tabular}
\end{center}
\caption{RG eigenvalues corresponding (by row) to the truncations in table \ref{table:LPA}.}
\label{table:LPA-crits}
\end{table}

\begin{table}[ht]
\begin{center}
\begin{tabular}{|c|c|c|c|c|c|c||c|c|c|c|c|c|}
\hline
$V_0$ & $V_2$ & $V_4$ & $V_6$ & $V_8$ & $V_{10}$ &  $V_{12}$ &$\lambda_1$ & $\lambda_2$ & $\lambda_3$ & $\lambda_4$ & $\lambda_5$ & $\lambda_6$\\
\hline
{ 1.57} & { 0.653} &0.838 &-1.56    & & & & &&&4&3.52 &-6.09\\ 
\hline
{1.25}  & 0.0939 &0.215 &0.372  &-0.375 &-8.10&&&11.1 &6.17 &4&3.87 &-2.71\\ 
\hline
1.22 & 0.0378 & 0.0908 & 0.198 & 0.235 & -1.50 & -15.6 
&\multicolumn{2}{|c|}{$10.7\pm1.80i$} &5.45&4&3.91&-2.13\\
\hline
\end{tabular}
\end{center}
\caption{Coefficients for the non-trivial fixed point polynomial truncations with potential unbounded below, which appear at rank $n=6,10$ and 12. The corresponding eigenvalues appear on the right of the table.}
\label{table:LPA2}
\end{table}

The RG eigenvalues corresponding to the truncations in table \ref{table:LPA} are displayed in table \ref{table:LPA-crits}, while the RG eigenvalues corresponding to the truncations in table \ref{table:LPA2} are  displayed in the same table \ref{table:LPA2} (apart for lack of space $\lambda_6 = -57.4,-29.4,-25.2$ in lines 1 -- 3 respectively). One eigenvalue is always found exactly, namely $\lambda=4$ corresponding to $\dV(\phi) = \dV_0$, so we use this to determine the order of all the other eigenvalues in the tables in order to judge convergence.  In table \ref{table:LPA-crits}, the non-trivial eigenvalue $\lambda_6$ seems clearly to be converging to four, or to a number close to four. Less convincingly the same may be true of $\lambda_5$ in table \ref{table:LPA2}.
The data for all the other eigenvalues suggests convergence but to numbers other than four. We also see that the number of relevant perturbations grows linearly with increasing rank, so that we would conclude for both sequences (tables \ref{table:LPA-crits} and \ref{table:LPA2}) that eventually at infinite rank we would have a discrete spectrum but with an infinite number of relevant directions. 

%

\subsection{Order derivative squared}

\begin{table}[ht]
\begin{center}
\begin{tabular}{|c||c|c|c||c|c||c|c|c|c|c|c|c|}
\hline
$\eta$& $V_0$ & $V_2$ & $V_4$ & $K_2$ & $K_4$ &$\lambda_1$ & $\lambda_2$ & $\lambda_3$ & $\lambda_4$ \\
\hline
1.70 & 0.875 &-0.270 &   & 0.289 & & 3.43 & 1.29 & &\\
\hline
2.05  & 0.869 & -0.104 & -0.279 & -0.0471  & 0.0383 & 2.00 & -1.18 & \multicolumn{2}{|c|}{$-1.34\pm1.28i$} \\ 
\hline
1.28 & 0.919 & -0.324 & 0.600 & 0.682 & 0.752 
&15.3 & 4.52
&1.78&1.07 \\    
\hline
\end{tabular}
\end{center}
\caption{Fixed point solutions and corresponding eigenvalues for rank 2 and 4 truncations. $K_0=1$ and the $\lambda=4$ eigenvalue is not listed. No attempt is made to match eigenvalues across different rank.}
\label{table:d2}
\end{table}

We briefly investigate the situation at $\mathcal{O}(\partial^2)$ for low rank truncations, to check if the situation is significantly different to the studies above. Now we also Taylor expand 
\be 
K_*(\phi) = \sum^n_{m=0} \frac{K_{2m}}{(2m)!} \phi^{2m}\,,
\ee
making the ansatz that the $\phi^{2n+2}$ coefficient vanishes (and likewise all higher coefficients).
Recall that we have the scaling symmetry \eqref{scalings}. We use that in this subsection to normalise $K_*(0)=K_0=1$ so as to be directly comparable to the previous subsections. At $\mathcal{O}(\partial^2)$ we can determine the anomalous dimension $\eta$. For polynomial truncations, the scaling symmetry \eqref{scalings} ensures that $\eta$ is in fact quantized. We exclude the Gaussian solution that was already treated in subsec. \ref{sec:truncGauss}. The results are displayed in table \ref{table:d2}. Now rank 2 already yields a non-trivial solution, however note that the fixed point potential is unbounded below. Such a property is not \textit{a priori} excluded for the exact solutions of \eqref{equ:sys-red-final} but it is not what we found for the exact solutions at the LPA level in sec. \ref{sec:planeLPA} or for the sample numerical (plus asymptotic) solutions found at $\mathcal{O}(\partial^2)$ in sec. \ref{sec:fpbeyondLPA}. At rank 4, we find already two fixed point solutions, one with potential unbounded below and one with potential unbounded above. It is interesting to note however that the kinetic term function $K(\phi)$ is bounded below for all these three cases. The polynomial truncations to the eigenoperators now have two exact solutions: $\dV(\phi)=\dV_0, \dK(\phi)=0$ with $\lambda=4$ as in LPA, and a redundant solution with $\lambda=0$,  following from an infinitessimal application of the scaling symmetry \eqref{scalings}\cite{Morris:1994ie,DietzMorris:2013-2}. We exclude both of these eigenoperators from the table. Even to the low level of truncation we have investigated, we already see evidence that the number of relevant directions is growing. It also interesting to see tentative evidence that $V_0$ is tending to the Gaussian fixed point value \eqref{GaussianVstar}. In the first rank 4 approximation, the other values for the fixed point itself are close to this (\ie $\eta=2$ $V_2=V_4=K_2=K_4=0$). To the level we have taken it, it seems clear that the $\mathcal{O}(\partial^2)$ results suffer the same problems we uncovered at the level of the LPA in subsec. \ref{sec:truncatedLPA}. 

\subsection{Interpretation}

Drawing together all the results, we have seen that at the Gaussian fixed point itself the continuum of non-polynomial eigen-perturbations is invisible to polynomial truncations. The  continuum of fixed point solutions themselves are also invisible in this approximation. However non-trivial fixed point solutions do emerge. When organised by increasing rank, they divide into two families depending on whether the fixed point potential is bounded above or below, and in each family these solutions appear to converge towards the Gaussian fixed point. Meanwhile the number of relevant RG eigenvalues keeps increasing with increasing rank. Although there is not much sign at this stage of the fact that the eigenvalues  actually form a continuum, it could be that at very high rank truncation, these eigenvalues move closer together as well as beginning to spread over the whole of the real line and thus form a better reflection of the true situation at the Gaussian fixed point, despite the fact that the Gaussian fixed point itself is not faithfully represented.
Of course none of this picture reflects the weight of evidence for asymptotic safety found in polynomial truncations in other studies \cite{Reuter:2012,Percacci:2011fr,Niedermaier:2006wt,Nagy:2012ef,Litim:2011cp}. We will return to this at the end of the conclusions.

\section{Summary, discussion and conclusions}
\label{sec:conclusions}

In this paper we take the $\mathcal{O}(\partial^2)$ system of flow equations and Ward identities for conformally reduced gravity derived in ref. \cite{Dietz:2015owa}, which involve no other approximation except to use the slow field limit for the background field $\chi$, and use these to investigate  thoroughly the structure of fixed points and corresponding spectra of eigenoperators in this model. As discussed in ref. \cite{Dietz:2015owa}  and sec. \ref{sec:sft} (see also sec. \ref{sec:sft2d}), the resulting background independent flow equations have a very close similarity to scalar field theory. This stems from the fact that the conformal factor field is a single component field with a wrong-sign kinetic term as in \eqref{equ:ansatzGamma}. Modified split Ward identities implement the background independence, and imply a change of variables which absorbs all dependence on the background field $\chi$. After this change to background-independent variables, the effective action becomes \eqref{equ:Gamma-hat}, which is precisely that of scalar field theory with a wrong-sign kinetic term, and similarly the flow equations become those of scalar field theory  adapted to this change in sign.

Nevertheless this one sign change has far reaching consequences for the exact RG flow, 
that have not until now been recognised.\footnote{They are separate from the issues associated their backward-parabolic nature (see ref. \cite{Bonanno:2012dg} and sec. \ref{sec:sft}), which means that the natural Wilsonian RG flow is towards the ultraviolet.} At the end of this discussion, 
we will show how these properties underlie the evidence for asymptotic safety in the literature.

For the eigenoperator spectrum, the consequences are already clear from studying the Gaussian fixed point, as we saw in sec. \ref{sec:gaussian}. The anomalous dimension $\eta$ of the conformal factor field $\phi$  is then fixed by the equations to be $\eta=2$, and thus its total scaling dimension is that of a scalar field at its Gaussian fixed point. The tower of polynomial eigen-perturbations ${\cal O}_n(\phi)$, which for the potential are associated to renormalised couplings with dimension\footnote{Throughout this discussion we are working in 
in $d=4$ dimensions, and $n$ is a non-negative integer. In the literature $\lambda$ is also called the `critical exponent' $\theta$.} $\lambda=4-n$, are the ones expected from scalar field theory except for the obvious sign changes induced by the wrong-sign propagator. However these polynomials are no longer orthonormal nor any longer do they form a complete set. Although a generic polynomial interaction can still be expanded in terms of them, it is no longer possible to approximate an interaction which is non-polynomial by a sum over the ${\cal O}_n(\phi)$ with suitable coefficients (as we saw in an explicit example). A related effect is that non-quantized eigen-perturbations can no longer be excluded by the large $\phi$ test, \ie excluded by their behaviour at large $\phi$ using the arguments developed in refs. \cite{Morris:1996nx,Morris:1996xq,Morris:1998,Bridle:2016nsu}. For any real $\lambda$, both the even and odd solution to the RG eigen-perturbation equation \eqref{gaussian-perturbations-V} now grow at most as a power of the field for large $\phi$, as in eqn. \eqref{GaussianEigenasymptotic}, while for $\lambda=5+n$ there is a tower of `super-relevant' eigen-perturbations that take the form of polynomials times $\exp (-a^2\phi^2)$, for example those displayed in eqn. \eqref{super-relevant}. By the analysis of refs. \cite{Morris:1996nx,Morris:1996xq,Morris:1998,Bridle:2016nsu}, we are thus led to conclude that there is thus a continuum of relevant couplings, in fact two independent relevant couplings for every real positive $\lambda$. 

In contrast to the one (Gaussian) fixed point we would find for scalar field theory \cite{Morris:1994ki}, we now find a continuous set of fixed points. Although the details of the spectrum of eigen-perturbations about the
non-perturbative fixed points in this continuous set, differs from the Gaussian case above, we again find that around each fixed point there is a continuous spectrum which includes relevant couplings. Although in this paper, we analyse only the perturbations with real RG eigenvalues $\lambda$, it is no longer possible to justify excluding complex $\lambda$. The analysis we presented here could be extended to this case. However it would of course not alter the discovery we have already made that a continuous spectrum of relevant perturbations exists around each of these fixed points.

In sec. \ref{sec:fpbeyondLPA} we show that $\eta$ can take a range of values that includes $\eta=0$. In sec. \ref{sec:confLPA} we set $\eta=0$ and analyse exactly the Local Potential Approximation (LPA) in this case.  We find that there exists a two-parameter set of fixed point potentials $V_*(\phi)$. All of these potentials have a minimum which thus rules out a breakdown of the LPA \cite{DietzMorris:2013-2} as an explanation for the continuous spectrum of eigenoperators. One parameter is accounted for by invariance under shifts in $\phi$, which can be used to set the minimum of $V_*$ at $\phi=0$. There then remains a line of $\phi\mapsto-\phi$ symmetric fixed point potentials ending at a Gaussian fixed point (whose detailed properties differ from the one above because we have imposed $\eta=0$). The eigen-operator spectrum about any of these fixed points however is again continuous. The spectrum around any of these fixed points includes the constant and linear perturbations \eqref{lambda4} with $\lambda=4$, and $\lambda\ne4$ perturbations that for large $\phi$ are either sinusoidal or grow exponentially. The sinusoidal perturbations correspond to super-relevant perturbations with $\lambda>4$ and survive the large $\phi$ tests, thus yielding two relevant renormalised couplings for every $\lambda>4$. The exponentially growing perturbations are associated to $\lambda<4$ (and also survive the large $\phi$ analysis since the scaling dimension of $\phi$ has been set to zero). However we see that at large $\phi$, these actually behave as relevant perturbations with a $t$-dependence that is characteristic of a dimension 4 coupling (independent of the value of $\lambda<4$). 
 
In this paper we have chosen to use a power-law cutoff profile \eqref{equ:cutoff-pwrlw} since this is required for the $\mathcal{O}(\partial^2)$ system of flow equations and Ward identities to be compatible when $\eta\ne0$  \cite{Labus:2016lkh}. However for $\eta=0$ any cutoff profile can be used. In ref. \cite{Labus:2016lkh}, it was shown that for the optimised cutoff profile \cite{opt1,opt3}, the LPA with $\eta=0$ system can again be solved in terms of background independent variables, leading to a flow equation that can be analysed with the methods in sec. \ref{sec:confLPA}. In fact the analysis in sec. \ref{sec:confLPA} depends only on qualitative features of the corresponding Newtonian potential $U$ and thus it is straightforward to verify that we obtain with the optimised cutoff precisely the same conclusions as above. At the same time it is also clear that the analysis for the potential perturbations around the Gaussian fixed point will reproduce exactly what we found in sec. \ref{sec:gaussian}. Indeed up to scaling the corresponding eigen-operator equation \eqref{gaussian-perturbations-V} is identical \cite{Bridle:2016nsu}. Finally, in sec. \ref{sec:LPA} the asymptotic analysis of the LPA fixed point equation for standard scalar field theory was carried out for general cutoff profile, space-time dimension and field dimension. There we saw precisely why the change in sign of the kinetic term turns the fixed point equation from one with only a discrete set of solutions into one with a continuum of solutions.

In sec. \ref{sec:fpbeyondLPA} we analyse the full background-independent $\mathcal{O}(\partial^2)$ equations. In our numerical investigation of the  fixed point equations we chose to restrict to $\phi\mapsto-\phi$ symmetric solutions, which thus provides two boundary conditions $V'_*(0)=K'_*(0)=0$. Power-law cutoff provides us with an extra scaling symmetry \eqref{scalings} which allows us to set a third condition; as we saw, it is convenient to set $K_*(0)=2$. This still leaves us with two parameters which we are free to take as $\eta$ and $V''_*(0)$, \cf sec. \ref{sec:numerical}. Although the equations have no fixed singularities, solutions for given choices of this pair could \textit{a priori} end at finite $\phi$ in a moveable singularity. However in the examples of $\eta$ that we chose, we did not find this restriction. We have confirmed numerically that solutions exist for $\eta=0$ and a range of $V''_*(0)$. Two example solutions with $\eta\ne0$ are displayed in fig. \ref{fig:confexsols}:  $\eta=-0.3, V''_*(0)=-2$ and $\eta=0.7, V''_*(0)=0.5$.  As in the $\eta=0$ LPA case analysed in sec. \ref{sec:confLPA}, we find in all cases that the potential is bounded below and is such that the quantum corrections, the right hand sides of \eqref{equ:sysfp}, cannot be neglected and in fact become ever more important for larger $\phi$. This numerical insight gives us the condition \eqref{limitv} that allows us to solve analytically for the asymptotic behaviour \eqref{solsasy} that applies to these solutions, in terms of two new functions $w=1/u^4$ and $v$, where $u$ and $v$ are defined in \eqref{uv}. As can be seen from fig. \ref{fig:solsasy} the numerical solutions match well onto this asymptotic behaviour and thus we can confirm that these solutions exist for all real $\phi$. For fixed $\eta$ such as the choices above, the asymptotic behaviour has four free parameters ($B_0,B_1,B_2$ and $B_3$) and thus imposes no further constraints on the solution. Thus from the counting arguments developed in refs. \cite{Morris:1994ie,Morris:1994jc,DietzMorris:2013-1} we would expect that the examples displayed in fig. \ref{fig:confexsols} are part of a continuous two-dimensional set of solutions. By varying $V''_*(0)$ and $\eta$ numerically and integrating out to large $\phi$ where we can again match into \eqref{solsasy}, we have confirmed that this is the case. 

We see that in moving from the LPA $\eta=0$ case in sec. \ref{sec:confLPA} to the full $\mathcal{O}(\partial^2)$ equations, the space of $\phi\mapsto-\phi$ symmetric fixed point solutions has gone from one-dimensional to two-dimensional reflecting the extra freedom to choose $\eta$. This should however be contrasted with the situation in scalar field theory \cite{Morris:1994ie,Morris:1994jc,Morris:1997xj,Morris:1998} where the extra scaling symmetry \eqref{scalings} afforded by a power-law cutoff, and constraints from the asymptotic behaviour, over-constrain the equations resulting in  $\eta$ (and $V''_*(0)$) taking quantized values. 

We saw that the LPA $\eta=0$ case also had fixed points corresponding to translating the minimum of $V_*(\phi)$ away from the origin, equivalently relaxing the condition $V_*'(0)=0$. Although we did not investigate it in this paper, it would be interesting to explore whether at $\mathcal{O}(\partial^2)$ there are in fact further fixed point solutions obtained after relaxing the conditions $V_*'(0)=K_*'(0)=0$. In (single-component) scalar field theory,  it turns out these latter conditions come for free once $\eta\ne0$, in the sense that then there are no known examples for fixed points where these conditions are violated. However we have already uncovered numerous differences with scalar field theory. 
 In fact, depending on $f(\phi)$, \ie the way in which the conformal factor is parametrised, the range of $\phi$ could be naturally restricted, for example to $\phi\ge0$ \cite{Dietz:2015owa}. This would be the case for example if we choose $f(\phi)=\phi^2$. Since in restricting the range of $\phi$, we lose no solutions, but have less opportunity for encountering moveable singularities, we might expect thus to find yet further fixed point solutions.

Given the mapping of the background independent equations to scalar field theory with a wrong sign kinetic term, stability in Minkowski space signature would appear to require the potential should be bounded \emph{above}. The flow equations themselves are subject to a weaker constraint that sets an upper bound on $V''(\phi)$, \cf eqn. \eqref{convboundfull}. In fact the non-perturbative solutions we found for the fixed point potentials, while still satisfying \eqref{convboundfull}, are bounded below and unbounded above \cf figs. \ref{fig:Potentials} and \ref{fig:confexsols}. At first sight this means that all the solutions we found lead to dynamical instability in Minkowski space. However this property should be determined from the physical potential and kinetic term, that is the unscaled background dependent objects in eqn. \eqref{equ:chvars}, and then only after the functional integral has been performed completely by taking the limit $k\to0$. In general this requires adding relevant perturbations, computing the full evolution as $k\to0$, and then studying stability of the result for ranges of the corresponding couplings. Even if we stick to the fixed point values, the $k\to0$ limit will result in physical potentials that either diverge ($\eta>16/(2-q)$), vanish ($0\le\eta<16/(2-q)$) or become independent of the physical conformal factor $\phi$ ($\eta<0$). (Interestingly only for the excluded point $\eta=16/(2-q)$ does the physical potential tend to a non-trivial finite limit.) Therefore even for the fixed points themselves, further analysis is required, which we do not describe further here. 

As we saw in sec. \ref{sec:asymptotics} there is in fact one interval $\eta \in \mathcal{R}=[\eta_c,\frac{16}{2-q})\approx[5.7003, 6.113)$ where the asymptotic behaviour does provide sufficient constraints to lead to such quantization. Furthermore, the single point $\eta=\frac{16}{2-q}$, where the cubics involved in the asymptotic analysis degenerate, was not analysed further in this paper. It is also intriguing that, as we saw in sec. \ref{sec:eigen-asymptotics}, there is an even smaller region $\qR=(\eta_q,\frac{16}{2-q})$, with $\eta_q\approx5.916$, where the eigenoperator spectrum may be fully quantized. However over such a small interval $\bar{\mathcal{R}}=[\eta_c,\frac{16}{2-q}]$, it seems unlikely that there are in fact fixed point solutions in this range. Even if such solutions exist, we have no dynamical principle for excluding the fixed points from the continuous set. 

Within the continuous set, we saw in sec. \ref{sec:eigen-asymptotics} that even amongst the legitimate linearised\footnote{\ie perturbations that stay sufficiently small to justify the linearisation step even for large $\phi$} eigenoperators, although we can expect a quantized spectrum over some ranges of $\lambda$, there is also a continuous spectrum of eigenoperators that covers at the least the range shown in fig. \ref{fig:continuous-spectrum} and in particular therefore includes a continuum of relevant directions at each such fixed point.

Our analysis has assumed no restriction on the space spanned by the eigenoperators other than that which comes out naturally. Thus in scalar field theory one finds that under any flow towards the infrared (no matter how small), solutions of the linearised flow equations are driven back into a Hilbert space spanned by the quantized perturbations \cite{Morris:1996nx,Morris:1996xq,Morris:1998,Bridle:2016nsu}. This does not happen here, but we can ask whether one could impose by hand a suitable restriction on the large field behaviour. This could be interpreted as a restriction on the function space arising as part of the definition of quantization \cite{Reuter:2008qx}. However such a restriction would have to be preserved by the full non-linear flow equations and  we note that even a restriction to exponentially decaying perturbations leaves the infinitely many `super-relevant' perturbations such as those in eqn. \eqref{super-relevant}.


Finally, in sec. \ref{sec:truncations} we considered polynomial truncations. By construction, such truncations can only give isolated fixed points with a quantized eigenoperator spectrum. We found indeed that at the Gaussian fixed point, the continuous spectrum is invisible to such an approximation, and beyond this the continuum of fixed points is similarly missing. However sequences of non-trivial fixed points emerge that appear to converge towards the Gaussian fixed point, but which support increasing numbers of relevant directions, and in this sense reflect some of the true situation at the Gaussian fixed point itself.

The picture that we have uncovered, of continuous sets of fixed points supporting both discrete and continuous spectra of eigenoperators, seems at first to be strongly at variance with the asymptotic safety literature where a single fixed point with a handful of relevant directions, typically three, is found (see \eg the reviews \cite{Reuter:2012,Percacci:2011fr,Niedermaier:2006wt,Nagy:2012ef,Litim:2011cp}). However the great majority of this work focusses on the single field approximation and/or polynomial truncations. 
Apart from the exceptions already discussed in the introduction \cite{Reuter:2008qx,Manrique:2009uh},
even when functional truncations are considered, these have utilised the single-field approximation. A space of constant background scalar curvature $R$ (usually a Euclidean four-sphere) is typically chosen, thus deriving a flow for the effective Lagrangian $f(R)$
\cite{Machado:2007,Codello:2008,Benedetti:2012,Demmel:2012ub,Demmel2015a,Demmel:2014fk,Falls:2014tra,Demmel:2014hla,Demmel2015b,Ohta:2015efa,Percacci:2015wwa,Ohta2016,Eichhorn:2015bna}, written in dimensionless variables. 

Note that the $R\to\infty$ regime corresponds to  fixed physical curvature and $k\to0$. This latter limit must exist because it must be possible to remove the infrared cutoff which was after all a technical device that was inserted by hand. Nevertheless it is unclear what significance should be attached the behaviour of $f(R)$ for $R\gg1$ since  in this case the size of the space is much smaller than the cutoff $1/k$ \cite{Demmel:2014fk,Ohta2016}. In fact in reality all results should be independent of the background, including the background curvature $R$, so the puzzle is actually an artefact of the single-metric approximation (more generally an artefact of the violation of the (modified) split Ward identities). If we 
set this puzzle aside, then the similarities with our own findings become  evident. 

Firstly we have just discussed why this situation would be hard to divine in polynomial approximations. Nevertheless, the very high order polynomial truncations considered in ref. \cite{Falls:2013,Falls:2014tra}, when plotted \cite{DanielTalk}, can be seen to track closely a partial solution to the exact $f(R)$ fixed point equations derived in ref. \cite{Codello:2008}. (An exact solution is impossible since the equations of ref. \cite{Codello:2008} have no global fixed point solutions \cite{DietzMorris:2013-1}.) We therefore see that existence of an asymptotically safe fixed point solution, even in polynomial truncations, is determined ultimately by the underlying functional solution, irrespective of the significance attached to the large $R$ behaviour. 

Secondly, it has been noted that the structure of the solutions is mainly governed by the conformal factor sector \cite{DietzMorris:2013-1,Demmel2015b}. Furthermore it is clear from much recent work \cite{DietzMorris:2013-1,
Demmel2015a,Demmel:2014fk,Falls:2014tra,Demmel:2014hla,Demmel2015b,Ohta:2015efa,Percacci:2015wwa,Ohta2016,Eichhorn:2015bna} 
that the presence of fixed singularities in the fixed point equations, induced by the form of the cutoff functions, plays an important r\^ole in yielding an isolated fixed point. Indeed if the type and magnitude of endomorphisms in the cutoffs are chosen carefully, sufficient numbers of fixed singularities can be arranged to ensure \cite{DietzMorris:2013-1} that only discrete fixed points are allowed \cite{Demmel2015b,Ohta2016}.
However the very freedom that exists in how and where almost all of these are introduced, 
suggests that these fixed singularites are unphysical artifacts and should be eliminated wherever possible.\footnote{See also refs. \cite{Benedetti:2012,DietzMorris:2013-1,Benedetti:2013jk}. For $f(R)$-type approximations with cutoffs of ``type I'' \cite{Codello:2008} the singularity at $R=0$ cannot be moved or eliminated and is there for a clear physical reason \cite{Benedetti:2012,DietzMorris:2013-1}.\label{R0 singularity}} 

We  already know that when these singularities are sufficiently eliminated \cite{Benedetti:2012}, the $f(R)$ approximation yields qualitatively the same conclusions as in this paper, in that it yields a continuum of fixed points supporting continuous spectra of eigenoperators \cite{DietzMorris:2013-1}. It is the lack of constraints from the large field behaviour that is ultimately responsible for this, 
as discussed in sec. \ref{sec:LPA}. For the $f(R)$ approximation, just as for the $\mathcal{O}(\partial^2)$ equations in \ref{sec:fpbeyondLPA},
it is enabled by the non-decoupling of the quantum part (the right hand side)  in the asymptotic expansion. And in fact we already noted in ref. \cite{DietzMorris:2013-1} that it is precisely the quantum fluctuations of the conformal factor part (called there ``the physical scalar'') that are responsible, leading us to tentatively suggest that these effects are a reflection of the conformal mode instability.  

Therefore, far from being at variance with the literature, we see that the properties we have found for fixed points in conformally reduced gravity have an analogue in the conformal factor sector  of $f(R)$ truncations, and in this way 
underlie the evidence for asymptotic safety that has been reported up to now. Indeed we see that evidence for asymptotic safety arises from the underlying continuum of solutions caused by this sector which are then constrained by the fixed singularities induced through choices of cutoff.


Now we address to what extent these conclusions could change in other approaches to asymptotic safety or through extending the approach of refs. \cite{Dietz:2015owa,Labus:2016lkh}. We have already noted in sec. \ref{sec:sft} that, following from the assumption of analyticity, we would be led to `Wick rotate' the conformal factor field $\phi=i\phi^{(s)}$, just as Gibbons, Hawking and Perry proposed \cite{Gibbons:1978ac}, and in so doing turn our background independent flow equations and effective action into precisely the flow equations and effective action for a real scalar field $\phi^{(s)}$. In this way we of course recover the Hilbert space structure of a complete orthonormal discrete set of eigenoperators \cite{Morris:1996nx,Bridle:2016nsu,Morris:1996xq,Morris:1998}, as we saw explicitly in sec. \ref{sec:gaussian}. However there is then no asymptotic safety, since in $d=4$ dimensions only the Gaussian fixed point exists (see \eg \cite{Wilson:1973,Morris:1994ki}). In view of this, it seems important to establish whether similar conclusions can be drawn for such a `Wick rotation' not just for the conformal factor on its own but also in the context of full quantum gravity. 

Although we have argued that topology change could in principle cure the problem \cite{DietzMorris:2013-1}, see also \cite{Demmel:2014fk}, and we further blamed the effect on a breakdown of the $f(R)$ approximation  \cite{DietzMorris:2013-2} and on the single field approximation \cite{Bridle:2013sra}, the latter two drawbacks are absent now, while it is no longer clear in this setting how topology change can be admitted. 

In ref. \cite{Benedetti:2013jk} it was shown that in the $f(R)$ approximation on a maximally symmetric space, if an $f(R)$-independent cutoff is used, then  about a fixed point with the expected properties, the spectrum is discrete with a finite number of relevant directions. However the conformal factor, called there the gauge-invariant trace mode $\bar{h}$, is treated differently. In order to compare, working in dimensionless variables, we take the limit of large mode number $n$ on a 4-sphere, where the eigenvalues of the scalar Laplacian are $\lambda_{n,0}\sim n^2 R/12$ \cite{Benedetti:2012}. Keeping ${p}^2:=\lambda_{n,0}$ finite, we thus have $R\to0$, so that the Hessian goes over to the one for flat space:
\be 
\label{barh-Hessian}
9 f''(0)\, {p}^4 + 3 f'(0)\, {p}^2 + 2 f(0)\,.
\ee
It is assumed that solutions exist such that $f''(R)$, and thus in particular $f''(0)$, is positive.  Although we are here dealing with fixed point values, on the full renormalised trajectory for sufficiently small cutoff $k$,  we must have $f'(0)$ large and negative, since this is required for small positive Newton's constant. This is where the conformal factor instability lies since it implies that the Hessian will be negative in some domain. As we have seen in our simpler setting, a negative $O(p^2)$ piece leads to a continuum of solutions. In  the other works on the $f(R)$ approximation \cite{DietzMorris:2013-1,Machado:2007,Codello:2008,Benedetti:2012,Demmel:2012ub,Demmel2015a,Demmel:2014fk,Falls:2014tra,Demmel:2014hla,Demmel2015b,Ohta:2015efa,Percacci:2015wwa,Ohta2016,Eichhorn:2015bna} an adaptive cutoff is used which in particular allows the sign to adapt to the sign of the Hessian. 
However in ref. \cite{Benedetti:2013jk}, adaptive cutoff functions were eschewed (as here), precisely to avoid the issues with fixed singularities discussed above, and also to avoid dependence on $f'''(R)$. Instead a cutoff profile $16 c_{\bar{h}} r(p^2)$ is added to \eqref{barh-Hessian}. If the free parameter $c_{\bar{h}}>0$ is chosen large enough we can then ensure that the regularised inverse Hessian is everywhere well defined as required. However it is not known whether a suitable asymptotically safe fixed point solution exists with these choices.

%
%

We have worked only at the LPA and $\mathcal{O}(\partial^2)$, equivalently $\mathcal{O}(p^2)$, levels. Including higher derivatives, for example already at $\mathcal{O}(\partial^4)$ as just discussed, could with suitable parametrisation provide sufficient stability, and therefore it is important to understand the implications for asymptotic safety in this case. Also at this level we would have to take into account the Weyl anomaly \cite{Capper1974b,Duff1977,Duff1994,Machado:2009ph,Codello2013}. 
Working with the full metric while respecting the Ward identities for background independence and diffeomorphism invariance, might also qualitatively alter the results.

\section*{Acknowledgments}
It is a pleasure to thank Maximilian Demmel, Astrid  Eichhorn,  Frank Saueressig, Omar Zanusso and especially Kevin Falls, for discussions on issues connected to this paper. TRM acknowledges support from STFC through Consolidated Grant ST/L000296/1.




\bibliographystyle{hunsrt}
\bibliography{references} 

\end{document}